\documentclass[12pt]{article}
\usepackage{epsfig,amssymb,amsmath,psfrag,subfigure}
\usepackage{color,colordvi}
\usepackage{rotating}
\usepackage{ulem}

\catcode`\@=11
\def \tr {\mathop{\rm tr}\nolimits}

\def \e  {\mathop{\rm e}\nolimits}
\newcommand\lr[1]{{\left({#1}\right)}}
\newcommand \widebar [1] {\overline{#1}}

\newcommand \ket [1] {|{#1}\rangle}
\newcommand \bra [1] {\langle {#1}|}
\newcommand\re[1]{(\ref{#1})}

\def \qqqquad {\qquad\qquad}

\newcommand{\cN}{{\cal N}}

\newcommand{\cO}{{\cal O}}
\newcommand{\cY}{{\cal Y}}

\newcommand{\nt}{\notag\\} 

\newcommand{\ep}{\epsilon}
\renewcommand{\a}{\alpha}
\renewcommand{\b}{\beta}

 \newcommand{\kernel}{K}
\newcommand{\shifta}{t_1}
\newcommand{\shiftb}{t_2}

\newcommand{\p}[1]{(\ref{#1})}

\newcommand \vev [1] {\langle{#1}\rangle}
\newcommand{\ft}[2]{{\textstyle\frac{#1}{#2}}}

\def\numberbysection{\@addtoreset{equation}{section}
                     \def\theequation{\thesection.\arabic{equation}}}

\numberbysection

\makeatletter
\def\timenow{\@tempcnta\time
  \@tempcntb\@tempcnta
  \divide\@tempcntb60
  \ifnum10>\@tempcntb0\fi\number\@tempcntb
  \multiply\@tempcntb60
  \advance\@tempcnta-\@tempcntb
  :\ifnum10>\@tempcnta0\fi\number\@tempcnta}
\makeatother

\begin{document}

\begin{titlepage}
 
\thispagestyle{empty}

\null\vskip-43pt \hfill
\begin{minipage}[t]{45mm}
CERN-PH-TH/2013-212\\
IPhT--T13--211\\
LAPTH-048/13
\end{minipage}

\vspace*{1cm}

\centerline{\large \bf  Event shapes in $\mathcal N=4$ super-Yang-Mills theory} 

\vspace*{1cm}

\centerline{\sc A.V.~Belitsky$^{a,b}$,  S.~Hohenegger$^c$, G.P.~Korchemsky$^b$, E.~Sokatchev$^{c,d,e}$, A.~Zhiboedov$^f$}

\vspace{5mm}

\centerline{\it $^a$Department of Physics, Arizona State University}
\centerline{\it Tempe, AZ 85287-1504, USA}

\vspace{3mm}

\centerline{\it $^b$Institut de Physique Th\'eorique\footnote{Unit\'e de Recherche Associ\'ee au CNRS URA 2306}, CEA Saclay}
\centerline{\it 91191 Gif-sur-Yvette Cedex, France}

\vspace{3mm}
\centerline{\it $^c$Physics Department, Theory Unit, CERN}
\centerline{\it CH -1211, Geneva 23, Switzerland}

\vspace{3mm}
\centerline{\it $^d$Institut Universitaire de France}
\centerline{\it  103, bd Saint-Michel
F-75005 Paris, France }

\vspace{3mm}
\centerline{\it $^e$LAPTH\,\footnote[2]{UMR 5108 du CNRS, associ\'ee \`a l'Universit\'e de Savoie},   Universit\'{e} de Savoie, CNRS}
\centerline{\it  B.P. 110,  F-74941 Annecy-le-Vieux, France}

\vspace{3mm}
\centerline{\it $^f$Department of Physics, Princeton University}
\centerline{\it Princeton, NJ 08544, USA}

\vspace{1cm}

\centerline{\bf Abstract}

\vspace{5mm}

We study event shapes in $\mathcal{N}=4$ SYM describing the angular distribution of energy and $R-$charge in the final states created by  the simplest half-BPS scalar 
operator. Applying the approach developed in the companion paper arXiv:1309.0769, we compute these observables using correlation functions of certain 
components  of the $\mathcal{N}=4$ stress-tensor supermultiplet: the half-BPS operator itself, the $R-$symmetry current and the stress tensor.  We present master formulas 
for the all-order event shapes as convolutions of the Mellin amplitude defining the correlation function of the half-BPS operators, with a coupling-independent 
kernel determined by the choice of the observable.  We find remarkably simple relations  between  various event shapes following from  $\mathcal{N}=4$ superconformal 
symmetry.  We perform thorough checks at leading order in the weak coupling expansion and show perfect agreement with the conventional calculations based on  amplitude 
techniques. We extend our results to strong coupling using the  correlation function of half-BPS operators obtained from the AdS/CFT correspondence. 

\newpage
  
\end{titlepage}

\setcounter{footnote} 0

\newpage

\pagestyle{plain}
\setcounter{page} 1

\tableofcontents

\newpage 
 
\section{Introduction}

Recently, significant  progress has been made in understanding properties of correlation functions and scattering amplitudes in
maximally supersymmetric Yang-Mills theory ($\mathcal N=4$ SYM). There is growing evidence that the theory possesses a hidden
integrability symmetry which is powerful enough to determine both quantities for an arbitrary value of the coupling constant, at least in the 
planar limit. Correlation functions and scattering amplitudes have different properties and carry complementary information about the dynamics
of $\mathcal N=4$ SYM. Unlike the correlation functions, the on-shell scattering amplitudes are not well defined in four dimensions
due to infrared (IR) singularities and, hence, they require regularization. This introduces a dependence on unphysical parameters (such as the 
dimensional regularization scale playing the role of the IR regulator) which break (super)conformal symmetry. At the same time, the correlation 
functions of protected (half-BPS) operators are well-defined functions of the coordinates of the  
operators in four-dimensional $\mathcal N=4$ SYM. As a consequence, they do not require regularization and 
enjoy the full unbroken $\mathcal N=4$ superconformal symmetry.

\begin{figure}[h!t]
\psfrag{em}[cc][cc]{$e^+$}  \psfrag{ep}[cc][cc]{$e^-$} \psfrag{g}[cc][cc]{$\gamma^*(q)$}
\centerline{\includegraphics[width = 0.5 \textwidth]{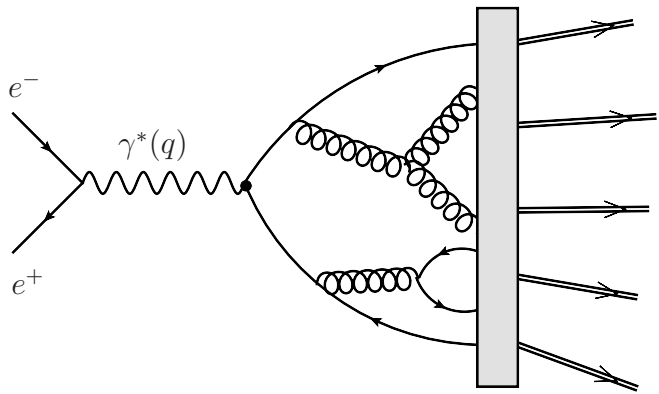}}
\caption{\small Final states in $e^+e^-$ annihilation in QCD. The electron and positron annihilate to produce a virtual photon $\gamma^*(q)$ that decays into an arbitrary 
number of quarks and gluons which go through a hadronization process (shaded rectangle) to become hadrons (double lines). The dot denotes the electromagnetic QCD current.}
\label{fig-ee}
\end{figure}

The main goal of this paper is to study a different class of gauge invariant quantities in $\mathcal N=4$ SYM which admit two equivalent representations:  They 
are given by  integrated correlation functions  and, at the same time, they can be expressed as (infinite) sums over absolute squares of scattering amplitudes. 
These quantities are closely related  to various observables which have been thoroughly studied in the  context of QCD for the final states produced in 
$e^+e^-$annihilation~\cite{Sterman:1977wj,Kunszt:1989km,Biebel:2001dm}. In the latter case, the  electron and positron annihilate to produce a virtual photon, 
which in turn  creates an 
energetic quark-antiquark pair from the vacuum. The outgoing particles move away from each other and emit a lot of radiation before fragmenting into 
hadrons (see Fig.~\ref{fig-ee}). The distribution of particles in the final state of  $e^+ e^-$ annihilation can be characterized by a set of observables, the so-called 
event shapes (see, e.g., \cite{Kunszt:1989km}). One of them, the energy-energy correlation \cite{Basham:1978bw}, plays a distinct role in our analysis. 
 
Needless to say, QCD is quite different from $\mathcal N=4$ SYM. Due to the presence of a mass gap in the hadron spectrum, QCD 
scattering amplitudes are free from IR singularities but their calculation is still impossible due to our inability to control the confining (hadronization) 
regime in the theory. A remarkable property of the event shapes is that, for asymptotically large values of the center-of-mass energy
 $q^2$, the hadronization corrections become negligible (for a review see, e.g., \cite{Dasgupta:2003iq}). As a consequence, the event shapes can be 
 approximated at high energy by a perturbative QCD expansion. It is in this context that $\mathcal N=4$ SYM arises as  a 
simpler model of gauge dynamics in four space-time dimensions. It shares many features with perturbative QCD, on the one hand, 
and can be studied analytically using its symmetries, on the other. In particular, the AdS/CFT correspondence 
opens up a possibility to explore the previously unreachable regime of strong coupling.

To generalize the QCD process shown in Fig.~\ref{fig-ee} to $\mathcal N=4$ SYM, we have to find an appropriate analogue of the QCD electromagnetic  
current. For this purpose we can choose any local protected operator in $\mathcal N=4$ SYM, e.g.,
the half-BPS operator $O_{\bf 20'}(x)$ built from two scalar fields. 
At weak coupling one can think about the state created by this operator as follows. The operator  $O_{\bf 20'}(x)$  produces out of the vacuum
a pair of scalars that propagate into the final state and radiate on-shell particles of $\mathcal N=4$ SYM 
-- scalars, gluinos and gluons. The fact that these particles are massless leads to a degeneracy of the final states. For instance,
a single-particle state is undistinguishable from the state containing an additional gluon with vanishing momentum and from the state
containing a pair of particles with aligned momenta and the same total charge. According to the Kinoshita-Lee-Nauenberg mechanism 
\cite{Kinoshita:1962ur,Lee:1964is}, the degeneracy of on-shell states leads to (soft and collinear) divergences in the perturbative expansion of  
the corresponding scattering amplitudes. 

This phenomenon is quite general when massless particles are present in the spectrum. An important issue in the early days of QCD  was 
whether one could define quantities that are free from infrared divergences at all orders of perturbation theory. 
The answer was found with the introduction of the so-called inclusive infrared safe observables. The latter  are given by a
sum over an infinite number of scattering amplitudes involving an arbitrary number of degenerate states \cite{Sterman:1977wj,Basham:1978bw}. Each amplitude
has infrared divergences but they cancel in the sum, so that infrared safe observables are well defined in four dimensions order-by-order in the coupling. 
The question arises whether there is another way to compute the same observables that bypasses the introduction of any regularization and, therefore, 
makes all symmetries of the theory manifest at each step of the calculation. 

As a simple example, consider the total probability of the transition $O_{\bf 20'}\to\text{everything}$.  This is an infrared safe quantity, but it is given by (an infinite) sum over 
all final states, with each individual term being infrared divergent. The optical theorem allows us to express the same quantity as the imaginary part of the 
two-point (time-ordered) correlation function of the half-BPS operators $O_{\bf 20'}(x)$  (see Eq.~\re{opt} below). This two-point function is well defined in four 
dimensions and its form is uniquely fixed by  $\mathcal N=4$ superconformal symmetry. In this way, we obtain a definite prediction for the total transition 
amplitude $O_{\bf 20'}\to\text{everything}$ that agrees with the result of an explicit calculation \cite{vanNeerven:1985ja} based on amplitudes. 

In this paper, we deal with a special class of event shape distributions related to the flow of various quantum numbers (energy, charge) in the final state.
A typical event contributing to such an observable is shown in Fig.~\ref{fig-EE} below. There, the particles propagate into the final state where the detectors
measure the flow of their quantum numbers per solid angle in the directions indicated by the unit vectors $\vec n,\vec n',\dots$.  
As was shown in Refs. \cite{Sveshnikov:1995vi,Korchemsky:1997sy,Korchemsky:1999kt,BelKorSte01} in the context of QCD, the optical theorem can be generalized to such 
differential distributions. For instance, the energy flow distributions
can be expressed in terms of the correlation functions $\vev{ O_{\bf 20'}(x) \mathcal E(\vec n)\mathcal E(\vec n') \dots   O_{\bf 20'}(0)}$ 
containing  additional energy flow operators $\mathcal E(\vec n), \mathcal E(\vec n'), \dots$  (one for each detector). Quantities of this type have 
been studied in the framework of conformal field theories in \cite{Hofman:2008ar}, 
particularly in connection with the AdS/CFT correspondence. An unusual feature of these correlation functions is that the operators are not time-ordered. In other words, we are 
dealing with correlation functions of the Wightman type defined on a space-time with Lorentzian signature. 
Notice, however, that significant advances have been made in the calculation of their Euclidean counterparts. The natural question arises whether we can make 
use of the latter to compute the weighted cross sections. The answer was presented in the companion paper \cite{PaperI}, where we explained in detail how
to obtain the charge flow correlators in a generic CFT, starting from the Euclidean correlation functions and making   a nontrivial analytic continuation.  We 
briefly review it below to make the exposition self-contained.  
In this paper, we apply the approach of   \cite{PaperI} to the particular case of  $\mathcal N=4$ SYM.

The paper is organized as follows. In Section 2, we consider the process $O_{\bf 20'}\to\text{everything}$ in $\mathcal N=4$ SYM
and introduce a set of infrared safe observables, determined by weighted cross sections, describing the flow of various quantum numbers
into the final state in this process. We also work out a representation for these observables in terms of Wightman correlation functions
involving insertions of flow operators. In Section 3, we consider  weighted cross sections with one or two flow operators
and evaluate them to lowest order in the coupling using the conventional amplitude techniques.
In Sections 4 and 5, we elaborate on the main result of this work and explain how the same observables can be obtained
from the known results for Euclidean correlation functions of half-BPS and other operators, such as the $R-$current and the energy-momentum tensor, in the same $\mathcal N=4$ supermultiplet. In particular, we derive a master formula which yields an all-loop
result for the weighted cross sections as a convolution of the Mellin amplitude defined by the Euclidean correlation function with a coupling-independent 
`detector kernel' corresponding to the choice of the flow operators. In Section 6, we demonstrate the efficiency 
of the formalism making use of the same examples as covered in Section 3, first at weak and then at strong coupling. Section 7 contains
concluding remarks.  Several technical details are deferred to appendices. 

\section{Correlations in $\mathcal N=4$ SYM}\label{sect2}

In the context of $\mathcal N=4$ SYM, we can introduce an analog of the quark electromagnetic 
current, the lowest-dimension half-BPS Hermitian operator  $O_{\bf 20'}^{IJ}(x)$ built from the six real
scalars $\Phi^I(x)$ (with  $SO(6)$ vector indices $I,J=1,\dots,6$),  
\begin{align}\label{O20}
 O_{\bf 20'}^{IJ}(x) =\tr \big[\Phi^I  \Phi^J  - \ft16 
 \delta^{IJ} \Phi^K \Phi^K \big] \,. 
\end{align}
Here $\Phi^I\equiv \Phi^{I a} T^a$ and  the  generators $T^a$ of the 
gauge group $SU(N_c)$ are normalized as
$\tr[T^a T^b]=\frac12 \delta^{ab}$ (with $a,b=1,\dots,N^2_c-1$). The operator \re{O20} possesses a protected scaling dimension, $\Delta=2$, very much like the QCD
electromagnetic  current. Moreover, it is the lowest-weight state of the $\cN=4$ stress-tensor supermultiplet and is related by supersymmetry  to the $R-$symmetry current, 
which can be viewed as a `cousin' of the electromagnetic current.

The operator \p{O20} 
belongs to the {real} irrep $\bf{20'}$ of the $R-$symmetry group $SO(6)\sim SU(4)$. To keep track of the isotopic structure, it is convenient to consider the projected operator
\begin{align} \label{OY}
O(x,Y)= Y^I Y^J O_{\bf 20'}^{IJ}(x)  =Y^I Y^J \tr [\Phi^I(x) \Phi^J (x)] \,,
\end{align}
where $Y^I$ is an auxiliary six-dimensional (complex) null vector, $Y^2=\sum_{I=1}^6 Y^I Y^I =0$, defining the 
orientation of the operator  in the isotopic space.~%
\footnote {We can always reveal the index structure of the $SO(6)$ tensor $O_{\bf 20'}^{IJ}(x)$ by differentiating 
the final expressions involving $O(x,Y)$ with respect to the $Y$'s,  bearing in mind the restriction $Y^2=0$.}   

Next, we can ask the question about the properties of the final states created by the operator \re{OY} from the vacuum. To lowest order in the coupling, the final state consists of  
a pair of scalars. For arbitrary coupling, the state $\int d^4 x \e^{iqx} O(x,Y)\ket{0}$ can be decomposed into 
an infinite tower of asymptotic on-shell  states, $ \ket{\rm ss}$, $\ket{\rm ssg}$, $ \ket{\rm s\lambda\lambda}$, $\dots$  involving an arbitrary number of scalars 
(${\rm s}$), gluinos ($\lambda$) 
and gauge fields (${\rm g}$). Each on-shell state  carries the same quantum numbers -- the total momentum $q^\mu$, zero (color) $SU(N_c)$ charge  and 
$R-$charges of the irrep $\bf 20'$.
We can then define the amplitude for creation of a particular final state $\ket{X}$ out of the vacuum,
\begin{align}
 \vev{X| \int d^4 x \e^{iqx} O(x,Y)|0} =  (2\pi)^4  \delta^{(4)}(q-k_X)\mathcal M_{O_{\bf 20'}\to X}\,,
\end{align}
where $k_X$ is the total momentum of the state $\ket{X}$. Defined in this fashion, the amplitude $\mathcal M_{O\to X}$ has the meaning of a form-factor, 
\begin{align}\label{M}
 \mathcal M_{O_{\bf 20'}\to X}=\vev{X|  O(0,Y)|0} \,.
\end{align}
For a given on-shell state $\ket{X}$, it suffers from IR divergences that require a regularization procedure. In addition, this quantity depends on the 
number of colors  $N_c$ and on the coupling 
constant $g_{\rm \scriptstyle YM}$. For our purposes it proves convenient to introduce the 't Hooft coupling $g^2=g_{\rm \scriptstyle YM}^2 N_c$ and the analog of 
the fine structure constant, $a=g_{\rm \scriptstyle YM}^2 N_c/(4\pi^2)$, familiar from QCD.

\subsection{Total  transition probability}

\begin{figure}[h!t]
\psfrag{q}[cc][cc]{$q$}  
\psfrag{k1}[cc][cc]{$k$}  
\psfrag{k2}[cc][cc]{$q-k$}  
\centerline{\includegraphics[width=.25\textwidth]{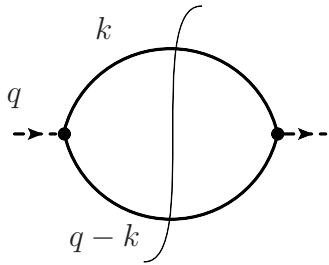}} 
\caption{\small The Feynman diagram contributing to $\sigma_{\rm tot}(q)$. The thin line stands for the unitarity cut.}
\label{fig:Born}
\end{figure}

In close analogy with the QCD process $e^+e^-\to\text{everything}$, we can examine the transition $O_{\bf 20'}\to\text{everything}$. The total probability
of this process is given by the sum over all final states  
\begin{align}\label{tot}
\sigma_{\rm tot}(q) = \sum_X (2\pi)^4 \delta^{(4)}(q-k_X)|\mathcal M_{O_{\bf 20'}\to X}|^2\,,
\end{align}
where the summation runs over the quantum numbers of the produced particles including their helicity, color, $SU(4)$ indices, etc.
To lowest order in the coupling, it describes the production of a pair of scalars as shown in Fig.~\ref{fig:Born}~\footnote{We use the 
Minkowski signature $(+,-,-,-)$ and the shorthand notation $\delta_+(k^2) = \delta(k^2)\theta(k^0)$. }, 
\begin{align}\label{tot-1}
\sigma_{\rm tot}(q) = \frac12 (N^2_c-1) (Y \widebar Y)^2\int {d^4 k\over (2\pi)^4} (2\pi)^2 \delta_+(k^2)\delta_+((q-k)^2) + \dots \,,
\end{align}
where the ellipsis stand for omitted higher order corrections and the integration in the first term goes over the phase space of the two massless particles carrying the total momentum 
$q^\mu$. The prefactor accompanying the integral is the $SO(6)$ invariant contraction of the auxiliary internal variables $(Y \widebar Y)=\sum_I Y^I \widebar Y^I$. 

Using the completeness condition on the asymptotic states, $\sum_X \ket{X}\bra{X}=1$,  we can rewrite \re{tot} as 
\begin{align}\notag 
\sigma_{\rm tot}(q) & = \int d^4 x\, \e^{iqx}  \sum_X   \vev{0|  O(0, \widebar Y)|X}  \e^{-ix k_X} 
\vev{X|  O(0, Y)|0} 
 \\
 & = \int d^4 x\, \e^{iqx}   \vev{0|  O(x,\widebar Y)  O(0, Y)|0} \,, \label{2.8}
\end{align}
where $ O(x,\widebar Y) = O_{\bf 20'}^{IJ} \widebar Y^I \widebar Y^J= \lr{O(x,  Y) }^\dagger$ (we recall that the operator $ O_{\bf 20'}^{IJ}$ is Hermitian). Notice that the operators in \p{2.8} are not time-ordered. 

The optical theorem allows us to rewrite \re{2.8}
as the imaginary part of the time-ordered correlation function
\begin{align}\label{opt}
\sigma_{\rm tot}(q)  & =  {\rm Im}\left[2 i \int d^4 x\, \e^{iqx}   \vev{0| {T}\,  O(x,\widebar Y)  O(0, Y)|0} \right].
\end{align}
The correlation function on the right-hand side is well defined in four dimensions and the same is true
for $\sigma_{\rm tot}(q)$. This is not the case however for each term on the right-hand side of \re{tot} 
since $\mathcal M_{O_{\bf 20'}\to X}$ suffers from IR divergences. In agreement with the
Lee-Nauenberg-Kinoshita theorem \cite{{Kinoshita:1962ur,Lee:1964is}},  the infrared finiteness of $\sigma_{\rm tot}(q)$
is restored in the infinite sum over the final states $\ket{X}$ in \re{tot}.

Another advantage of the representation \re{opt} is that it allows us to compute $\sigma_{\rm tot}(q)$ exactly, to all orders in the coupling. Indeed, in ${\cal N}=4$ SYM the two-point correlation function of the half-BPS operators $O(x,Y)$
is protected from loop corrections\footnote{See, e.g., \cite{Howe:1996rb,D'Hoker:1998tz,Howe:1998zi}
for non-renormalization theorems and Refs.\ \cite{Lee:1998bxa} and \cite{Penati:1999ba} for explicit one- and two-loop perturbative tests, respectively.} and is given by the Born level expression
\begin{align}\label{2.10}
\vev{0| {T}\,  O(x,\widebar Y)  O(0, Y)|0} = \frac12 (N^2_c-1) (Y \widebar Y)^2[D_F(x)]^2\,,
\end{align}
where  $D_F(x)=1/(4\pi^2(-x^2+i0))$ is the Feynman propagator of the scalar field. Its substitution into \re{opt} yields the leading tree-level
term in \re{tot-1}. Performing the integration in \re{tot-1} we arrive at 
\begin{align}\label{tot-ex} 
\sigma_{\rm tot}(q) 
=\frac1{16\pi} (N^2_c-1) (Y \widebar Y)^2\theta(q^0)\theta(q^2)\,.
\end{align}
The fact that $\sigma_{\rm tot}(q)$ does not depend on the coupling constant implies that the perturbative corrections cancel order by order in $\mathcal N=4$ SYM. To two loop accuracy, this 
property has been verified in Ref.~\cite{Penati:1999ba} by an explicit calculation. The product of the two step functions on the right-hand side of \re{tot-ex} ensures that the cross section is different 
from zero for the total momentum  $q^0>0$ and $q^2>0$. In what follows we tacitly assume that this condition is satisfied and we do not display the step functions in any formulas that follow.

\subsection{Weighted cross section}\label{sect-wcs}

The quantity \re{tot} is completely inclusive  with respect to the final states. We can define a less inclusive quantity by
assigning a weight factor $w(X)$ to the contribution of each state $\ket{X}$
\begin{align}\notag\label{wcs}
\sigma_{W}(q) &= \sigma_{\rm tot}^{-1}\sum_X (2\pi)^4 \delta^{(4)}(q-k_X)w(X)|\mathcal M_{O_{\bf 20'}\to X}|^2 
\\
 & = \sigma_{\rm tot} ^{-1} \int d^4 x\, \e^{iqx}  \sum_X   \vev{0|  O(x,\widebar Y)|X}  w(X)
\vev{X|  O(0, Y)|0} \,,
\end{align}
where the additional factor of $1/\sigma_{\rm tot}$ is inserted to obtain the normalization condition $\sigma_{W}(q)=1$ for
$w(X)=1$.  Appropriately choosing the weight factors $w(X)$ and evaluating the
corresponding weighted cross section $\sigma_{W}(q)$, we can get a more detailed description of the flow of   various
quantum numbers of particles (energy, charge, etc) in the final state $\ket{X}$. 

For a generic final state $\ket{X}$, the scattering amplitude $\mathcal M_{O_{\bf 20'}\to X}$ contains soft and collinear divergences 
as we reviewed in the introduction. They arise from the integration over the loop momenta of virtual particles and appear as poles in $\epsilon$ 
in dimensional regularization with $D=4-2\epsilon$. Taken by itself,   
each term in the sum in the first relation in \re{wcs} vanishes as $|\mathcal M_{O_{\bf 20'}\to X}|^2\sim \e^{-f(g^2)/\epsilon^2}
\to 0$ for $\epsilon\to 0$ (with a positive-definite function $f(g^2)$ related to the cusp anomalous dimension), due to the exponentiation 
of infrared singularities (see, e.g., Ref.\ \cite{Collins:1989bt}). 
However, these are not the only divergences that we encounter in the calculation of the cross section. Namely,  additional poles in $1/\epsilon$
come from the integration over the phase space of soft and collinear massless particles in the final state $\ket{X}$. For the cross section to be IR finite, 
the two effects, i.e., virtual and real singularities should cancel each other, thus producing a finite net result. In the case of the  total transition probability $\sigma_{\rm tot}$,
the cancellation of IR divergences follows from the Kinoshita-Lee-Nauenberg theorem. For weighted cross sections, the 
condition of infrared finiteness imposes a severe restriction on the weights $w (X)$ \cite{Kunszt:1992tn}. Namely, the weight should be insensitive to 
adding  one particle to the final state $\ket{X}$, with the momentum either soft, or collinear to the momenta of  the parent
particles in the state $\ket{X}$.
 
\subsubsection{Energy flow}

One of the well-known examples of a weight factor, which was introduced in the context of $e^+e^--$
annihilation and which is very useful for our purposes, corresponds to the energy flow. 
For a given final on-shell state $\ket{X} = \ket{k_1,\dots,k_\ell}$, consisting of $\ell$ massless particles, $k_i^2=0$, with 
the total momentum $\sum_i k_i^\mu = q^\mu$, it is defined in the rest frame $q^\mu=(q^0,\vec 0)$ as 
\begin{align}\label{w-energy}
w_{\mathcal E}(k_1,\dots,k_\ell) = \sum_{i=1}^\ell k_i^0 \,\delta^{(2)}(\Omega_{\vec k_i} - \Omega_{\vec n}) \,,
\end{align}
where $k_i^\mu=(k_i^0,\vec k_i)$  and $\Omega_{\vec k_i}=\vec k_i/|\vec k_i|$ is the solid angle in the direction of $\vec k_i$.
The corresponding weighted cross section has a simple physical meaning -- it measures the distribution of energy in the final
state  that flows in the direction of the vector $\vec n$. Most importantly, the  
weight  \re{w-energy} can be identified with the eigenvalue of the energy flow operator,
\begin{align}\label{E-spectr}
\mathcal E(\vec n) \ket{X}=w_{\mathcal E}(X)   \ket{X}\,.
\end{align}
As we show below, this relation allows us to simplify \re{wcs}  along the same lines as \re{2.8} and to express the cross section $\sigma_{\mathcal E}$ in terms of the 
correlation function $\vev{0|O(x,\widebar Y) \mathcal E(\vec n)  O(0, Y)|0}$ with an insertion
of the energy flow operator.

The explicit expression for the operator $\mathcal E(\vec n)$ is given in terms of
the energy-momentum tensor  of $\mathcal N=4$ SYM \cite{Sveshnikov:1995vi,Korchemsky:1997sy,Korchemsky:1999kt,BelKorSte01} (see also \cite{Hofman:2008ar}) 
\begin{align} \label{E-flow}
\mathcal E(\vec n) =  \int_0^\infty dt \, \lim_{r\to\infty} r^2\, n^i\, T_{0 i}(t,r\vec n)\,,
\end{align}
where the unit vector $\vec n=(n^1,n^2,n^3)$ (with $\vec n^2=1$)   indicates the {spatial} direction of the energy flow. 
To get a better understanding 
of the action of the operator $\mathcal E(\vec n) $ on the asymptotic states, we replace
the energy-momentum tensor $T_{\mu\nu}(x)$  in \re{E-flow}  by its expression in terms of free fields and obtain the following representation
for $\mathcal E(\vec n)$ in terms of creation and annihilation operators: 
\begin{align}\label{E-osc}
 \mathcal E(\vec n) =\int {d^4 k\over (2\pi)^4} 2\pi \delta_+(k^2)  \, k_0 \, 
\delta^{(2)}(\Omega_{\vec n}-\Omega_{\vec k})  { \sum_{p={\rm s,\lambda,g}} {a}_p^{b\,\dagger}(k) a_p^b(k) }\,,
\end{align} 
where the sum goes over all on-shell states (scalars, helicity $(\pm 1/2)$ gluinos and helicity $(\pm 1)$ gluons)
carrying an $SU(N_c)$ index $b=1,\ldots,N^2_c-1$. To simplify the formulas,  in what follows we do not display the $SU(N_c)$ indices of the creation/annihilation
operators.
Making use of the (anti)commutation relations between $a_i^\dagger(k)$ and $a_i(k)$ (see, e.g., Eq.\ \p{A3}), it is 
straightforward to verify the relations \re{E-spectr} and \re{w-energy}, as well as the commutativity condition
\begin{align}\label{E-com}
 [  \mathcal E(\vec n) ,  \mathcal E(\vec n') ] = 0\,,\qquad  {\rm for} \ \vec n \neq \vec n'\,. 
\end{align}
The latter equality states that the energy flows in two different directions $\vec n$ and $\vec n'$ are independent from each other
and can be measured separately. 

Making use of \re{E-com}, we can define a weight 
which measures the energy flows in various directions
$\vec n_1,\dots,\vec n_\ell$ simultaneously: 
\begin{align}
 \mathcal E(\vec n_1)\dots  \mathcal E(\vec n_\ell)\ket{X}=w_{\mathcal E(\vec n_1)}(X)  \dots w_{\mathcal E(\vec n_\ell)}(X) \ket{X}
 \equiv w(X) \ket{X}\,.
\end{align}
Substituting this relation into \re{wcs} we can apply the completeness relation $\sum_X \ket{X} \bra{X}=1$ and obtain the following representation of 
the corresponding weighted cross section\footnote{Below we show that dividing by $\sigma_{\rm tot}$, Eq.~\p{tot-ex}, the $Y-$dependence drops out from \re{sigma-E}.}  
\begin{align}\notag\label{sigma-E}
\vev{\mathcal E(\vec n_1)\dots \mathcal E(\vec n_\ell)}  & \equiv    \sigma_{\mathcal E}(q;\vec n_1,\dots,\vec n_\ell) 
\\
&= \sigma_{\rm tot}^{-1} \int d^4 x\, \e^{iqx}  \vev{0|  O(x,\widebar Y) \, \mathcal E(\vec n_1)\dots  \mathcal E(\vec n_\ell)\, O(0, Y)|0} \,,
\end{align}
which has the meaning of an energy flow correlation. Notice that the product of operators on the right-hand side 
of \re{sigma-E} is not time-ordered and, therefore, their correlation function is of the Wightman type.

\subsubsection{Charge flow}
\label{charge}

In close analogy with \re{w-energy} we can define a weight that measures the flow of  the $R-$charges through the detector.
We recall that in $\mathcal N=4$ SYM only the scalars and gluinos are charged with respect to the $R-$symmetry group $SU(4)$.
The flow of the $R-$charge is defined by the operator 
\begin{align}\label{Q-flow}
\mathcal  Q^B_{A}(\vec n)   & =  \int_0^\infty dt \, \lim_{r\to\infty} r^2 \, (J_0)^B_{A} (t,r\vec n)  \,, 
\end{align}
 involving the time component of the $R-$current $(J_\mu)^B_{A}(x)$. An important difference as compared with \re{E-flow}
is that  the operator transforms under  $SU(4)$. We shall come back to this point in a moment. 

Replacing the $R-$current in \re{Q-flow} by its expression in terms of the free fields, we obtain 
\begin{align}\notag\label{Q-osc}
\mathcal  Q^B_{A}(\vec n)    
&=  \int {d^4 k\over (2\pi)^4} 2\pi \delta_+(k^2)  \,   
\delta^{(2)}(\Omega_{\vec n}-\Omega_{\vec k})
\\
&\qqqquad\times   
 \big[ a_{AC}^\dagger(k) a^{CB}(k)+  a_{A,1/2}^\dagger(k) a_{-1/2}^{B}(k)-  a^{B,\dagger}_{-1/2}(k)a_{A,1/2}(k)\big] - \text{(trace)},
\end{align}
where (trace) denotes terms proportional to $\delta_A^B$ that are needed to ensure  the tracelessness  condition
 $\delta_B^A \mathcal  Q^B_{A}(\vec n) =0  $.
 Here $a_{A,1/2}^\dagger$ and $a^{B \, \dagger}_{-1/2}$ are the creation operators of gluinos with helicity $\pm 1/2$, respectively, 
and $a_{AC}^\dagger(k)$ are the creation operators of the scalars in $SU(4)$ notation (see Appendix \ref{appA}).
Using \re{Q-osc} we can work out the action of the operator  $\mathcal  Q^B_{A}(\vec n)$ on the asymptotic states. For instance,
for a single-particle gluino state  $\ket{k}_E\equiv a_{E,1/2}^\dagger(k) \ket{0}$ we get  
\begin{align}\label{222}
\mathcal   Q^B_{A }(\vec n) \,\ket{k}_E =    \delta^{(2)}(\Omega_{\vec k}-\Omega_{\vec n})
\left[{\delta_E^B\,\ket{k}_A -\ft14 \delta_A^B\,\ket{k}_E }\right] \,,
\end{align}
from where we conclude that the operator $\mathcal  Q^B_{A}(\vec n)$ does not change the momentum of the particle but it rotates
its $SU(4)$ index. The same is true for the scalar states whereas the gluon state has zero $R-$charge and, therefore, is
not affected by $\mathcal  Q^B_{A}(\vec n)$. 

Relation \re{222} seems to contradict our definition of the flow operator \re{E-spectr}, according to which the on-shell states should diagonalize
the operator $\mathcal  Q^B_{A}(\vec n)$. 
To restore the diagonal action of the operator  $\mathcal  Q^B_{A}(\vec n)$ on the asymptotic states we introduce an 
 auxiliary traceless matrix $Q_B^A$ and consider the following linear combination of the operators \re{Q-osc}
\begin{align}\label{Qq}
 \mathcal  Q (\vec n;Q)= Q_B^A\,\mathcal  Q^B_{A}(\vec n) \,.
\end{align}
To preserve the reality condition on
the eigenvalues of the flow operator \re{Qq}, the matrix $Q_B^A$ should be Hermitian. 
This allows us to decompose it  over its eigenvectors,
\begin{align}\label{}
Q_B^A =   \sum_{\a=1}^4 Q_\a \bar u^A_\a u^\a_B\,, \qquad \sum_{A=1}^4\bar u^A_\a u^\b_A=\delta_\b^\a\,, \qquad \sum_{\a=1}^4\bar u^A_\a u^\a_B=\delta_B^A\,.
\end{align} 
Its real eigenvalues 
satisfy the tracelessness condition $\sum_{\a=1}^4 Q_\a=0$.~\footnote{The eigenvectors define a unitary matrix,  $\bar u^A_\a=(u^\a_A)^*$   
and $uu^\dagger= \mathbb{I}$, which diagonalizes the Hermitian projection matrix $Q_B^A$. They can be interpreted as harmonic variables 
on the coset $SU(4)/[U(1)]^3$ \cite{Galperin:1984av,Howe:1995md}.}

Using the eigenvectors of $Q_B^A$, we can define the projected on-shell states $\ket{k}_\alpha = \bar u^A_\a\ket{k}_A$. 
Such states allow us to rewrite the action of the $SU(4)$ flow operator \re{Qq} on, e.g., a single-particle gluino state $\ket{k}_A$, in a form 
analogous to the diagonal action of the energy flow operator in \p{E-spectr}. Indeed, the charge flow operator 
\re{Qq} acts diagonally on the projected states $\ket{k}_\alpha$\,,
\begin{align}\label{2.24}
  \mathcal  Q (\vec n;Q) \ket{k}_\alpha = Q_\alpha\, \delta^{(2)}(\Omega_{\vec k}-\Omega_{\vec n})\,\ket{k}_\alpha\,.
\end{align} 
The contribution of the gluino state to \re{wcs} can be written in two equivalent forms, $\sum_{A}\ket{k}_A \bra{k}^A =  \sum_{\alpha} \ket{k}_A \bar u^A_\a u^\a_B\bra{k}^B 
=  \sum_{\alpha}\ket{k}_\alpha \bra{k}^\alpha$. Then, the action of the charge flow operator $\mathcal  Q (\vec n;Q)$ takes a diagonal form in the new basis:
\begin{align}\notag\label{Q-dec}
& \mathcal  Q (\vec n;Q)\sum_{A=1}^4\ket{k}_A \bra{k}^A   =\left( \delta^{(2)}(\Omega_{\vec k}-\Omega_{\vec n})\, Q^B_A \ket{k}_B \right)   \bra{k}^A  
\\[-2mm]
&= \mathcal  Q (\vec n;Q) \sum_{\alpha}\ket{k}_\alpha \bra{k}^\alpha 
= \sum_{\alpha=1}^4 \ket{k}_\alpha \left(Q_\alpha\, \delta^{(2)}(\Omega_{\vec k}-\Omega_{\vec n}) \right) \bra{k}^\alpha
\,.
\end{align}
According to \re{Q-dec}, the charge detector, described by the flow operator $\mathcal  Q (\vec n;Q)$, decomposes the on-shell state
of each particle, propagating in the direction of the vector $\vec n$, over the four basis vectors $u^\a_A$ in the $SU(4)$ space and 
assigns a charge $Q_\alpha$  to each component. \footnote{This decomposition corresponds to introducing the Cartan basis for the Lie algebra $su(4)$. 
The charges  $Q_\a$ can be interpreted as linear combinations of the three Cartan charges.}

Like in  \re{E-com}, the charge flow operators depending on two distinct vectors $\vec n\neq \vec n'$ commute with each other and with the energy flow operators,
\begin{align}
[\mathcal  Q (\vec n;Q),\mathcal  Q (\vec n';Q')]=[\mathcal  Q (\vec n;Q),\mathcal  E (\vec n')]=0\,.
\end{align}
Along the same lines as before, we can define the charge flow along various directions $\vec n_1,\dots,\vec n_\ell$ and
express the corresponding weighted cross section as 
\begin{align}\label{sigma-Q}
\vev{\mathcal Q(\vec n_1)\dots \mathcal Q(\vec n_\ell)} &\equiv \sigma_{{\cal Q}}(q; \vec n_1, \ldots, \vec n_\ell; Q_1, \ldots, Q_\ell; Y)\nt
&=\sigma_{\rm tot}^{-1} \int d^4 x\, \e^{iqx}  \vev{0|  O(x,\widebar Y) \, \mathcal Q(\vec n_1,Q_1)\dots  \mathcal Q(\vec n_\ell,Q_\ell)\, O(0, Y)|0} \,.
\end{align}
Here each detector is specified by the unit vector $\vec n_i$ and by the Hermitian matrix $(Q_i)_{A_i}^{B_i}$, where $i=1,\ldots \ell$.  Unlike the case of the energy correlations in 
\p{sigma-E}, this weighted cross section has a non-trivial dependence on the isotopic variables $Q$ and $Y$ (see Eq.~\p{Q-single} below). 

\subsubsection{Scalar flow} \label{sect-OO}

The definition of  the energy and charge flow, Eqs.~\re{E-flow} and \re{Q-flow}, respectively, involves two of the conserved currents of the ${\cal N}=4$ SYM theory,  
the energy-momentum 
tensor $T_{\mu\nu}(x)$ and the $R-$current $(J_\mu)_A^B$.  As was already mentioned, they belong  to the same $\mathcal N=4$ stress-tensor supermultiplet
 whose lowest-weight state is the  
half-BPS scalar operator $O_{\bf 20'}^{IJ}(x)$, Eq.~\re{O20}. 
This suggests to introduce, in addition to the energy and charge flow,  a `scalar flow' operator corresponding to $O_{\bf 20'}^{IJ}(x)$,
\begin{align}\label{O-flow}
\mathcal  O^{IJ}(\vec n)   & =   \int_0^\infty dt \, \lim_{r\to\infty} r^2 \, O_{\bf 20'}^{IJ}(t,r\vec n)  \,.
\end{align}
These three flow operators  have scaling dimensions
\begin{align}\label{dimop}
\Delta_{\mathcal O}=-1\,, \qquad   \Delta_{\mathcal Q}=0\,, \qquad \Delta_{\mathcal E}=1\,,
\end{align} 
respectively, as follows from the dimensions $\Delta_O=2$, $\Delta_J=3$, $\Delta_T=4$ of the defining operators $O^{IJ}(x)$, $(J_\mu)^B_{A}(x)$ and $T^{\mu\nu}(x)$. 
The fact that the scaling dimension of  $\cO$ is negative has important consequences, as we demonstrate below. Yet another basic difference  of $O_{\bf 20'}^{IJ}$ compared 
to $(J_\mu)^B_{A}$ and $T^{\mu\nu}$  is that it is a Lorentz scalar and is not a conserved current.
Since the operators $O^{IJ}$, $(J_\mu)^B_{A}$ and $T^{\mu\nu}$ belong to the same supermultiplet,
we anticipate that the correlations of the corresponding flow operators $\cO$, ${\cal Q}$ and ${\cal E}$ should be related to each other by supersymmetry. In 
Section~\ref{Sect:GenMasterFormulae}, we provide a lot of evidence for such relations, but the 
precise mechanism will be worked out in our future work.

The expression for the scalar flow operator \re{O-flow} in terms of the free scalar fields 
looks as  \footnote{The additional factor of $1/2$ on the right-hand side is due to our normalization of the gauge group generators $\tr[T^a T^b]=\delta_{ab}/2$.} 
\begin{align} \label{O-osc}
\mathcal  O^{IJ}(\vec n) 
&=  \frac12  \int {d^4 k\over (2\pi)^4} 2\pi \delta_+(k^2)  \,k_0^{-1} \,
\delta^{(2)}(\Omega_{\vec n}-\Omega_{\vec k})\, a^{\dagger\,\{I}(k) a^{J\}}(k)\,,
\end{align}
where $a^{I\,\dagger}(k)$ and $a^J(k)$ are the creation and annihilation operators of scalars in $SO(6)$ notations
(see Appendix \ref{appA}) and $\{IJ\}$ denotes traceless symmetrization of the pair of $SO(6)$ indices $I$ and $J$. The operator 
$\mathcal  O^{IJ}(\vec n)$ acts non-trivially only on the scalar on-shell states, 
$\ket{k}^I\equiv  a^{\dagger\, I}(k) \ket{0}$, by rotating them in the isotopic $SO(6)$ space. What is the most  unusual about \p{O-osc} is the inverse power
of the energy (in the rest frame of the source $q^\mu=(q^0,\vec 0)$). Its presence is a consequence of the negative  dimension $(-1)$ of the 
operator $\cO(\vec n)$ (see \p{dimop}). 

To define the 
corresponding scalar flow operator we introduce, in close analogy with \re{Qq}, the following projection
of the operators \re{O-osc},
\begin{align}\label{231}
\mathcal  O(\vec n; S) = S_{IJ} \,\mathcal  O^{IJ}(\vec n) \,.
\end{align}
Since $\mathcal  O^{IJ}(\vec n)$ is symmetric and traceless, the projection matrix $S_{IJ}$ has to have the same properties.
In addition, the reality condition on the detector measurement leads to the {reality}    condition $S_{IJ}= S^*_{IJ}$. {This allows us to decompose
$S_{IJ}$ over its real eigenvectors, 
\begin{align}\label{C}
S_{IJ} = \sum_{i=1}^6 S_i\,  \phi_I^{i} \phi_J^{i}\,,\qquad  \sum_{I=1}^6  \phi_I^{i} \phi_I^{j} =\delta^{ij} \,,\qquad 
 \sum_{i=1}^6 \phi_I^{i} \phi_J^{i} =\delta_{IJ}\,,
\end{align}
with  real eigenvalues $S_i$.  The condition for $S_{IJ}$ to be traceless leads to 
$\sum_{i=1}^6 S_i=0$. \footnote{The eigenvectors define an $SO(6)$ matrix,  $\phi \phi^T=\mathbb{I}$, which diagonalizes the projection matrix $S_{IJ}$. They play the role of harmonic variables on the coset $SO(6)/[SO(2)]^3$.}

Then, the scalar flow operator \re{O-osc} takes the following form
\begin{align}\label{Os}
\mathcal  O(\vec n; S) = {\frac12} \sum_{i=1}^6 S_i \int {d^4 k\over (2\pi)^4} 2\pi \delta_+(k^2) \, k_0^{-1} \,
\delta^{(2)}(\Omega_{\vec n}-\Omega_{\vec k}) (\phi^{i} a^\dagger(k))(\phi^{i} a(k))\,,
\end{align}
and it is diagonalized by the projected on-shell scalar states $\ket{k}^i= \phi_I^{i} \ket{k}^I$:
\begin{align}\notag\label{O-dec}
& \mathcal  O (\vec n;S)\sum_{I=1}^6\ket{k}^I \bra{k}^I = 
(2 k_0)^{-1}\, \delta^{(2)}(\Omega_{\vec k}-\Omega_{\vec n})   \sum_{I,J=1}^6  S_{IJ}\ket{k}^J  \bra{k}^I 
 \\[-2mm]
 &= \mathcal  O (\vec n;S) \sum_{i=1}^6\ket{k}^i \bra{k}^i 
 = 
 \sum_{i=1}^6\ket{k}^i\,   \left ( {S_i\over 2k_0}\,   \delta^{(2)}(\Omega_{\vec k}-\Omega_{\vec n}) \right)\,   \bra{k}^i
\,.
\end{align}
The interpretation of \re{O-dec} is similar to that of \re{Q-dec} for the charge flow. The scalar flow operator decomposes the on-shell scalar
state moving in the direction of the vector $\vec n$ over the basis of eigenvectors (or $SO(6)$ harmonics) $\phi^{i}_I$ and assigns
to each component an eigenvalue $S_i$ divided by (twice) the energy of the particle.

We emphasize the appearance of the inverse energy factor in the last relation in \re{O-dec}. It leads to some unusual properties of the scalar 
flow operator \re{Os} as compared to its  energy and charge counterparts. To show this, we examine the commutator 
$[\mathcal  O (\vec n;S),\mathcal  O (\vec n';S')]$. Since each operator receives  
contributions from particles propagating along two different directions $\vec n$ and $\vec n'$, we may expect that 
\begin{align}
\mathcal  O (\vec n;S)\mathcal  O (\vec n';S')\ket{k}\sim \delta^{(2)}(\Omega_{\vec k}-\Omega_{\vec n})\delta^{(2)}(\Omega_{\vec k}-\Omega_{\vec n'}) \ket{k}=0\,.
\end{align}
Hence, the commutator should vanish since the same particle cannot go through the two detectors simultaneously. 
This is correct unless the momentum of the particle vanishes. Indeed, for $\vec n\neq \vec n'$, the conditions imposed by the two 
delta functions, $\vec k = k^0\, \vec n=k^0 \, \vec n'$, are verified only if  $k^0=\vec k=0$. Thus, the commutator 
$[\mathcal  O (\vec n;S),\mathcal  O (\vec n';S')]$ can receive contributions only from particles with zero momentum. 
The corresponding Feynman diagrams are shown in Fig.~\ref{fig:cross}.
\begin{figure}[h!t]
\psfrag{q}[cc][cc]{} 
\psfrag{1}[cc][cc]{$1$}\psfrag{2}[cc][cc]{$2$}\psfrag{3}[cc][cc]{$3$}\psfrag{4}[cc][cc]{$4$}
\psfrag{Cor}[lc][cc]{$ \Longrightarrow \, \quad  {\rm Limit_{2,3}} \ \vev{0| O(1) O(2) O(3) O(4)|0}$}
\psfrag{T1}[cc][cc]{} 
\psfrag{T2}[cc][cc]{} 
\psfrag{Null}[cc][cc]{}
\centerline{\includegraphics[height=45mm]{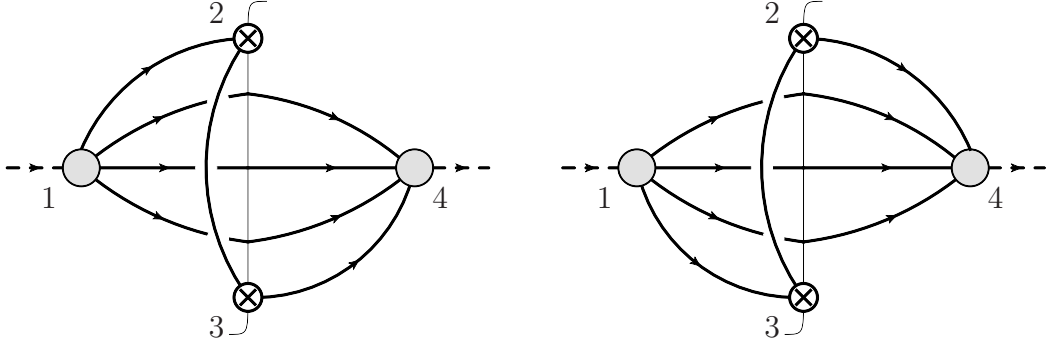}}
\caption{\small Cross-talk between the two detectors. The thin line stands for the unitarity cut. The shaded blobs (vertices 1 and 4) stand for the source and sink. The 
crosses denote the two detectors  (vertices 2 and 3) oriented
along the vectors $\vec n$ and $\vec n'$. The detectors interact with each other by exchanging a particle with zero momentum. }
\label{fig:cross}
\end{figure}
Carefully examining their contributions (see Appendix~\ref{app-cross}),    we find that, precisely due to the factor of $1/k_0$ in \re{Os}, the commutator
is different from zero,
\begin{align}\label{nonzero}
[\mathcal  O (\vec n;S),\mathcal  O (\vec n';S')] \sim  
  {a^{\dagger\, I} (0) (S S'- S'S)_{IJ} a^J(0) 
 \over 1-(\vec n\vec n')}   \,.
\end{align}  
At the same time, the scalar flow operator commutes with those of the energy and charge flow,
\begin{align}
[\mathcal  O (\vec n;S),\mathcal  E (\vec n')]=[\mathcal  O (\vec n;S),\mathcal  Q (\vec n';Q)]=0\,.
\end{align}
The non-vanishing commutator \re{nonzero} leads to a divergence in certain weighted cross sections, as explained later in the paper. 

We observe that  the expression on the right-hand side of \re{nonzero} involves the {commutator} of the matrices defining the two
scalar detectors. Therefore, we can restore the commutativity of the operators $\mathcal  O (\vec n;S)$ and $\mathcal  O (\vec n';S')$
by requiring 
\begin{align}\label{cross-talk}
[S,S']=0\,.
\end{align}
{In physical terms, this condition prohibits the cross-talk between the two detectors mediated by the exchange of particles with zero momentum.
Notice that while this is a necessary condition on the detector matrices, which eliminates potentially divergent contributions due to particle exchanges
with zero energy, it is not sufficient when we try to match the weighted cross sections with the integrated correlation functions. The latter require further 
constraints on the projection matrices $S$, see Appendix~\ref{appE} for details.}

Having defined the scalar flow operator \re{Os}, we can introduce the corresponding multiple-detector weighted cross section
\begin{align}\label{sigma-O}
\vev{\mathcal O(\vec n_1)\dots \mathcal O(\vec n_\ell)}  &\equiv \sigma_{{\cal S}}(q; \vec n_1, \ldots, \vec n_\ell; S_1, \ldots, S_\ell; Y)\nt
&=\sigma_{\rm tot}^{-1} \int d^4 x\, \e^{iqx}  \vev{0|  O(x,\widebar Y) \, \mathcal O(\vec n_1,S_1)\dots  \mathcal O(\vec n_\ell,S_\ell)\, O(0, Y)|0} \,.
\end{align}
It depends on a set of vectors $\vec n_i$ determining the spatial orientation of the detectors, as well as on the projection matrices $S_i$, Eq.~\re{C}.  By 
construction, the scalar flow operators $\mathcal O(\vec n_i,S_i)$ should commute with each other.  

In addition to \re{sigma-E}, \re{sigma-Q} and \re{sigma-O}, we can also define mixed correlations 
involving all three flow operators, $\mathcal O(\vec n_i,S_i)$, $\mathcal Q(\vec n_i,Q_i)$ and 
$\mathcal  E (\vec n_i)$. As was already mentioned,  they are effectively related to each other by supersymmetry. The scalar correlations \re{sigma-O} play a special role 
in our analysis below. Firstly, they can be expressed in terms of the correlation 
function involving $\ell+2$  copies of the same half-BPS operator \re{O20}. Secondly, in the special case $\ell=2$   the correlation function of the half-BPS operators $O(x)$ 
uniquely determines, by means of $\cN=4$ supersymmetry transformations, all four-point correlation functions of the other operators from the stress-energy multiplet that generate 
charge and energy flow correlations. This is not the case for $\ell>2$ since the solution to the corresponding $\cN=4$ superconformal Ward identities
is not unique anymore due to the appearance of nontrivial $\cN=4$ superconformal invariants depending on $2+\ell\ge 5$ points.

We would like to emphasize that, in virtue of the definition of the flow operators \re{E-flow}, \re{Q-flow} and \re{O-flow}, the weighted cross sections \re{sigma-E}, \re{sigma-Q} and 
\re{sigma-O} are related to the integrated Wightman correlation functions defined at spatial infinity. This makes the issue of infrared finiteness of the 
flow correlations extremely nontrivial.  It is believed that the energy flow correlations are IR finite both at weak and strong coupling whereas for the scalar and $R-$charge flow 
observables the situation remains unclear. As a counterexample, we can recall that similar problem also arose in QCD, where the flavor observables in jet physics (closely related to 
charge flow correlations) while perfectly well-defined at leading order of the perturbative expansion, cease to stay finite once higher order corrections are accounted for \cite{Salam}. The 
problem requires further studies and will be addressed elsewhere. 

\section{Weighted cross sections from amplitudes}  

In this section we employ the conventional approach based on the scattering amplitudes to evaluate the 
weighted cross sections  introduced in the previous section to lowest order in the coupling in $\mathcal N=4$ SYM.

We start by computing the matrix elements \re{M} involving the operator defined in \re{OY}. At tree level, the final state consists of a pair of scalars denoted by $\ket{{\rm s}^I(k_1) {\rm s}^J(k_2)}$: 
\begin{align}\label{tree}
\sigma_0= | \mathcal M_{O_{\bf 20'}\to {\rm ss}} |^2 = |\vev{{\rm s}(k_1) {\rm s}(k_2) |  O(0,Y)|0}|^2 = \frac12(N^2_c-1)(Y \widebar Y)^2\,.
\end{align} 
To first order in the coupling, the final state also contains three-particle states $\ket{X}=\ket{\rm s,s,g}$ and $\ket{X}=\ket{\rm s,\lambda,\lambda}$. The corresponding transition amplitudes are  
\begin{align}\notag\label{mat-el}
& |\mathcal M_{O_{\bf 20'}\to {\rm ssg}}|^2 = \big|\vev{{\rm s}(k_1) {\rm s}(k_2) {\rm g} (k_3) |  O(0,Y)|0} \big|^2
= g^2 \sigma_0  \frac{4 s_{12}}{s_{13}s_{23}}  \,,
\\
& |\mathcal M_{O_{\bf 20'}\to {\rm s}\lambda\lambda}|^2= \big|\vev{\lambda(k_1)\lambda (k_2) {\rm s}(k_3) |  O(0,Y)|0} \big|^2
=g^2 \sigma_0  \frac{8}{s_{12}}  \,,
\end{align}
where the notation was introduced for the Mandelstam invariants $s_{ij}=(k_i+k_j)^2$ with $k_i^2=0$.

The total transition probability  \re{tot} is given to order $O(g^2)$ by
\begin{align}\notag\label{tot-1loop}
 \sigma_{\rm tot}(q) & =\int \text{dPS}_{2}\,| \mathcal M_{O_{\bf 20'}\to {\rm ss}} |^2+  
\int \text{dPS}_{3}\,\lr{|\mathcal M_{O_{\bf 20'}\to {\rm ssg}}|^2+ |\mathcal M_{O_{\bf 20'}\to {\rm s}\lambda\lambda}|^2}+ O(g^4)
\\[1.5mm] &  
= {\sigma_0\over 8\pi}\,  \left[ 1 + g^2 F_{\rm virt}(q^2)\right] 
 + {4} g^2 \sigma_0  
\int \text{dPS}_{3}\,  \frac{s_{12}^2+2 s_{13}s_{23}}{s_{12}s_{13}s_{23}}
 + O(g^4)\,,
\end{align}
where $F_{\rm virt}(q^2)$ describes the one-loop (virtual) correction to the transition amplitude \re{tree} and the notation was introduced
for the Lorentz invariant integration measure on the  phase space of $\ell$ massless particles with the total momentum $q^\mu$,
\begin{align} \label{LIPS}
\int \text{dPS}_{\ell} &=\int  \left({\prod_{i=1}^\ell {d^4 k_i\over (2\pi)^4}\,2\pi\delta_+(k_i^2)}\right) \, (2\pi)^4 \delta^{(4)}\big(q-\sum_{i=1}^\ell k_i\big)\,.
\end{align}
The symmetry of this integration measure under the exchange of any pair of particles allows us to rewrite \re{tot-1loop} as
\begin{align}\label{tot-zero}
 \sigma_{\rm tot}(q) = {\sigma_0\over 8\pi} + g^2 {\sigma_0\over 8\pi}  \left[ F_{\rm virt}(q^2)+
\int \text{dPS}_{3}\,  \frac{{32} \pi(q^2)^2}{3\,s_{12}s_{13}s_{23}}\right]
 + O(g^4)\,,
\end{align}
where we symmetrized the one-loop integrand with respect to the particle momenta and used the identity
$q^2=s_{12}+s_{23}+s_{13}$.

As was already mentioned, the total transition amplitude $\sigma_{\rm tot}(q)$ is protected from perturbative corrections and, therefore,
the terms proportional to $g^2$ on the right-hand side of \re{tot-zero} should vanish. 
This allows us to fix the virtual correction $F_{\rm virt}(q^2)$.\footnote{ Strictly speaking, both terms in  the square brackets in \re{tot-zero} are IR divergent and require regularization.
Their sum vanishes in the dimensional regularization scheme.
}
Taking into account \re{tree} we verify that the resulting expression for $\sigma_{\rm tot}(q)$ coincides with \re{tot-ex}.

\subsection{Single detector} \label{31}

Let us now examine the weighted cross sections involving a single detector. We start with the energy flow. According to the definition \re{wcs}, the corresponding cross section can be obtained from the first relation in 
\re{tot-1loop} by inserting the energy weight factor \re{w-energy} into the phase space integrals. 
The weight factor 
$w_{{\mathcal E}(\vec n)}(k_1,k_2)$ for the transition $\mathcal M_{O_{\bf 20'}\to \rm{ss}}$ 
depends on the energy of the two scalars (in the rest frame of the source), whereas for $\mathcal M_{O_{\bf 20'}\to \rm{ssg}}$ and $\mathcal M_{O_{\bf 20'}\to \rm{s}\lambda\lambda}$ it is given by
$w_{{\mathcal E}(\vec n)}(k_1,k_2,k_3)$ that receives additive contributions from all produced particles, as seen from Eq.\ \re{w-energy}}. In this way, we obtain 
\begin{align}\notag\label{EC-1}
\sigma_{{\mathcal E} }(q;\vec n)  &= \sigma_{\rm tot}^{-1} \int \text{dPS}_{2}\,w_{{\mathcal E}}(k_1,k_2) \,| \mathcal M_{O_{\bf 20'}\to {\rm ss}} |^2
\\ & +  
\sigma_{\rm tot}^{-1} \int \text{dPS}_{3}\,w_{{\mathcal E}}(k_1,k_2,k_3)\lr{|\mathcal M_{O_{\bf 20'}\to {\rm ssg}}|^2+ |\mathcal M_{O_{\bf 20'}\to {\rm s}\lambda\lambda}|^2}+ O(g^4)\,.
\end{align}
Replacing the matrix elements on the right-hand side by their explicit expressions \re{tree} and \p{mat-el}, we symmetrize the integrands with respect to the particle momenta and find, after a simple calculation,  
\begin{align}\label{E-single} 
 \vev{\mathcal E(\vec n)}\equiv\sigma_{{\mathcal E} }(q;\vec n)={1 \over  8\pi} \int \text{dPS}_{2} \sum_{i=1,2} k_i^0\,\delta^{(2)}(\Omega_{\vec k_i}-\Omega_{\vec n})
 = \frac1{4\pi}  q^0\,.
\end{align}
Notice that this expression does not depend on the coupling constant. In Sect.~4 we show that $ \vev{\mathcal E(\vec n)}$ is indeed protected from loop corrections.\footnote{This property is related to the well-known fact that the three-point functions of half-BPS multiplets are protected \cite{Howe:1996rb,D'Hoker:1998tz,Lee:1998bxa,Howe:1998zi,Penati:1999ba}, see Sect.~4. \label{foot1}}
We verify that the integral of \re{E-single} over $\vec n$ correctly reproduces the total energy, $\int d^2 \Omega_{\vec{n}}\, \vev{\mathcal E(\vec n)} = q_0$, in agreement with the results in \cite{Hofman:2008ar}.

For the charge flow the analysis goes along the same lines, with the only difference that the non-trivial weight factor $w_{\mathcal Q}$ is 
different from zero only for scalars and gluinos carrying  non-zero $R-$charges. As explained in Sect.~\ref{charge}, the charge flow operator introduces the
projection (or `polarization') matrix  $Q_A^B$ for the $SU(4)$ states of these particles. More precisely, at tree level, the contribution of the transition 
$\mathcal M_{O_{\bf 20'}\to {\rm ss}}$ to the total transition probability involves the $R-$symmetry factor (in $SU(4)$ notation, see Appendix \ref{appA}) 
 $| \mathcal M_{O_{\bf 20'}\to {\rm ss}} |^2\sim y_{A_1B_1} y_{A_2 B_2} \bar y^{A_1B_1} \bar y^{A_2B_2} =\tr( y\bar y )^2$. Here the sum
 over the $SU(4)$ indices corresponds to the summation over the quantum numbers of the particles propagating in the final state.
The contribution to the weighted cross section reads
\begin{align} \notag\label{Q-weight}
    y_{A_1B_1} \big[ 2 Q_{A_1'}^{A_1} \delta^{(2)}(\Omega_{\vec k_1}-\Omega_{\vec n}) \big] \bar y^{A_1'B_1}  y_{A_2 B_2} \bar y^{A_2B_2} 
+
 y_{A_1 B_1} \bar y^{A_1B_1}   y_{A_2B_2} \big[2Q_{A_2'}^{A_2} \delta^{(2)}(\Omega_{\vec k_2}-\Omega_{\vec n}) \big]\bar y^{A_2'B_2}    
 \\[2mm]
 = 2 \tr( y \bar y) \tr( y Q \bar y)\left[\delta^{(2)}(\Omega_{\vec k_1}-\Omega_{\vec n}) + \delta^{(2)}(\Omega_{\vec k_2}-\Omega_{\vec n}) \right] ,
\end{align}
where the last expression is the corresponding weight factor.\footnote{The factor of 2 takes into account 
that the charge flow operator rotates both indices of $y_{AB}$ (we recall that
 $y_{AB}=-y_{BA}$).}  Inserting this result in  the phase space integral $\int \text{dPS}_{2}\,| \mathcal M_{O_{\bf 20'}\to ss} |^2$ and taking into account \p{tree} and \p{bas}, we obtain  
\begin{align}\label{Q-single}
\vev{\mathcal Q(\vec n)} \equiv  \sigma_{{\mathcal Q} }(q;(\vec n,Q){ ; y})  =  {\vev{Q} \over 4\pi} \int \text{dPS}_{2} \sum_{i=1,2}\delta^{(2)}(\Omega_{\vec k_i}-\Omega_{\vec n}) 
=  \frac{1}{\pi}  \vev{Q}\,,
\end{align}
where $ \vev{Q}$ is an isotopic factor keeping track of the $R-$charges,
\begin{align}\label{3.10}
  \vev{Q}= \frac{{\tr( y Q \bar y)}}{ {\tr( y \bar y)}}\,.
\end{align} 
To first order in the coupling constant, the correction to \re{Q-single} has the same form as \re{EC-1} with the only difference that
the corresponding charge weight factor is given by an expression analogous to \re{Q-weight}. Calculating the $O(g^2)$ correction 
to \re{Q-single} we find that it vanishes (see footnote \ref{foot1}).    

Finally, for the scalar flow, the weight factor $w_{\mathcal O}$ is different from zero only for scalar particles. According to \re{O-dec}, 
for a scalar with momentum $k_i$ in the final state, the insertion of the weight factor modifies the $SO(6)$ tensor structure as follows:
\begin{align}
Y^I\delta^{IJ} \widebar Y^J \ \to  \
Y^I \left( {(k_i^0)}^{-1}\, \delta^{(2)}(\Omega_{\vec k_i}-\Omega_{\vec n}) {\frac12}  S^{IJ} \right) \widebar Y^J,
\end{align}
where $S^{IJ}$ is the projection matrix of the scalar detector.
Repeating the calculation at Born level we obtain  
\begin{align}\label{O-single}
\vev{\mathcal O(\vec n)}\equiv  \sigma_{{\mathcal O} }(q;(\vec n,S) ; Y) =  {\vev{S}\over {16}\pi}   \int \text{dPS}_{2} \sum_{i=1,2} (k_i^0)^{-1}
 \delta^{(2)}(\Omega_{\vec k_i}-\Omega_{\vec n}) 
 = \frac{1}{{2} \pi} \frac{\vev{S}}{q^0}  
 \,,
\end{align}
where  $\vev{S}$ is an isotopic factor keeping track of the tensor structure,
\begin{align}\label{3.13}
 \vev{S} = \frac{( Y S \,\widebar Y)}{ ( Y \widebar Y) }\,,
\end{align}
with  $( Y S \,\widebar Y)\equiv Y^I S_{IJ} \widebar Y^J$. 
The calculation of  the $O(g^2)$ correction to this expression shows 
that it vanishes in the same manner as for $ \vev{{\mathcal E} (\vec n)}$ and  $\vev{{\mathcal Q}(\vec n)}$  (see footnote \ref{foot1}).

We recall that the expressions for the one-point correlations
\re{E-single}, \re{Q-single} and \re{O-single} were obtained in the rest frame of the source, $q^\mu=(q^0,\vec 0)$. We present 
the same expressions in  Lorentz covariant form in Sect.~\ref{sect42}.

\subsection{Double correlations}

In the previous subsection  we have shown that the {single-detector weighted cross sections} \re{E-single}, \re{Q-single}  and \re{O-single}  do not depend 
on the spatial orientation of the detector $\vec n$. For  {weighted cross sections involving two detectors }oriented along the vectors
$\vec n$ and $\vec n'$, the rotation symmetry implies that they can  only depend on the
relative angle $0\le \theta\le\pi$ between the vectors, $(\vec n \cdot\vec n')=\cos\theta$.  We call such observables `double correlations'. For $\theta=0$ the 
orientations of the two detectors coincide so that the same particle can go through them sequentially. For $\theta=\pi$
the detectors capture particles moving back-to-back in the rest frame of the source. 

\begin{figure}[h!t]
\psfrag{theta}[cc][cc]{$\theta$}
\psfrag{T1}[cc][cc]{$\mathcal E(\vec n)$}
\psfrag{T2}[cc][cc]{$\mathcal E(\vec n')$}
\centerline{\includegraphics[height=55mm]{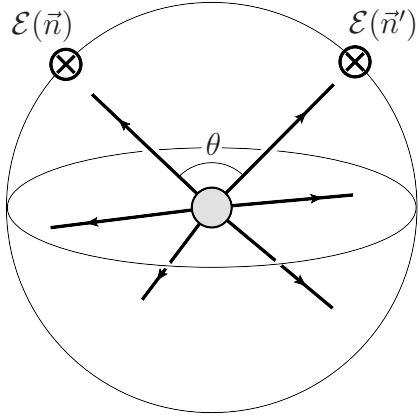}}
\caption{\small Graphical representation of the double energy correlation: particles produced out of the vacuum by
the source are captured by the two detectors located at spatial infinity in the directions of the unit vectors $\vec n$ and $\vec n'$.
}
\label{fig-EE}
\end{figure}

The calculation of the double correlations at one loop goes along the same lines as in the previous subsection. For the double energy
correlation $\vev{\mathcal E(\vec n)\mathcal E(\vec n')}$, the only difference as compared with \re{EC-1} is that the weight factor $w_{\mathcal E}(\vec n)$ gets 
replaced by the product 
$w_{\mathcal E}(\vec n)w_{\mathcal E}(\vec n')$, which fixes the spatial orientation of the momenta of the two particles 
in the final state.

At Born level, the final state consists of two scalar particles moving collinearly in the rest frame $q^\mu=(q^0,\vec 0)$, each carrying energy $q^0/2$. As a consequence, their contribution 
to $\vev{\mathcal E(\vec n)\mathcal E(\vec n')}$ is localized at
$\theta=0$ and $\theta=\pi$,
\begin{align}\label{EE-born}
\vev{\mathcal E(\vec n)\mathcal E(\vec n')}^{(0)}  = {q_0^2\over 8\pi}\left[\delta(\theta)+\delta(\pi-\theta) \right]\,.
\end{align}
Here the first delta-function describes the situation of aligned detectors capturing particles moving into the same direction, while the second delta-function corresponds to 
two particles moving back-to-back. For $0<\theta<\pi$, the double-energy correlation receives a non-zero contribution starting from one loop. It comes from
the three-particle transitions $\mathcal M_{O_{\bf 20'}\to {\rm ssg}}$ and $\mathcal M_{O_{\bf 20'}\to {\rm s\lambda\lambda}}$,
\begin{align}\notag\label{EE-aux}
\vev{\mathcal E(\vec n)\mathcal E(\vec n')}=\sigma_{\rm tot}^{-1} \int & \text{dPS}_{3}\,\sum_{i,j=1}^3 k_i^0\, k_j^0 \,
\delta^{(2)}(\Omega_{\vec k_i}-\Omega_{\vec n})\delta^{(2)}(\Omega_{\vec k_j}-\Omega_{\vec n'})
\\
& \times\big({|\mathcal M_{O_{\bf 20'}\to {\rm s}(k_1){\rm s}(k_2){\rm g}(k_3)}|^2+ |\mathcal M_{O_{\bf 20'}\to {\rm s}(k_1)\lambda(k_2)\lambda(k_3)}|^2}\big)\,.
\end{align}
Using the 
explicit expressions for the matrix elements \re{mat-el} we find (see Appendix~\ref{app:1loop} for details)
\begin{align}\label{EE-1}\notag
\vev{\mathcal E(\vec n)\mathcal E(\vec n')} & = {g^2\over {2} (2\pi)^4} {q_0^2\over\sin^2\theta} 
\int_0^1 {d\tau_1 \over 1-\tau_1(1-\cos\theta)/2} +O(g^4)
\\
& =  {g^2\over (2\pi)^4}\,  q_0^2 \,{ 1+\cos\theta \over  \sin^4\theta } \ln {2\over 1+\cos\theta} +O(g^4)\,, 
\end{align}
where the logarithmic correction arises from the integration over the energy fraction of one of the particles, $\tau_1=2k_1^0/q^0$.  For $\theta\to 0$,
the expression in the right-hand side of \re{EE-1} scales as $O(\theta^{-2})$, whereas for $w=\pi-\theta\to 0$ it has the well-known Sudakov behavior  $O(w^{-2}\ln(w^{-2}))$. 
Both asymptotics are modified at higher loops in a controllable way \cite{Collins:1981uk}.

It is convenient to rewrite \re{EE-1} by introducing  the scaling variable 
\begin{align}\label{3.16}
z=(1-\cos\theta)/2\,,
\end{align}
where $0<\theta<\pi$ is the angle between the detector vectors $\vec n$ and $\vec n'$.
Then, the double-energy correlation takes the following form at one loop
 \begin{align}\label{EE}
\vev{\mathcal E(\vec n)\mathcal E(\vec n')} =   {a\over 4\pi^2}\, {q_0^2 \over 8 z^3} \lr{-{z \ln (1-z)\over  1-z}} + O(a^2)\,,
\end{align}
with $z$ varying in the range $0<z<1$ and $a=g^2/(4\pi^2)$, as defined earlier.  
It is instructive to compare \re{EE} with the analogous expression in QCD. In that case, the final state is created by an electromagnetic
current  and it consists of quarks and gluons. To lowest order in the coupling, the energy-energy correlation looks as \cite{Basham:1978bw}
\begin{align}\notag\label{QCD}
\vev{\mathcal E(\vec n)\mathcal E(\vec n')}_{\rm QCD} =   {a_{\rm {\scriptscriptstyle QCD}}\over 4\pi^2}\, {q_0^2 \over 8 z^3}  
\bigg[\lr{-{z\over 1-z}+\frac{9}{z^2}-\frac{15}{z}+3}\ln(1-z)& 
\\
 +\lr{\frac{9}{z}-\frac{3}{2 (1-z)}-9}\bigg]& + O(a_{\rm {\scriptscriptstyle QCD}}^2)\,,
\end{align}
where $a_{\rm {\scriptscriptstyle QCD}}=  g^2_{\rm {\scriptscriptstyle QCD}}C_2/(4\pi^2)$ is the QCD fine structure coupling constant
(with $C_2=(N_c^2-1)/(2N_c)$ being the
quadratic Casimir in the fundamental representation of the $SU(N_c)$).

We observe that the $\mathcal N=4$ SYM result \re{EE} can be obtained from the QCD expression \re{QCD} by discarding the rational
term inside the square brackets in \re{QCD} and by retaining only the leading singularity for $z\to 1$ in the $\ln(1-z)$ term. In other words, the two expressions have the same leading asymptotic Sudakov behavior as $z\to 1$.
According to \re{3.16}, this limit corresponds to the two detectors capturing particles moving back-to-back in the rest frame of the source (see Fig.~\ref{fig-EE}).

For the double-scalar correlation  $\vev{\mathcal O(\vec n)\mathcal O(\vec n')}$ the weights corresponding to the particles 
in the final state of $\mathcal M_{O_{\bf 20'}\to X}$ also depend on the detector matrices $S$ and $S'$.
For $0<\theta<\pi$, to one-loop order, it reduces to the product of isotopic factors  $\vev{S}\vev{S'}$  from the single detector correlation   \re{O-single}: 
\begin{align}\notag\label{SSC-1}
\vev{\mathcal O(\vec n)\mathcal O(\vec n')}= \sigma_{\rm tot}^{-1} \int   \text{dPS}_{3}  (k_1^0\, k_2^0)^{-1} \,
2\, \delta^{(2)}(\Omega_{\vec k_1}-\Omega_{\vec n})\delta^{(2)}(\Omega_{\vec k_2}-\Omega_{\vec n'})
\\
 \times  (Y S \widebar Y)(Y S' \widebar Y)  |\mathcal M_{O_{\bf 20'}\to {\rm s}(k_1){\rm s}(k_2){\rm g}(k_3)}|^2 \,,
\end{align}
  where the additional factor of $2$ comes from the symmetry of the weight factor under exchanging the detectors, $\vec n \leftrightarrow \vec n'$. 
Performing the integration over the phase space of the three particles in the final state of
$\mathcal M_{O_{\bf 20'}\to \rm{ssg}}$, we obtain (see Appendix~\ref{app:1loop})
\begin{align}\label{OO}
\vev{\mathcal O(\vec n)\mathcal O(\vec n')}  = {a \over  4\pi^2}  {\vev{S}\vev{S'}\over 2 q_0^2 z}\lr{- {z \ln(1-z)\over 1-z}}
 + O(a^2)\,,
\end{align}
with $0< z<1$.
 
For the double-charge correlation $\vev{\mathcal Q(\vec n)\mathcal Q(\vec n')}$, with $0<\theta<\pi$, we have to take into account all possible correlations
between the charges of scalars and gluinos in the final state
 \begin{align}\notag
\vev{\mathcal Q(\vec n)\mathcal Q(\vec n')}=2\sigma_{\rm tot}^{-1} \int   \text{dPS}_{3}\, \bigg[ 4  \tr( y Q \bar y)   \tr( y Q' \bar y)
\delta^{(2)}(\Omega_{\vec k_1}-\Omega_{\vec n})\delta^{(2)}(\Omega_{\vec k_2}-\Omega_{\vec n'})
\\\notag
  \times\big({|\mathcal M_{O_{\bf 20'}\to {\rm s}(k_1){\rm s}(k_2){\rm g}(k_3)}|^2+ |\mathcal M_{O_{\bf 20'}\to {\rm s}(k_1)\lambda(k_2)\lambda(k_3)}|^2}\big)
\\
 +\tr[y\bar y]\tr[y Q  \bar y Q' ]\delta^{(2)}(\Omega_{\vec k_2}-\Omega_{\vec n})\delta^{(2)}(\Omega_{\vec k_3}-\Omega_{\vec n'})|\mathcal M_{O_{\bf 20'}\to {\rm s}(k_1)\lambda(k_2)\lambda(k_3)}|^2\bigg],
\end{align}
where the last line describes the correlations between two gluinos. Note that $\tr[y Q \bar y Q']=\tr[y Q' \bar y Q]$ due to the
antisymmetry of the matrices $y, \bar y$. The integration over the final state phase space yields
\begin{align}
\vev{\mathcal Q(\vec n)\mathcal Q(\vec n')}=&- {a \over   \pi^2}  { \ln(1-z)\over 4 z^2}
\left({2z\over 1-z} \vev{Q}\vev{Q'}+ \vev{Q,Q'}  \right) + O(a^2)\,,
 \label{C.6}
\end{align}
where $\vev{Q}$ is given by \re{3.10} and the notation was introduced for the (non-factorizable) correlation between the matrices of the two detectors
\begin{align}\label{Q-factor}
\vev{Q,Q'} \equiv {\tr \left[y Q \bar y Q' +{\tilde {\bar y} Q \tilde y  Q' } \right] \over {2} \tr[y\bar y]} =  {\tr\left[y Q' \bar y Q +{\tilde {\bar y} Q' \tilde y  Q } \right]  \over {2}\tr[y\bar y]} \,,
\end{align}
with $\tilde {\bar y}_{AB} =\frac12 \epsilon_{ABCD} \bar y^{CD}$ and $\tilde y^{AB} = \frac12 \epsilon^{ABCD} y_{CD}$.
In the above calculation we have tacitly assumed that the detectors measure two different particles. 
In general, we also have to examine the possibility for the same particle to go sequentially through the two detectors. 
As was already explained, for $\vec n \neq \vec n'$ (or equivalently $\theta\neq 0$) the momentum of the particle should be necessarily zero in this case and, as a result, its
contribution to the charge correlations vanishes. The same is true for the scalar correlations provided that the projection matrices of the detectors
satisfy the `no-cross-talk' condition \re{cross-talk}.
 
In a similar manner, we can also define mixed correlations involving various flow operators:  
\begin{align}\notag
& \vev{\mathcal Q(\vec n)\mathcal E(\vec n')} =  {a \over 4\pi^2}\,{\vev{Q} q^0}\lr{-{\ln (1-z)\over z^2 (1-z)}}+ O(a^2) \,,
\\[2mm]\notag
& \vev{\mathcal O(\vec n)\mathcal E(\vec n')} =  { a \over 16\pi^2}\, \vev{S} \lr{- {\ln(1-z)\over z^2 (1- z)}}+ O(a^2)\,,
\\[2mm]
& \vev{\mathcal Q(\vec n)\mathcal O(\vec n')} =   {a\over 4\pi^2}\, {  \vev{Q}\vev{S'}   (q^0)^{-1}}\lr{-\frac{ \ln (1-z)}{z(1-z)}}+ O(a^2)\,.  \label{3.18}
\end{align}
Notice that the dependence of these expressions on the total energy $q^0$ is uniquely fixed by the scaling dimension of the flow operators.
The non-trivial dynamical information resides in the $z-$dependence. Quite remarkably, the obtained one-loop 
expressions are all proportional
to the same function $\ln(1-z)/(1-z)$. Its appearance is not accidental, of course, since it ensures the universal Sudakov  
behavior of the correlations for $z\to 1$.  

The approach described in this section can be extended to higher loops. Namely, to any order in the coupling constant,  following \re{wcs} we can
express the correlations as a sum over all possible final states, evaluate the corresponding transition amplitudes 
\re{M} and, then, perform the integration over the phase space. However, such an approach becomes very cumbersome and inefficient
beyond one loop for the following reasons. The number of production channels $O_{\bf 20'}\to X$ grows  rapidly at higher loops and, therefore,
we have to deal with an increasing number of terms in the sum over the final states. Secondly, with many particles in the final state the integration 
over their phase space becomes very complicated and cannot be done analytically. Finally, each individual transition amplitude
$|\mathcal M_{O_{\bf 20'}\to X}|^2$ suffers from infrared divergences and requires regularization. Infrared divergences cancel however
in the sum over all final states between contribution involving different number of particles.\footnote{See however the last paragraph in Sect.~2.}

In the next section we describe another approach to computing the various double correlations in $\mathcal N=4$ SYM. It makes efficient use of the superconformal symmetry 
of the theory. It also allows us to go to higher loops and even to strong coupling, via the AdS/CFT correspondence.

\section{Weighted cross sections from correlation functions} 

\begin{figure}[h!t]
\psfrag{q}[cc][cc]{$q^\mu$}
\psfrag{1}[cc][cc]{$1$}\psfrag{2}[cc][cc]{$2$}\psfrag{3}[cc][cc]{$3$}\psfrag{4}[cc][cc]{$4$}
\psfrag{Cor}[lc][cc]{$ \Longrightarrow \, \quad  {\rm Limit_{2,3}} \ \vev{0| O(1) O(2) O(3) O(4)|0}$}
\psfrag{T1}[cc][cc]{} 
\psfrag{T2}[cc][cc]{} 
\psfrag{Null}[cc][cc]{}
\centerline{\includegraphics[height=50mm]{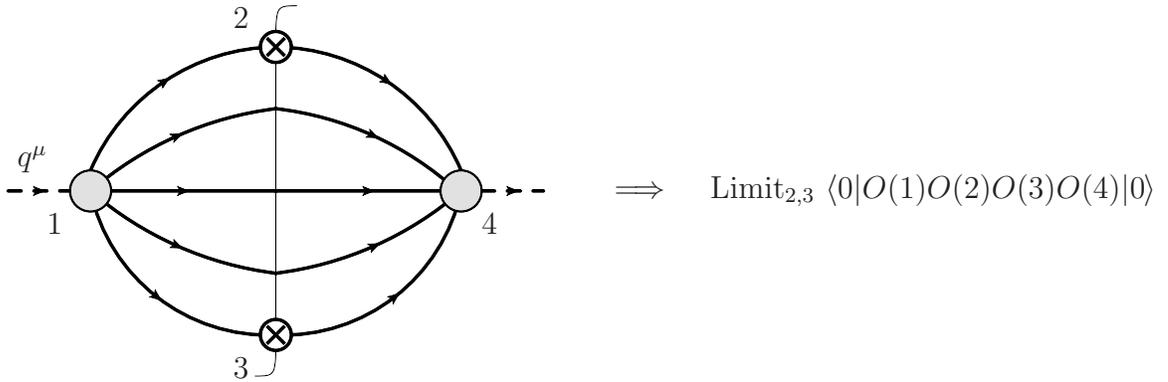}}
\caption{
\small The relation between the weighted cross section and Wightman correlation function. The operators at points $1$ and $4$ describe the source and sink, 
respectively. The operators at points 
$2$ and $3$ define the flow operators shown by crosses. `Limit${}_{2,3}$' stands for the  detector limit which amounts to sending the operators at point $2$ and $3$ at null infinity
with subsequent integration over their light-cone coordinates (see Eqs.~\re{E-new} and \re{4.4'} below).}
\label{fig:detectors}
\end{figure}

In this section we shall exploit the relation between physical observables and correlation functions involving two half-BPS operators as the source and sink, and 
various flow operators, Eqs.~\re{sigma-E}, \re{sigma-Q} and \re{sigma-O}, as the detectors. Unlike  the 
more familiar  Euclidean correlation functions, widely discussed in the $\mathcal N=4$ SYM literature, the operators on the right-hand side of 
\re{sigma-E}, \re{sigma-Q} and \re{sigma-O}  are essentially Minkowskian and are non-time ordered. This means that we will be dealing with 
Wightman correlation functions.  As we have shown in the previous section, they define  the charge flow correlations in the
detector limit (see Fig.\ \ref{fig:detectors}). 

In this section we explain how the Wightman correlation functions in \re{sigma-E}, \re{sigma-Q} and \re{sigma-O} 
can be obtained from their Euclidean counterparts by analytic continuation. The Euclidean correlation functions have singularities at $x_{ij}^2=0$, corresponding to short-distance
separation between the operators, $x_i\to x_j$. In Minkowski space, additional singularities appear when the
operators become light-like separated. In this case the analytic properties of the correlation function crucially depend on the 
ordering of the operators (time ordering versus Wightman). Therefore, performing the analytic continuation of the
correlation functions from Euclid to Minkowski we have to pay special attention to their analytic properties.

\subsection{Lorentz covariant definition of the detectors }\label{Lcdd}

In Section~\ref{sect-wcs} we have defined the flow operators \re{E-flow}, \re{Q-flow} and \re{O-flow} in the rest frame of the source
$q^\mu=(q^0,\vec 0)$. To make use of the conformal symmetry of the correlation functions, we have to restore the Lorentz covariance and extend the  definitions to any reference frame.
 
We recall that the time integral in the definition of the flow operators \re{E-flow}, \re{Q-flow} and \re{O-flow}  
has the interpretation of the working time of the detector located at the position  $r\vec n$ relative to the collision point. The space-time
coordinate of the detector $x^\mu=(t,r\vec n)$ can be decomposed in the basis of  two light-like vectors,
\begin{align}\label{x}
& x^\mu = x_+ n^\mu + x_- \bar n^\mu \,,\qquad n^\mu=(1,\vec n)\,,\qquad \bar n^\mu=(1,-\vec n)\,,
\end{align} 
with $x_+=\frac1{2} (t+r) ={(x\bar n)/2}$ and $x_- = \frac1{2} (t-r)={(xn)/2}$.
We can restore manifest Lorentz covariance by rescaling each light-like vectors by an {\it arbitrary} positive scale,
\begin{align}\label{rhon}
n^\mu \to \rho\, n^\mu\,, \qquad \bar n^\mu \to \rho'\, \bar n^\mu \,, 
\end{align}
with $ \rho, \rho' >0$.
This lifts the restriction that the time component of $n^\mu$ is equal to $1$. 
Then  the covariant definition of the light-cone coordinates in \p{x} looks as
\begin{align}\label{xpm}
& x_+={(x\bar n)\over (n \bar n)}\,,\qquad x_-  ={(xn)\over (n \bar n)} \, .
\end{align}
Notice, however, that covariance can only be maintained if all the expressions we encounter are homogeneous under such {\it local} rescalings, allowing us 
to go back to the original 
form of vectors $n$ and $\bar n$ in \p{x}. By `local' we mean that the coordinates of each flow operator should rescale with their own, independent parameter $\rho$. 

The next step is to reformulate the detector limit, $r\to\infty$ and $0\le t <\infty$, in terms of the light-cone variables $x_{\pm}$. We illustrate the correct procedure 
relying on the example of the energy 
flow. If we just take $r\to\infty$ but keep $t$ fixed, as in the original definition \re{E-flow}, we would have $x_\pm \to \pm \infty$ which is clearly too strong. 
We need to keep one of these variables 
fixed while taking the other one to infinity. In physical terms, the flow operator admits the following interpretation. We can think of a massless particle captured 
by the detector as of a wave front propagating
in the direction $n^\mu$ and spreading along the direction  $\bar n^\mu$.   This suggests to first send the detector to future infinity along, say, the light-cone direction $n^\mu$
and then integrate over the position of the massless particles on the wave front along the direction $\bar n^\mu$. In terms of the light-cone coordinates this means that we first take the 
limit $x_+\to\infty$, whereas $x_-$ remains finite. The time integral in  \re{E-flow} then becomes an integral over $-\infty < x_- < \infty$. This brings us to the new definition 
\begin{align}\label{E-new}
\mathcal E(n) =  (n\bar n) \int_{-\infty}^\infty  dx_-  \lim_{x_+\to\infty} x_+^2 \,T_{++}(x_+ n  + x_- \bar n)\,,
\end{align}
in terms of the covariant light-cone component of the stress tensor
$T_{++}\equiv \bar n^\mu\bar n^\nu T_{\mu\nu}(x)/(n\bar n)^2$.  Under the rescaling \p{rhon} the flow operator transforms homogeneously with the 
weight $(-3)$, $\mathcal E \to  \rho^{-3}\,  \mathcal E$, 
as required for maintaining Lorentz covariance. 

In a similar manner, we can define the Lorentz covariant generalization of the charge and scalar flow operators, Eqs.~\re{Q-flow} and \re{O-flow},
\begin{align}\notag
\mathcal Q_A^B(n) & =   (n\bar n)  \int_{-\infty}^\infty  dx_-  \lim_{x_+\to\infty} x_+^2 \,( J_+  )_A^B(x_+ n  + x_- \bar n)  \,,
\\
\mathcal  O^{IJ}(n)   & =  (n\bar n) \int_{-\infty}^\infty  dx_- \,  \lim_{x_+\to\infty} x_+^2 \, O_{\bf 20'}^{IJ}(x_+ n  + x_- \bar n) \,, \label{4.4'}
\end{align}
where $ J_+(x)\equiv \bar n^\mu J_\mu(x)/(n\bar n)$ is the light-cone component of the $R-$current. They have rescaling weights $(-2)$ and $(-1)$, respectively. 

The expressions in the right-hand sides of \re{E-new}  and \re{4.4'} involve the two light-like vectors $n$ and $\bar n$, 
but the dependence on the latter is redundant. To illustrate  this point, it is sufficient to rewrite these operators in terms of creation and 
annihilation operators.   For instance, we can start with \re{E-new} and repeat the calculation of \re{E-osc} to find, with the help of  \re{useful},
\begin{align}\label{E-tau}
 \mathcal E(n) = 
{1\over 2(2\pi)^3}   \int_0^\infty d \tau\,\tau^2   { \sum_{i=s,q,g} a_i^\dagger(n \tau) a_i(n \tau) }\,,
\end{align}
so that the dependence on $\bar n^\mu$ drops out. Notice that the annihilation/creation operators in \re{E-tau} depend on 
the momentum $k^\mu =n^\mu \tau$, collinear with the light-like vector defining the space-time orientation of the detector.
The charge and scalar flow operators admit similar representations (see Eq.~\re{new-rep} for the latter). An important difference is, however, that the corresponding $\tau-$integral measure becomes $\int d\tau\, \tau$ and $\int d\tau$, respectively.

Relation \re{E-tau} (together with the analogous relations for the scalar and charge flow operators) confirms  the transformation 
properties of the flow operators under the rescaling \p{rhon}  derived above.  Changing the integration variable 
$\tau\to \tau/\rho$, we find
\begin{align}\label{spin}
 \mathcal E(\rho\, n) = \rho^{-3}\,  \mathcal E(n)\,,\qquad
\mathcal Q_A^B(\rho\, n) = \rho^{-2}\, \mathcal Q_A^B(n)\,,\qquad \mathcal  O^{IJ}(\rho\,n) =  \rho^{-1}\,\mathcal  O^{IJ}(n) \,,
\end{align}  
where the factors of $\rho$ are related to the power of $\tau$ in the corresponding integral representation. As we show in the next subsection, the requirement of homogeneity under the {\it independent} rescalings \p{spin} of each flow operator imposes a strong constraint on the possible form of the correlations of the flow operators, Eqs.~\re{sigma-E}, \re{sigma-Q} 
and \re{sigma-O}.

\subsection{Symmetries}\label{se4.3}

Let us  introduce a generic notation $\mathcal D_i(n)$ for all flow operators (scalar, charge and energy).
We are interested in the symmetry properties of their correlations
\begin{align}\label{DD}
\vev{\mathcal D_1(n_1)\dots \mathcal D_\ell(n_\ell)}_q 
&=\sigma_{\rm tot}^{-1} \int d^4 x\, \e^{iqx}  \vev{0|  O(x,\widebar Y) \, \mathcal D_1( n_1)\dots  \mathcal D_\ell(n_\ell)\, O(0, Y)|0}\,.
\end{align}
Here we introduced the subscript $q$ on the left-hand side to indicate the dependence on the total momentum
transferred (see Fig.~\ref{fig:detectors}). 
By construction, the correlation function \re{DD} is  invariant under permutations of any pair of detectors.
In addition, $\vev{\mathcal D_1(n_1)\dots \mathcal D(n_\ell)}_q$ is invariant under Lorentz transformations 
of the total momentum $q^\mu$ and of the light-like vectors $n_i^\mu$ defining the detector orientation,
\begin{align}\label{Lor}
\vev{\mathcal D_1(\Lambda n_1)\dots \mathcal D_\ell(\Lambda n_\ell)}_{\Lambda q} = 
\vev{\mathcal D_1(n_1)\dots \mathcal D_\ell(n_\ell)}_q \,.
\end{align}

The correlations \p{DD} have two additional symmetries related to the independent rescaling of $n_i^\mu$ and $q^\mu$. The flow operators are defined in terms of the operators  $O_{\bf 20'}$, $J^\mu$ and $T^{\mu\nu}$ carrying Lorentz spins $0$, $1$ and $2$, respectively. According to
\re{spin}, the spin controls their transformation properties under the rescaling $n ^\mu\to \rho\,  n^\mu$, leading to 
\begin{align}\label{boost}
\vev{\mathcal D_1(\rho_1 n_1)\dots \mathcal D_\ell(\rho_\ell n_\ell)}_q =
  \rho_1^{-1-s_1}\dots \rho_\ell^{-1-s_\ell} \vev{\mathcal D_1(n_1)\dots \mathcal D_\ell(n_\ell)}_q\,,
\end{align}
where $s_{\mathcal O}=0$, $s_{\mathcal Q}=1$ and $s_{\mathcal E}=2$.

The other  scaling property is related to the rescaling $q^\mu \to \lambda q^\mu$. Since the vectors $n_i^\mu$ are dimensionless, 
this transformation can be compensated by the dilatations $x^\mu \to \lambda^{-1} x^\mu$ of the coordinates,
\begin{align}\label{dil}
\vev{\mathcal D_1(n_1)\dots \mathcal D_\ell(n_\ell)}_{\lambda q} = \lambda^{\Delta_1+\dots+\Delta_\ell} \vev{\mathcal D_1(n_1)\dots \mathcal D_\ell(n_\ell)}_{q} \,,
\end{align}
where $\Delta_i$ is the scaling dimension of the flow operator, $\Delta_{\mathcal O}=-1$, $\Delta_{\mathcal Q}=0$ and $\Delta_{\mathcal E}=1$.  Notice that all three flow operators 
have the same twist $\Delta-s=-1$.

Relations \re{Lor} -- \re{dil} allow us to fix the single-detector correlation of all flow operators up to an overall normalization factor.
 Denoting by $\mathcal D(n;s)$ the flow operator with the spin $s$ and scaling dimension $\Delta=s-1$,  we can write
\begin{align}\label{O-D}
\mathcal O(n)=\mathcal D(n;0)\,,\qquad \mathcal Q(n)=\mathcal D(n;1)\,,\qquad \mathcal E(n)=\mathcal D(n;2)\,.
\end{align}
By virtue of \re{Lor},
the single-detector correlation $\vev{\mathcal{D}(n;s)}$
depends on the two Lorentz invariants $(qn)$ and $q^2$. The dependence on the former is controlled by the spin $s$, i.e., by the required degree of homogeneity under the 
rescalings \p{boost}.  Then the $q^2-$dependence
is fixed by the scaling dimension $\Delta$, resulting in
\begin{align}\label{D-1point}
\vev{\mathcal D(n;s)} = c_{\mathcal D} {(q^2)^s \over (qn)^{s+1}}\,.
\end{align}

For the charge and energy flow correlations, we can fix the normalization constant $c_{\mathcal D}$ by making use of the fact that the flow operator
in both cases is determined by a conserved current. As a consequence, integrating over the position of the 
detector on  the sphere $n^\mu=(1,\vec n)$ with $\vec n^2=1$ in the rest frame of the source $q^\mu=(q^0,\vec 0)$, we should find  the corresponding conserved
quantity  --  the total energy $ \int d \Omega_{\vec n} \vev{\mathcal E(\vec n)}=q^0$ and the total charge $ \int d \Omega_{\vec n} \vev{\mathcal Q(\vec n)}=4\vev{Q}$  of the 
state.\footnote{Equivalently, the three-point functions of two scalars with a current or a stress-energy satisfy a Ward identity which relates their normalization to that of the two-point functions.} In this way we get
\begin{align}\label{4.14}
c_{\mathcal E} = 1/  {\textstyle \int} d \Omega_{\vec n}  = {1\over 4\pi}\,,\qquad  c_{\mathcal Q} =  {\vev{Q}\over \pi}\,.
\end{align}

Let us now consider the double-detector correlation $\vev{\mathcal D(n;s) \mathcal D(n';s')}$. The main difference,
compared to the previous case, is that it is not fixed anymore by  symmetries. The reason for this is that we can define
a ratio invariant under the transformations \re{Lor} -- \re{dil},
\begin{align}\label{z}
z = {q^2 (n n') \over 2 (q n) (q n')}\,.
\end{align}
Here we introduced the factor of $2$ to ensure that $0 \leq z \leq 1$ for a time-like vector $q$ and  light-like vectors $n$ and $n'$. 
In particular, in the rest frame of the source $q^\mu=(q^0,\vec 0)$, the variable $z$ is related to the angle between
the two detectors, $z=(1-\cos\theta)/2$. This variable already appeared in the amplitude context, see \p{3.16}.

Thus, all the symmetries mentioned above allow us to determine the double-detector correlation up to an arbitrary function of the invariant variable 
$z$,\footnote{For certain  detectors $\mathcal{D}$ the function $\mathcal{F}_{ss'}$ will also depend on  the auxiliary `isotopic' variables, see below.}
\begin{align}\label{F(z)}
\vev{\mathcal D(n;s) \mathcal D(n';s')} =   { (q^2)^{s'-1} (qn')^{s-s'} \over (nn')^{s+1}} 
  {\mathcal F_{ss'}(z)\over  4\pi^2} \,.
\end{align}
{In what follows we shall refer to $\mathcal F_{ss'}(z)$ as the  \textit{event shape function}. 
The symmetry of the correlator in the left-hand side of \re{F(z)} under the exchange of the flow 
operators leads to the crossing symmetry $\mathcal F_{ss'}(z)=(2z)^{s'-s}\mathcal F_{s's}(z)$.} 
Equation \re{F(z)} is consistent with the fact that $\vev{\mathcal D(n;s) \mathcal D(n';s')}$ is related to the four-point correlation 
function which is fixed by conformal symmetry up to a function depending on two conformally invariant cross-ratios. In the detector limit, i.e., 
after sending $r \to \infty$ and the subsequent time integration, this function effectively depends on a single variable. It is straightforward to extend 
\re{F(z)} to the multi-detector correlations \re{DD}
involving an arbitrary number of flow operators. In that case, $\vev{\mathcal D_1(n_1)\dots \mathcal D_\ell(n_\ell)}$ is given by the product
of single-detector correlations \re{D-1point} times an arbitrary function depending on the  relative angles between the  detectors. 

We present below a list  of double-detector correlations involving the three types of detectors considered in this paper. Combining together \re{F(z)} and \re{O-D}, we get 
\begin{align}\label{diag}\notag
&\vev{{\cal E}(n) {\cal E}(n') } = {q^2\mathcal F_{{\cal E}{\cal E}}(z)\over  4\pi^2 (nn')^3} \,, &&\vev{{\cal E}(n) {\cal Q}(n') } = {(qn')\mathcal F_{{\cal E}{\cal Q}}(z)\over  4\pi^2   (nn')^{3}} \,, 
\\ \notag
&\vev{ {\cal Q}(n) {\cal Q}(n') } = {\mathcal F_{{\cal Q}{\cal Q}}(z) \over 4\pi^2 (nn')^2} \,, &&\vev{ {\cal E}(n) {\cal O}(n') } = {(qn')^2\mathcal F_{{\cal E}{\cal O}}(z) \over 4\pi^2 q^2 (nn')^{3}} \,, 
\\
&\vev{ {\cal O}(n) {\cal O}(n') } = {\mathcal F_{{\cal O}{\cal O}}(z) \over 4\pi^2 q^2 (nn')}\, ,&&\vev{ {\cal Q}(n) {\cal O}(n') } = {(qn') \mathcal F_{{\cal Q}{\cal O}}(z) \over 4\pi^2  q^2 (nn')^2}\, ,
\end{align}
where  $\mathcal F(z)$ are arbitrary functions of $z$. Matching these relations with the amplitude results 
\re{EE}, \re{OO} and \re{C.6}, we find to the lowest order in the coupling  
\begin{align}\notag \label{funs}
\mathcal F_{{\cal E}{\cal E}}(z) &= - a {z \ln (1-z)\over  1-z} + O(a^2)\,,
\\\notag
\mathcal F_{{\cal Q}{\cal Q}}(z) &= - { 4} a 
\left({2z \ln(1-z) \over 1-z} \vev{Q}\vev{Q'}+\ln(1-z) \vev{Q,Q'}  \right) + O(a^2)\,,
\\
\mathcal F_{{\cal O}{\cal O}}(z) &=- {a} {z \ln(1-z)\over 1-z}\vev{S}\vev{S'}
 + O(a^2)\,,
\end{align}
and similarly for the mixed correlations from \p{3.18},
\begin{align} \notag
\mathcal F_{{\cal E}{\cal Q}}(z) &=-  {  8} {a} {z\ln(1-z)\over  1-z }\vev{Q'}
 + O(a^2)\,, \phantom{aaaaaaaaaaaaaaaaaaaaaa}
\\\notag
\mathcal F_{{\cal E}{\cal O}}(z) &=-   {  2} {a } {z\ln(1-z)\over  1-z }\vev{S'}
 + O(a^2)\,,
\\
\mathcal F_{{\cal Q}{\cal O}}(z) &=- { 4} {a} {z \ln(1-z)\over  1-z }\vev{Q}\vev{S'}
 + O(a^2)\,.  \label{4.40}
\end{align}
Notice that $\mathcal{F}_{\mathcal OO} (z)$ agrees (up to an overall factor) with the result presented in \cite{Engelund:2012re}.
We observe that the lowest order corrections to all event shape functions (except $\mathcal F_{{\cal Q}{\cal Q}}(z)$) involve the same function $ z\ln(1-z)/(1-z)$. 
We come back to  this issue in Section~\ref{Sect:GenMasterFormulae}.
    
\subsection{Single-detector correlations}\label{sect42}

As an illustration of the general considerations above, in this subsection we revisit the calculation of the detector  correlations with a single detector 
insertion, $\vev{\mathcal O(n)}$, $\vev{\mathcal Q(n)}$ and $\vev{\mathcal E(n)}$, by using their relation to the three-point correlation functions of the 
scalar half-BPS operators $O_{\bf 20'}$, the $R-$symmetry current and the energy-momentum tensor.  The latter was discussed previously in Ref.\ \cite{Hofman:2008ar}. 
Here we illustrate the main steps needed to obtain the cross section from the correlation function: analytic continuation from Euclidean to Wightman functions, 
taking the detector limit, performing the time integrals and the Fourier transform. 

\subsubsection{Single scalar detector}\label{4.2.1}

According to \re{sigma-O}, the single scalar detector correlation is given by 
\begin{align}\label{4.5}
\vev{\mathcal O( n)}  
=\sigma_{\rm tot}^{-1}  \int d^4 x_1\, \e^{iqx_1}  \vev{0|  O(x_1, \widebar Y) \, S^{IJ}\mathcal O^{IJ}(n) \, O(0, Y)|0}  \,,
\end{align}
with the normalization factor defined in \re{tot-ex}. At the lowest order in the coupling, $\vev{\mathcal O(\vec n)}$ was 
computed in \p{O-single} from the amplitudes in the rest frame of the source. 
Let us obtain the same result from the three-point Wightman function of the half-BPS operators 
by inserting \re{4.4'} into \re{4.5}:
\begin{align}\notag \label{det-lim}
\vev{\mathcal O( n)}  
 & =   {S^{IJ}   \over\sigma_{\rm tot}}   \int d^4 x_1\, \e^{iqx_1}   
 \int_{-\infty}^\infty  dx_{2-}  (n\bar n) 
 \\[2mm]
 & \times
  \lim_{x_{2+}\to\infty} x_{2+}^2 \,  
\vev{0|  O(x_1, \widebar Y) \, O_{\bf 20'}^{IJ}(x_{2+} n  + x_{2-} \bar n) \, O(0, Y)|0}_W \ .
\end{align}
We do this in three steps: 
\begin{enumerate}
\item[(i)] start with the expression for the Euclidean three-point correlation function
of  the half-BPS operators  $O_{\bf 20'}^{IJ}(x)$ and project the $SO(6)$ indices according to \re{det-lim}; 
\item[(ii)] perform an analytic 
continuation to obtain the Wightman function in Minkowski space-time; 
\item[(iii)] take the detector limit (by sending $x_+\to\infty$ and 
integrating over $-\infty < x_-<\infty$) and make the Fourier transform with respect to the position of the source to introduce the total momentum $q^\mu$. 
 \end{enumerate}
 
The correlation functions of the half-BPS operators $\mathcal O^{IJ}_{\bf 20'}$ have a number of remarkable properties in
$\mathcal N=4$ SYM. First of all, the conformal weight of $\mathcal O^{IJ}_{\bf 20'}$ is protected from quantum corrections and, as a 
consequence, their three-point correlation function in Euclidean space,
\begin{align}\label{4.6}
G_E(1,2,3) &= \vev{0|O(x_1,Y_1) O(x_2,Y_2) O(x_3,Y_3)|0 }_E  \,,
\end{align}
 is fixed by conformal symmetry up to an overall normalization 
factor. Furthermore, this factor also does not receive quantum corrections, meaning  that the three-point correlation
function \re{4.6} is given by its Born-level expression. Since $\mathcal O(x_i,Y_i)$ is bilinear in the
scalar fields, the latter expression reduces to the product of three free scalar propagators $D_E(x) \sim 1/x^2$:
\begin{align}\label{4.7}
G_E(1,2,3) =  (N^2_c-1) (Y_1 Y_2)(Y_2 Y_3)(Y_1 Y_3) D_E(x_{12}) D_E(x_{23})
 D_E(x_{13})\,.
 \end{align}
Here $(Y_i Y_j) = \sum_{I=1}^6 Y_i^I Y_j^I$ are the invariant contractions of the auxiliary $SO(6)$ complex null vectors $Y^I_i$ which help us to keep track of the $R-$symmetry structure.  In the context of the correlation function, they play the role of the coordinates in the internal (isotopic) 
$SU(4)$ space and are treated on equal footing with the space-time coordinates $x_i$ of the operators.
 As explained above, the definition of the single-detector correlation \p{4.5} involves the Wightman function $G_W(1,2,3)$ in Minkowski
 space-time.
 Knowing the expression in Euclidean space \p{4.7}, we can obtain $G_W(1,2,3)$  by simply replacing the Euclidean scalar propagators by their Wightman counterparts, 
 \begin{align}\label{an3}
G_W(1,2,3) =G_E(1,2,3) \bigg|_{D_E(x)\to D_W(x) }\,.
 \end{align}
 Here $D_W(x)$ is given by the two-point (non-time ordered!) correlation function of free scalar fields in Minkowski space-time
\begin{align}
 \vev{0|\Phi^I(x_i) \Phi^J(x_j)|0} =\delta^{IJ} D_W(x_{ij}) = -{\delta^{IJ}\over 4\pi^2} {1\over x_{ij}^2-i\epsilon x_{ij}^0}\,.
\end{align} 
We would like to emphasize that the analytic continuation in \re{an3} relied on the simplicity of the three-point correlation function \re{4.7}.
Later in the paper we shall consider the four-point correlation function of half-BPS operators $G_E(1,2,3,4)$. It is not protected anymore
and the perturbative corrections to $G_E(1,2,3,4)$ are described by a complicated function of the conformal cross-ratios. It then becomes 
a non-trivial task to find its analytic continuation $G_W(1,2,3,4)$.

The Wightman function \re{an3} depends on three auxiliary isotopic null vectors $Y_i$ (with $i=1,2,3$). In the case of the scalar flow correlation \p{4.5} the choice of the analogous auxiliary variables 
was dictated by the requirement for $\vev{\mathcal O(n)}$ to have real values. This is why we associate the complex null vector $Y$ with the source and its conjugate $\widebar Y$ with the sink, while the detector matrix $S_{IJ}$ is chosen real. In the case of the correlation function, the variables $Y_i$ are {\it holomorphic coordinates} of the half-BPS operators in the internal space. However, what really matters is that we 
can always remove the auxiliary variables $Y_i$ to reveal the $R-$tensor structure of the correlation function, and then project it
with the new set of auxiliary variables $Y, \overline Y, S$ appearing in \p{4.5}. In the case of the Wightman function \re{an3}, this amounts to 
the  substitution rule  for the isotopic variables $Y_1^I \to  \widebar Y^I$, $Y_3^I \to Y^I$ and $Y^I_2 Y^J_2 \to S^{IJ}$,  
leading to  
\begin{align}
(Y_1 Y_2)(Y_2 Y_3)(Y_1 Y_3) \ \to \ (Y\widebar Y) (Y S \widebar Y) \equiv  (Y\widebar Y)^2 \vev{S} \,.
\end{align}
For a more detailed discussion of the two sets of auxiliary isotopic variables and their matching see Appendix~\ref{appE}.  

The next step is to take the detector limit of the Wightman function $G_W(1,2,3)$. To match the correlation function on the right-hand side
of \re{det-lim}, we put $x_3=0$ and identify $x_2^\mu = x_{2+} n^\mu + x_{2-}\bar n ^\mu$ to be the detector coordinate. Then, for $x_{2+}\to\infty$
we make use of the relations   $x^2_{12} -i 0  x_{12}^0 \to 2x_{2+} (  x_{2-} (n\bar n)-(x_{1}n) +i 0)$ and $x^2_{2} -i\ep x_{2}^0 \to 2x_{2+} (x_{2-} (n\bar n)- i\ep) $
to find from \re{an3} and \re{4.7}
\begin{align}\label{4.11}
\lim_{x_{2+} \to \infty}  x^2_{2+} G_W(1,2,3)  \sim \frac{(Y\widebar Y)^2 \vev{S} }{(x_{2-}(n\bar n)-(x_{1}n)+i\ep) (x_{2-}(n\bar n) - i\ep)(x^2_1 -i\ep x^0_1)}\,.
\end{align}
According to \re{det-lim}, we have to integrate this expression over the detector light-cone coordinate $x_{2-}$.   The expression 
on the right-hand side of \p{4.11} has two poles  in $x_{2-}$ located on the opposite sides of the real axis. Closing the integration contour 
in, say, the upper half-plane, we get
\begin{align}\label{det3}
(n\bar n) \int_{-\infty}^\infty dx_{2-}   \lim_{x_{2+} \to \infty}  x^2_{2+} G_W(1,2,3)  \sim \frac{(Y\widebar Y)^2 \vev{S}}{ ((x_1n) -i\ep) (x_1^2 -i\ep x_1^0)}\,.
\end{align}
Notice that,  as announced in Sect.~\ref{Lcdd}, the second light-like vector $\bar n$ has dropped out of the right-hand side. 
Finally, we perform the  Fourier transform of \p{det3} and collect various factors to obtain 
\begin{align}\label{4.13}
\vev{\mathcal O(n)}=\sigma_{\rm tot}^{-1}\int d^4 x\, \e^{iqx}   \int_{-\infty}^\infty dx_{2-}  (n\bar n)  \lim_{x_{2+} \to \infty}  x^2_{2+} G_W(1,2,3) = {1\over 2\pi} {\vev{S} \over (qn)}  \,.
\end{align}
In the rest frame of the source $q^\mu=(q^0,\vec 0)$ this relation coincides with \re{O-single} and agrees with the general form \p{D-1point}.

\subsubsection{Single charge detector}\label{4.2.2}

Let us repeat the above analysis for the single-detector correlation of the charge flow operator \re{sigma-Q},
\begin{align}\label{4.17}
\vev{\mathcal Q(n)}  ={Q^B_A\over \sigma_{\rm tot}} \int d^4 x\, \e^{iqx}  \vev{0|  O(x,\widebar Y) \, \mathcal Q^A_B(n)\, O(0, Y)|0}\,.
\end{align}
Replacing $\mathcal Q^A_B(n)$ with its definition \re{4.4'}, we find that it is related to the three-point Wightman correlation function of two half-BPS 
operators $O_{\bf 20'}$ and the $R-$symmetry current $J$,
\begin{align}\label{3-W}
Q^B_A\bar n_\mu  \vev{0|O(x_1, \widebar Y) (J^\mu)_B^A(x_2) O(x_3, Y)|0}_W\,,
\end{align}
upon an appropriate identification of the coordinates.

As before, we start with the Euclidean version of three-point function \re{3-W}. This is is another example of a protected correlation function, hence it is given by the Born level result  
\begin{align}\label{4.18} \notag
G_E{}^\mu & \equiv \vev{0|O^{I_1 J_1}(x_1) (J^\mu)_B^A(x_2) O^{I_3 J_3}(x_3)|0}_E 
\\
& =- i { N^2_c-1 \over 16\pi^6} \frac{ \left (\Gamma^{\{I_1 \{I_3}\right)_B^A\ \delta^{J_1\} J_3\}}}{x^2_{12} x^2_{23} x^2_{13}}  \, \left(\frac{x_{12}^\mu}{ x^2_{12}}+
\frac{x_{23}^\mu}{ x^2_{23}}\right) \,,
\end{align}
where $\Gamma^{IJ}$ is the generator of the fundamental representation of $SU(4)$ defined in \p{SM} and $\{IJ\}$ denotes the traceless symmetrization of the pairs 
of indices $I_1 J_1$ and $I_3 J_3$, as required by the index structure of the operators $O_{\bf 20'}$.
It is easy to check that the expression in the right-hand side of \re{4.18} satisfies the current conservation condition at point 2 and is conformally
covariant with the relevant scaling weight at each point.  

The expression for $\vev{\mathcal Q(n)}$ in \re{4.17}   involves a projection of the $SU(4)$ tensor structure in 
\re{4.18} with  $\widebar Y^{I_1}\widebar Y^{J_1}\, Q^B_A\, Y^{I_3} Y^{J_3}$, resulting in  \begin{align}\label{Q-str}
(Y \widebar Y)   \ \widebar Y^I (\Gamma^{IJ})^A_B\,  Q^B_A  Y^J   = \frac14\tr(y\bar y) \tr( y Q\bar y) \,,
\end{align}
where $ \tr( y Q\bar y) =  y_{AB} Q^B_C \bar y^{CA}$ and the variables  $y_{AB}$ and $\bar y^{AB}$ are defined in \re{def1}.
 
As in the previous case, the Wightman correlation function \re{3-W} can be obtained from the Euclidean one in \re{4.18}
by replacing the intervals (see footnote \ref{foot0} below) as $x^2_{ij} \ \to \ -(x^2_{ij} - i\ep x^0_{ij})$ (for $i<j$) in the right-hand side of \re{4.18}.  Further, 
we specify the coordinates of the operators as $x_3=0$ and $x_2^\mu = x_{2+} n^\mu + x_{2-}\bar n ^\mu$ and obtain in the detector limit $x_{2+}\to\infty$
\begin{align}\label{4.19}
&\lim_{x_{2+} \to \infty}  x^2_{2+} G_W{}^\mu  \sim \frac{n^\mu (x_{1}n)}{ ((x_{1}n)- x_{2-}(n\bar n) -i\ep) ^2(x_{2-} (n\bar n) - i\ep)^2(x^2_1 -i\ep x^0_1)} \,.
\end{align}
Notice that this expression is proportional to the light-like vector $n^\mu$ determining the orientation of the detector. Next, the integral over the detector time produces  
\begin{align}\label{}
(n\bar n)\int_{-\infty}^\infty dx_{2-}   \lim_{x_{2+} \to \infty}  x^2_{2+}\bar n_\mu  G_W{}^\mu  \sim \frac1{ ((x_1n) -i\ep)^2 (x_1^2 -i\ep x_1^0)}\,.
\end{align} 
Finally, performing the Fourier integral and reinstating the $y/Q-$structure from \p{Q-str}, we obtain
\begin{align}\label{4.22}
\vev{\mathcal Q(n)}    &= \sigma_{\rm tot}^{-1}\int d^4 x_1\, \e^{iqx_1}   \int_{-\infty}^\infty dx_{2-}   (n\bar n)  \lim_{x_{2+} \to \infty}  x^2_{2+}\bar n_\mu  G_W{}^\mu
  =  \frac{\vev{Q}}{\pi} \frac{ q^2 }{(qn)^2} \,,
\end{align}
where $\vev{Q}$ was defined in \re{3.10}. In the rest frame of the source, this relation coincides with \p{Q-single} and agrees with  \re{4.14}.

\subsubsection{Single energy detector}\label{4.2.3}

Here we essentially repeat the calculation from  \cite{Hofman:2008ar}.  The single-detector correlation of the energy-flow operator \re{E-new} takes the form
\begin{align}\label{E-1}
\vev{{\cal E}(n)} = \sigma_{\rm tot}^{-1}  \int d^4 x\, \e^{iqx}  \vev{0|  O(x,\widebar Y) \, \mathcal E(n)\, O(0, Y)|0}\,.
\end{align}
As follows from  \re{E-new} and \re{E-1}, the energy correlator $\vev{{\cal E}(n)}$ is determined by the  three-point function of 
two half-BPS operators and one energy-momentum tensor. In Euclidean space it is given by
\begin{align}\label{4.24}\notag
(G_E)^{\mu\nu}(1,2,3) &= \vev{0|O(x_1, \widebar Y) T^{\mu\nu}(x_2) O(x_3, Y)|0 }_E 
\\[2mm] &
 = -  \frac{ N_c^2 - 1}{16 \pi^6}  \frac{(Y \widebar Y)^2}{ x^2_{12} x^2_{23} x^2_{13}}  \, \left(\frac{x_{12}^\mu}{ x^2_{12}}+\frac{x_{23}^\mu}{ x^2_{23}}\right) \left(\frac{x_{12}^\nu}{ x^2_{12}}+\frac{x_{23}^\nu}{ x^2_{23}}\right).  
\end{align}
Since the energy-momentum tensor is an $SO(6)$ singlet,  the $R-$symmetry tensor structure in \p{4.24} is much simpler than before. 
Moreover, the isotopic factor in \p{4.24}  is canceled by that of the total transition probability $\sigma_{\rm tot}$ in the right-hand side of \re{E-1}.
Repeating the steps outlined above, analytic continuation to the Wightman function, detector limit and time integration (see \p{E-new}), followed by a Fourier transform, we obtain
\begin{align}\label{HM}
\vev{{\cal E}(n)} =   {1\over 4\pi}\frac{(q^2)^2}{(qn)^3} \,.
\end{align} 
For $q^\mu=(q^0,\vec 0)$, we reproduce the result of the amplitude calculation \p{E-single}, in accord  with  \re{4.14}.
 
\section{Double-detector correlations from four-point correlation functions}

As was explained in the previous section, starting with two (or more) detector insertions we have to deal with Wightman 
correlation functions involving four (or more) operators -- two half-BPS operators serving as the source/sink,  and more than one flow operators serving as detectors.  
Unlike the single-detector case discussed above, such functions are not protected from quantum corrections and
we expect them to have a complicated form order-by-order in the coupling constant. As a consequence, the question
about the analytic continuation of the Wightman functions from their Euclidean counterparts becomes very non-trivial.
In this section we explain the procedure on the example of  the double-detector correlations.
 
\subsection{Energy-momentum tensor supermultiplet} \label{s5.1}

We recall that the flow operators \re{E-new} and \re{4.4'} are built from three protected operators -- the 
half-BPS scalar operator, the $R-$current and the energy-momentum tensor. It is well known that in $\mathcal N=4$ 
SYM these operators
belong to the same supermultiplet $\mathcal T(x,Y,\theta,\bar\theta)$,
\begin{align}\label{T}
\mathcal T(x,Y,\theta,\bar\theta) = O(x,Y) + \dots + (\theta \sigma^\mu \bar\theta )J_{\mu }(x,Y) + \dots + 
(\theta \sigma^\mu \bar\theta)(\theta\sigma^\nu \bar\theta) T_{\mu\nu}(x)+\ldots\,,
\end{align}
where we do not display the $SU(4)$ indices. 
Its lowest component is the half-BPS operator \re{OY} and the higher components are obtained
by successive use of $\mathcal N=4$ supersymmetry transformations. 

The supermultiplet \re{T} has a  number of remarkable properties. 
First of all, it is annihilated by half of the $\mathcal N=4$ supercharges
and, as a consequence, its expansion in powers of the Grassmann variables  $(\theta,\bar\theta)$ is shorter than one might 
expect.  In particular, one corollary of supersymmetry is the conservation of the $R-$current $J_\mu$ and of the energy-momentum tensor $T_{\mu\nu}$. 
Secondly, $\mathcal N=4$ superconformal symmetry imposes strong constraints on the correlation
functions of the superfield \re{T}. To compute the double-detector correlations, we need the four-point Euclidean 
(super)correlation function
\begin{align}\label{superG}
\mathcal G_E(1,2,3,4)= \vev{0|\mathcal T(x_1,Y_1,\theta_1,\bar\theta_1)\dots\mathcal T(x_4,Y_4,\theta_4,\bar\theta_4)|0}\,.
\end{align} 
Setting $\theta_{1,4}=\bar\theta_{1,4}=0$ in \re{superG}, we select  the  lowest component $O$ at  points 1 and 4, to play the role of the source/sink.  The flow operators  \re{E-new}
or \re{4.4'} are then given by the relevant components of the superfields  at points 2 and 3. Setting all odd variables to zero, we obtain the four-point correlation function of half-BPS operators  
\begin{align}\label{lowG}
G_E(1,2,3,4)=\vev{0|O(x_1,Y_1)O(x_2,Y_2)O(x_3,Y_3) O(x_4,Y_4)|0}\,.
\end{align}
This is the starting point for the analysis of the scalar-scalar flow correlation. 
It is known \cite{Eden:1999gh,Eden:1999kw} that $\mathcal N=4$ superconformal symmetry allows one to reconstruct the complete super-correlation function \re{superG}
in terms of its lowest component \re{lowG}. Each term in the $(\theta,\bar\theta)$ expansion of  \re{superG} is given by a particular differential operator acting on the coordinates $x_i$ 
of the scalar operators in \re{lowG}.
Most importantly, these differential operators do not depend on the coupling constant.
Their explicit form for the cases of interest (the components involving one or two $R-$currents and/or energy-momentum tensors) will be worked out in a forthcoming paper. Thus, to 
compute the quantum corrections to the super-correlation function \re{superG} it
suffices to know the correlation function of the four half-BPS operators \re{lowG}.

\subsection{Four-point correlation function of half-BPS operators}\label{s5.2}

The Euclidean four-point function of the half-BPS operators $O_{\mathbf{20'}}$ has been studied  extensively in 
\cite{GonzalezRey:1998tk,Eden:1998hh,Eden:1999kh,Eden:2000bk,Eden:2000mv,Bianchi:2000hn}. It is convenient to split 
it into a sum of two terms, $G_E=G_E^{\rm (Born)}+G_E^{\rm (loop)}$, describing the Born approximation and the quantum corrections, respectively.

At Born level, the correlation function \re{lowG} is given by products of free scalar propagators with the appropriate isotopic factors,  $Y_1^I Y_2^J\vev{\Phi^I(x_1) \Phi^J(x_2)} \sim y^2_{12}/x^2_{12}$, 
where we have introduced the 
shorthand notation $y_{ij}^2=(Y_i  Y_j)$. Thus, we  obtain
\begin{align}\label{8.15}
G_E^{\rm (Born)}(1,2,3,4) 
&=  {N_c^2-1 \over (4\pi^2)^4}\left(\frac{y^2_{12}}{ x^2_{12}} \frac{y^2_{23}}{ x^2_{23}} \frac{y^2_{34}}{ x^2_{34}} \frac{y^2_{14}}{ x^2_{14}}  
+\frac{y^2_{13}}{ x^2_{13}} \frac{y^2_{23}}{ x^2_{23}} \frac{y^2_{24}}{ x^2_{24}} \frac{y^2_{14}}{ x^2_{14}}
+ \frac{y^2_{12}}{ x^2_{12}} \frac{y^2_{24}}{ x^2_{24}} \frac{y^2_{34}}{ x^2_{34}} \frac{y^2_{13}}{ x^2_{13}}  \right)\nt
&  +  {(N_c^2-1)^2 \over 4(4\pi^2)^4}\left(\frac{y^4_{12}}{ x^4_{12}} \frac{y^4_{34}}{ x^4_{34}}  + \frac{y^4_{13}}{ x^4_{13}} \frac{y^4_{24}}{ x^4_{24}}  + \frac{y^4_{14}}{ x^4_{14}} \frac{y^4_{23}}{ x^4_{23}}\right)\,.
\end{align}
It contains six distinct terms corresponding to the six channels in the tensor product of $SU(4)$ irreps $\mathbf{20'} \times \mathbf{20'} = \mathbf{1} + \mathbf{15} + \mathbf{20'} + \mathbf{84} + \mathbf{105} + \mathbf{175}$ (see Appendix~\ref{appE}).
A remarkable feature of the correlation function \p{lowG} is that the loop corrections in all six channels are given by a single function  of two variables \cite{Eden:2000bk},
\begin{align}\label{1.1}
G_E^{\rm (loop)}(1,2,3,4) \nonumber
=  {2(N_c^2-1) \over (4\pi^2)^4}\bigg[  \frac{y_{12}^2y_{23}^2y_{34}^2y_{41}^2}{x_{12}^2x_{23}^2x_{34}^2x_{41}^2}(1-u-v)+\frac{y_{12}^2y_{13}^2y_{24}^2y_{34}^2}{x_{12}^2x_{13}^2x_{24}^2x_{34}^2}(v-u-1)
\\[2mm]
+   \frac{y_{13}^2y_{14}^2y_{23}^2y_{24}^2}{x_{13}^2x_{14}^2x_{23}^2x_{24}^2}(u-v-1)+\frac{y_{12}^4y_{34}^4}{x_{12}^4x_{34}^4}u+\frac{y_{13}^4y_{24}^4}{x_{13}^4x_{24}^4}  +\frac{y_{14}^4y_{23}^4}{x_{14}^4x_{23}^4}v \bigg]\Phi_E(u,v)\,. 
\end{align}
Here $\Phi_E(u,v) $ depends on the two conformal cross-ratios and admits a perturbative expansion, 
\begin{align}
\Phi_E(u,v) = \sum_{\ell=1}^\infty a^\ell\, \Phi_{\ell}(u,v)\,,\qquad u=\frac{x_{12}^2x_{34}^2}{x_{13}^2x_{24}^2}\,, \qquad v=\frac{x_{23}^2x_{14}^2}{x_{13}^2x_{24}^2} \,.
\end{align}
The functions $\Phi_{\ell}(u,v)$ are currently known in terms of Euclidean scalar Feynman integrals up to six loops ($\ell=6$)  \cite{Usyukina:1992wz}. The one-loop correction is given
by the so-called one-loop box integral 
\begin{align} \label{2loops}
 \Phi_E(u,v) ={}&  a\, \Phi^{(1)}(u,v) + O(a^2) = -{ a \over 4\pi^2}\, \int    {d^4 x_0\,x_{13}^2 x_{24}^2 \over x_{10}^2 x_{20}^2 x_{30}^2 x_{40}^2}+ O(a^2)\,.
\end{align}
In Euclidean space, the function $\Phi_E(u,v)$ has logarithmic
singularities in the limit $x_{ij}^2\to 0$, that corresponds to short-distance separations between the operators.

Let us apply \re{8.15} and \re{1.1} to obtain the correlation of two scalar flow operators $\vev{ {\cal O}(n){\cal O}(n')}$, Eq.~\re{sigma-O}, from the correlation
function $G_E(1,2,3,4)$. At the first step, we associate the operators at points 4 and 1 with the source $O(0,Y)$ (we set $x_4=0$) and the conjugate sink
$O(x_1,\widebar Y)$, respectively. Further, the operators  at points 2 and 3 are associated with the two detectors, i.e., the scalar flow operators $\mathcal O(n,S)$ 
and $\mathcal O(n',S')$, involving the real detector matrices $S_{IJ}$ and $S'_{IJ}$. The description of the correlation function in \p{8.15} and \p{1.1} is holomorphic in the 
auxiliary null vectors $Y_i$ at all four points. So, to match the isotopic structures of the two objects, we have to make the  substitutions $\widebar Y \to Y_1$, 
$Y \to Y_4$ and $S_{IJ} \to Y_{2I} Y_{2J}$, $S'_{IJ} \to Y_{3I} Y_{3J}$. 

Next, we recall that the scalar flow operators $\mathcal O(n,S)$ and $\mathcal O(n',S')$ do not commute for generic detector matrices $S$ and $S'$ because of 
a possible cross-talk between the detectors. To avoid this unwanted cross-talk, when defining the scalar detector correlations we had to impose the additional 
condition \re{cross-talk} on the detector matrices.  Making the substitution $S_{IJ} \to Y_{2I} Y_{2J}$ and $S'_{IJ} \to Y_{3I} Y_{3J}$ described above, we find that the condition  \p{cross-talk} on the scalar flow 
detectors  is translated into the orthogonality condition 
$y_{23}^2=(Y_2 Y_3)=0$ for the two half-BPS operators at positions 2 and 3 in the internal $SO(6)$ space. 
This condition eliminates half of the structures in \re{8.15} and \p{1.1}, leaving us with
\begin{align}\label{hat-Born}
\widehat G_E^{\rm (Born)} 
&=  {N_c^2-1 \over (4\pi^2)^4}  \frac{y^2_{12}}{ x^2_{12}} \frac{y^2_{24}}{ x^2_{24}} \frac{y^2_{13}}{ x^2_{13}}  \frac{y^2_{34}}{ x^2_{34}}+  {(N_c^2-1)^2 
\over 4(4\pi^2)^4}\left(\frac{y^4_{12}}{ x^4_{12}} \frac{y^4_{34}}{ x^4_{34}}  + \frac{y^4_{13}}{ x^4_{13}} \frac{y^4_{24}}{ x^4_{24}} \right)\,,
\end{align}
and
\begin{align}\label{5.3}
 \widehat  G_E^{\rm (loop)}  
= {2(N_c^2-1) \over (4\pi^2)^4}\bigg[y_{12}^2y_{13}^2y_{24}^2y_{34}^2(v-u-1)  + y_{12}^4y_{34}^4  +y_{13}^4y_{24}^4\,u\bigg]{\Phi_E(u,v)\over x_{12}^2x_{13}^2x_{24}^2x_{34}^2}\,.  
\end{align}
Here we dressed the symbols in the left-hand side of these equations with hats to indicate that both expressions have been evaluated for $y_{23}^2=0$. Comparing 
these relations with the general
expressions \re{8.15} and \p{1.1}, we observe that the suppression of the cross-talk between the detectors has a simple interpretation in 
terms of the correlation function. Namely, the condition $y_{23}^2=0$ eliminates the most singular terms in \re{8.15} and \p{1.1} 
in the short-distance limit $x_{23}^2\to 0$. 

To conclude the discussion of how to match the isotopic structures of the correlation function of four half-BPS operators in \re{hat-Born} and \re{5.3} and that of the double scalar flow correlation, 
we give the translation table between the $SO(6)$ invariant $y-$ and $Y/S-$structures, after imposing the condition $y_{23}^2=0$:~\footnote{See Appendix~\ref{appE} for more details.}
\begin{align}\notag\label{y-S}
  y_{12}^2y_{13}^2y_{24}^2y_{34}^2 &\to  (Y S \widebar Y) (Y S' \widebar Y) 
  \,,
\\[2mm]  \notag
(y_{12}^2 y_{34}^2)^2 &\to (Y S Y) (\widebar Y S' \widebar Y)\,,  
\\[2mm]  
 (y_{13}^2 y_{24}^2)^2 &\to (Y S' Y) (\widebar Y S \widebar Y) 
 \,.
\end{align}

The next step is to use  relations \re{hat-Born} and \re{5.3} to obtain the Wightman correlation function of four half-BPS operators.  

\subsection{Analytic continuation}
\label{AnalyticCont}

Let us first examine the contribution to $\vev{ {\cal O}(n){\cal O}(n')}$ coming from the Born level approximation to the correlation function \re{hat-Born}.

A distinguishing feature of $\widehat G_E^{\rm (Born)}$ is that, like the three-point function \re{4.7}, it is a rational function of the distances $x_{ij}^2$.
This suggests that we can obtain the Wightman function  $\widehat G_W^{\rm (Born)}(1,2,3,4)$ by replacing $x_{ij}^2\to (x_{ij}^2-i\epsilon x_{ij}^0)$ (for $i<j$) on 
the right-hand side of \re{hat-Born}.  Notice that the Wightman function is not invariant under the exchange of points and the `$-i\epsilon x_{ij}^0$' prescription indicates
that the operator at point $x_i$ stands to the left from the operator at point $x_j$  inside $\widehat G_W^{\rm (Born)}(1,2,3,4)$.
Since we assigned the points $4$ and $1$ to the source and to its complex conjugated image (sink), respectively, and the points $2$ and $3$ to the detectors, 
the above rule should be applied to all factors of $1/x_{ij}^2$ with $i<j$ except $1/x_{23}^2$. Notice that, in virtue of the commutativity of the flow 
operators, the Wightman function should be insensitive to the ordering of the operators located at points $x_2$ and $x_3$. Indeed, it is
easy to see from \re{hat-Born} that $\widehat G_W^{\rm (Born)}$ is regular at $x_{23}^2=0$. The same is true for the expression
in the square brackets in \re{5.3}.

To compute $\vev{ {\cal O}(n){\cal O}(n')}$, we convert \re{hat-Born} into a Wightman function and go to the detector limit by putting $x_4=0$ and identifying the coordinates of points 2 and 3 as
$x_2=x_{2+}n+x_{2-}\bar n$ and $x_3=x_{3+}n'+x_{3-}\bar n'$.  Then, for $x_{2+}, x_{3+}\to\infty$, we notice that the second term
on the right-hand side of \re{hat-Born} gives rise to an expression containing double poles in $x_{2-}$ and $x_{3-}$. These poles are located in the same half-plane and,
therefore, vanish upon integration over $x_{2-}$ and $x_{3-}$. In this way, with the help of \re{y-S} we obtain 
\begin{align} \notag
\vev{ {\cal O}(n){\cal O}(n')}^{\rm (Born)}   =\sigma_{\rm tot}^{-1} & \int d^4 x_1 \e^{iqx_1}\int_{-\infty}^\infty dx_{2-}dx_{3-} 
\lim_{x_{2+}, x_{3+}\to\infty} (x_{2+} x_{3+})^2\ \widehat G_W^{\rm (Born)} 
\\ \notag
 \sim & \int d^4 x_1 \e^{iqx_1}
\int_{-\infty}^\infty 
{dx_{2-}dx_{3-}\vev{S}\vev{S'}\over (x_{12-}-i\epsilon)(x_{2-}-i\epsilon)(x_{13-}-i\epsilon)(x_{3-}-i\epsilon)}
\\
\sim & \int d^4 x_1 \e^{iqx_1} {\vev{S}\vev{S'} \over ((x_1 n)-i\epsilon) ((x_1 n')-i\epsilon)}\,,
\end{align}  
where in the second relation  $x_{12-} = (x_1n)-x_{2-}$, $x_{13-}=(x_1n')-x_{3-}$  and $\vev{S}$ was defined in \re{3.13}.
The Fourier integral in the last relation reduces to $\int_0^\infty dt dt' \delta^{(4)}(q-n t- n't')$. It is easy to see
that, in the rest frame of the source, for $q^\mu=(q^0,\vec 0)$, $n^\mu=(1,\vec n)$ and $n'^\mu=(1,\vec n')$,
it vanishes unless $\vec n=-\vec n'$. The latter case corresponds to detecting particles moving back-to-back  and
it produces a contribution proportional to $\delta(1-z)$. This result is in 
agreement with the calculation of $\vev{ {\cal O}(n){\cal O}(n')}^{\rm (Born)}$ based on the amplitudes, Eq.~\re{EE-born}. Namely, 
at Born level, for $q^\mu=(q^0,\vec 0)$, the final state in $O_{\mathbf{ 20'}}\to \text{everything}$ consists of two scalar particles 
moving back-to-back. They can be detected only for $\vec n= -\vec n'$, corresponding to $z=1$ in \re{z}. Thus,
for $z<1$, the correlation function contributes to the event shape function $\mathcal F_O(z)$ defined in \re{diag} only starting
from order $O(a)$.

Let us now repeat the same analysis for the Euclidean correlation function \re{5.3}. The main difficulty in this case is that it involves a
complicated function of the two cross-ratios. How to convert it to the Wightman function $\Phi_W(u,v)$? The answer to this
question was proposed in \cite{Luscher:1974ez} and a particularly elegant formulation was given in Ref.\ \cite{Mack:2009mi}. It relies on the Mellin representation of the correlation function,
\begin{align}\label{Mellin}
\Phi_E(u,v) = \int_{-\delta-i\infty}^{-\delta+i\infty} {d j_1 dj_2 \over (2\pi i)^2} M(j_1,j_2)\, u^{j_1} v^{j_2}\,,
\end{align}
where $u,v>0$ in Euclidean space and $M(j_1,j_2)$ is a meromorphic  function of the Mellin parameters. 

At weak coupling, the Mellin amplitude is given by a perturbative expansion in  $a=g^2/(4\pi^2)$, 
\begin{align}\label{5.15}
M(j_1,j_2;a)= \sum_{\ell \geq 1} a^\ell M^{(\ell)}(j_1,j_2)\,.
\end{align}
For our purposes here, we will only need the lowest-order term,
\begin{align}\label{Mellin-1loop}
M^{(1)}(j_1,j_2) = {- \frac{1}{4}}  \left[\Gamma(-j_1) \Gamma(-j_2)\Gamma(j_1+j_2+1)\right]^2 \,,
\end{align}
which is obtained by rewriting the one-loop box integral in \p{2loops} in the Mellin form \cite{Usyukina:1992wz}.
At strong coupling  in planar $\mathcal N=4$ SYM, the Mellin amplitude was computed in \cite{Arutyunov:1999fb,Arutyunov:2000py,Arutyunov:2000ku} via the AdS/CFT correspondence, 
\begin{align}\label{Mellin-strong}
M^{(\infty)}(j_1,j_2;a) = - \left[\Gamma(1-j_1) \Gamma(1-j_2)\Gamma(j_1+j_2+1)\right]^2{1+j_1+j_2\over  2 j_1j_2}
\,.
\end{align}

The invariance
of the correlation function $G(1,2,3,4)$ under the exchange of any pair of points $(x_i,Y_i) \leftrightarrow (x_j,Y_j)$ leads to the crossing symmetry relations 
\begin{align}
\Phi_E(u,v) = \Phi_E(v,u) ={1\over v}\Phi_E\left({u\over v},{1\over v}\right)  \,,
\end{align}
which translate into relations for the Mellin amplitude,
\begin{align}\label{Mellin-cross}
M(j_1,j_2) =M(j_2,j_1) = M(j_1,-1-j_1-j_2)\,.
\end{align}
{ It is easy to check that the Mellin amplitudes \re{Mellin-1loop} and \re{Mellin-strong} verify these relations.} 

According to \cite{Mack:2009mi}, the Mellin amplitude $M(j_1,j_2)$ is a universal function, depending neither on the space-time signature, nor on the ordering of the operators in the four-point correlation function. Extending \re{Mellin} to Minkowski space,
we have to specify the analytic continuation of $(x_{ik}^2)^j$ to negative $x_{ik}^2$ for arbitrary complex $j$. The choice
of the prescription is determined by the ordering of the operators. In the special case of the Wightman function
 $G_W(1,2,3,4)$, the prescription is $(x_{ik}^2)^j\to (-x_{ik}^2+i\epsilon x_{ik}^0)^j$ for $i<k$.
 \footnote{For time-ordered operators, the same prescription looks as $(x_{ik}^2)^j\to (-x_{ik}^2+i\epsilon)^j$ with the additional minus
 sign due to our choice of signature $(+,-,-,-)$ for Minkowski and $(+,+,+,+)$ for Euclidean space.\label{foot0}}
  In this way, we obtain from \re{Mellin}
\begin{align}\notag \label{Phi-W}
\Phi_W(1,2,3,4) =& \int_{-\delta-i\infty}^{-\delta+i\infty} {d j_1 dj_2 \over (2\pi i)^2} M(j_1,j_2;a)  \,(-x_{13}^2+i\epsilon x_{13}^0)^{-j_1-j_2} (-x_{24}^2+i\epsilon x_{24}^0)^{-j_1-j_2}
\\[2mm]
&\times  {(-x_{12}^2+i\epsilon x_{12}^0)^{j_1} (-x_{34}^2+i\epsilon x_{34}^0)^{j_1}} 
 {(-x_{23}^2+i\epsilon x_{23}^0)^{j_2} (-x_{14}^2+i\epsilon x_{14}^0)^{j_2} } \,.
\end{align}
Notice that $\Phi_W$ is locally conformally invariant. It coincides with $\Phi_E(u,v)$ for all $x_{ij}^2<0$ and differs otherwise.  
 
For our purposes we will only need the Wightman function \re{Phi-W} in the detector limit. As before, we assign points $4$ and $1$
to the source and sink, respectively, and points $2$ and $3$ to the detectors. We put $x_4=0$, 
$x_2=x_{2+}n+x_{2-}\bar n$, $x_3=x_{3+}n'+x_{3-}\bar n'$ and take the limit $x_{2+}, x_{3+}\to\infty$
to get \footnote{Here we have assumed that the detector limit commutes with the Mellin integral.
}
\begin{align} \label{Mellin-limit}
\lim_{x_{2+}, x_{3+}\to\infty}\Phi_W(1,2,3,4) = \int_{-\delta-i\infty}^{-\delta+i\infty} {d j_1 dj_2 \over (2\pi i)^2} M(j_1,j_2;a)  
f(j_1,j_2+1)\,,
\end{align}
where the notation was introduced for the function
\begin{align}\notag\label{f}
 f(j_1,j_2) & =\big( (nn')/2\big)^{-j_1-j_2} \big(-x_{1}^2+i\epsilon x_{1}^0\big)^{-j_1-j_2}  
\\[1.5mm] \notag
& \times  \big((x_1n)-x_{2-}(n\bar n)-i\epsilon\big)^{j_1} \big(-x_{2-}(n\bar n)+i\epsilon\big)^{j_2}
 \\[1.5mm] &\times
 \big((x_1n')-x_{3-}(n'\bar n')-i\epsilon\big)^{j_2}   \big(-x_{3-}(n'\bar n')+i\epsilon\big)^{j_1}\,.
\end{align}
In what follows the dependence of $f(j_1,j_2)$ on the coordinates $x_1$, $x_{2-}$ and $x_{3-}$ is tacitly assumed.
To obtain \re{Mellin-limit},  we replaced the integration variable $j_2\to -1-j_1-j_2$ and made use of \re{Mellin-cross}.  Viewed as a function 
of the detector times $x_{2-}$ and $x_{3-}$, the expression in the right-hand side of \re{f} has
singularities located on both sides of the real axis. This ensures that the integrals over the detector times
$x_{2-}$ and $x_{3-}$  do not vanish.

\subsection{Master formulas} 
 
We are now ready to perform the detector limit  of the four-point Wightman correlation function. Combining together
\re{5.3}, \re{y-S}  and \re{Mellin-limit}, we find
\begin{align}\notag\label{G4-limit}
\lim_{x_{2+}, x_{3+}\to\infty} (x_{2+}x_{3+})^2 \,\widehat  G_W^{\rm (loop)}   
= {{4} \sigma_{\rm tot}\over (2\pi)^7 (nn')^2} \big(-x_{1}^2+i\epsilon x_{1}^0\big)^{-2}
  \int_{-\delta-i\infty}^{-\delta+i\infty} {d j_1 dj_2 \over (2\pi i)^2} M(j_1,j_2;a)   
\\[2mm]
 \times \bigg[\vev{S}\vev{S'}f(j_1-1,j_2-1)
+   \vev{S,S'} f(j_1-1,j_2) +\widebar{  \vev{S,S'}} f(j_1,j_2-1)\bigg] 
\,, 
\end{align}
where $\sigma_{\rm tot}$ is given by \re{tot-ex}.  Here we see  three independent $R-$symmetry  structures,
\begin{align}\label{S-factor}\notag
& \vev{S,S'} = { (YSY)(\widebar Y S' \widebar Y) - (Y S\widebar Y) (Y S' \widebar Y) \over (Y\widebar Y)^2} \,, \qquad
 \\
&  \widebar{  \vev{S,S'}}= \vev{S,S'}^*\,,\qquad
 \vev{S} = { (Y S\widebar Y)  \over (Y\widebar Y)}\,.
 \end{align}
Each of them is  accompanied by the function $f(j_1, j_2)$ with shifted arguments.

Finally, to obtain the correlation $\vev{\mathcal O(n) \mathcal O(n')}$ we have 
to integrate both sides of \re{G4-limit} over the detectors times, $x_{2-}$ and $x_{3-}$, and perform the Fourier transform
with respect to $x_1$. Assuming that  the order of integrations can be exchanged, we find from \re{G4-limit} 
\begin{align}\notag\label{OO-dec}
 \vev{\mathcal O(n) \mathcal O(n')}   = { {{1} \over 4 \pi^2 q^2 (nn')} }
  \int_{-\delta-i\infty}^{-\delta+i\infty} {d j_1 dj_2 \over (2\pi i)^2} M(j_1,j_2;a)   \bigg[\vev{S}\vev{S'}K(j_1,j_2;z) &
\\[2mm]
+   \vev{S,S'} K(j_1,j_2+1;z) +\widebar{\vev{S,S'}} K(j_1+1,j_2;z)\bigg] & ,
\end{align} 
where we introduced a notation for
\begin{align}\label{K-fun}\notag
 K(j_1,j_2;z) =&  {{1 \over 8 \pi^5}} {q^2\over (nn')}\int { d^4x_1 \e^{iqx_1} \over (-x_{1}^2+i\epsilon x_{1}^0)^{2}} 
 \\
 &\times
 \int_{-\infty}^\infty  dx_{2-}  (n\bar n)  \int_{-\infty}^\infty  dx_{3-}  (n'\bar n')\,f(j_1-1,j_2-1; x_1, x_{2-},x_{3-})\,.
\end{align} 
Changing the integration variables, $x_{2-}\to x_{2-}/(n\bar n)$ and $x_{3-}\to x_{3-}/(n'\bar n')$,  
we verify using \re{f} that $K(j_1,j_2;z)$ is dimensionless and  invariant under 
(independent) rescalings of the light-like vectors $n$ and $n'$. Therefore,  $K(j_1,j_2;z)$ can depend
on the four-dimensional vectors
only through the scaling variable $z$ defined in \re{z}. 
 Going through the calculation
of \re{K-fun} we find (see Appendix \ref{app:K} for details) 
 \begin{align} \label{KO}
K(j_1,j_2;z) & = \lr{z\over 1-z}^{1-j_1-j_2} {2\pi \over \sin(\pi(j_1+j_2))[\Gamma(j_1+j_2)\Gamma (1 - j_1)\Gamma (1 - j_2)]^2} \,,
\end{align} 
where $0< z< 1$. The function $K (j_1,j_2;z)$ characterizes the scalar detectors. 
It is a symmetric function of $j_1$ and $j_2$ and, most importantly,  it is independent of the coupling constant. In what follows we refer to this functions as the {\it detector kernel}.
  
Equation \re{OO-dec} agrees with the general expression for the scalar detector correlation in \re{diag}. It leads to the following remarkable {\it master formula}  
for the event shape function $\mathcal F_{\mathcal \mathcal {O O}}(z)$
\begin{align}\label{master}
 \mathcal F_{\mathcal {O O}}(z) = \vev{S}\vev{S'} \mathcal F^{+}_{\mathcal {O O}}(z) +\left(\vev{S,S'}   +\widebar{\vev{S,S'}} \right) \mathcal F^{-}_{\mathcal {O O}}(z)\,,
\end{align}
where  $\mathcal F^{\pm}_{\mathcal {O O}}(z)$  are given by the Mellin integrals
\begin{align}\notag \label{scal-fun}
\mathcal F^{+}_{\mathcal {O O}}(z) &=   \int_{-\delta-i\infty}^{-\delta+i\infty} {d j_1 dj_2 \over (2\pi i)^2} M(j_1,j_2;a)K(j_1,j_2;z)\,,
 \\[2mm]
 \mathcal F^{-}_{\mathcal {O O}}(z) &=  \int_{-\delta-i\infty}^{-\delta+i\infty} {d j_1 dj_2 \over (2\pi i)^2} M(j_1,j_2;a)K (j_1,j_2+1;z)\,.
\end{align}
These relations establish the correspondence between the four-point Euclidean correlation function of half-BPS operators
and the double-scalar detector correlation. Given the correlation function $G_E(1,2,3,4)$ in the form \p{1.1}, we start by extracting the Mellin 
amplitude $M(j_1,j_2;a)$ from \p{Mellin}. Then we obtain the event shape function $\mathcal F_{\mathcal O}(z)$ by integrating the Mellin amplitude 
in the Mellin space with the kernel $K(j_1,j_2;z)$, 
which is uniquely fixed by the choice of the detectors. In this representation, the dependence on the coupling constant resides in
the Mellin amplitude  $M(j_1,j_2;a)$, whereas the $z-$dependence comes from the detector kernel  $K(j_1,j_2;z)$. 
  
\subsection{One-loop check} 
\label{OneLoopCheck} 
  
To illustrate the power of the master formula  \re{master}, let us use the known one-loop expression for the Mellin amplitude \re{Mellin-1loop} 
to compute the event shape functions \re{scal-fun} to the lowest order in the coupling. Taking into account \re{KO}, we find 
\begin{align} \notag  
& M(j_1,j_2;a)K(j_1,j_2;z) =   - {a \over 2} \lr{z\over 1-z}^{1-j_1-j_2} {\pi (j_1+j_2)^2\over (j_1 j_2)^2\sin(\pi(j_1+j_2)) }  + O(a^2)\,,
\\ \label{K-minus}
& M(j_1,j_2;a)K(j_1,j_2+1;z) = {a \over 2} \lr{z\over 1-z}^{-j_1-j_2} {\pi \over j_1^2 \sin(\pi(j_1+j_2))  }+ O(a^2)\,.
\end{align}
To perform the Mellin integration in \re{scal-fun} it is useful to shift  the integration variable as $j_2\to j_2-j_1$. In
this way, we obtain  
\begin{align}\label{Kplus}
\mathcal F^{+}_{\mathcal {O O}}(z) &=  - \frac{a}{2} 
\int_{-\delta-i\infty}^{-\delta+i\infty} {d j_1 dj_2 \over (2\pi i)^2}  {\pi \ j_2^2\over (j_1 (j_2-j_1))^2\sin(\pi j_2) } \lr{z\over 1-z}^{1-j_2}+ O(a^2) \,,
\end{align}
where the $j_2-$integration contour goes to the left from the one over $j_1$. The integral can be easily computed by residue method after
closing the $j_1-$integration contour in the right half-plane and the $j_2-$integration contour  in the left half-plane. Then, the integral
is given by the residues at $j_1=0$ and $j_2=-k$ (with $k=1,2,\dots$)
\begin{align}\label{Fp1}
\mathcal F^{+}_{\mathcal {O O}}(z) &=  - a  
\sum_{k=1}^{\infty}{(-1)^k\over k}\lr{z\over 1-z}^{1+k} + O(a^2)=  { - a} {z\over 1-z} \ln (1-z)+ O(a^2)\, .
\end{align} 
Going through the same steps for \re{K-minus}  we find
\begin{align}
\mathcal F^{-}_{\mathcal {O O}}(z) &= {  \frac{a}{2}} \int_{-\delta-i\infty}^{-\delta+i\infty} {d j_1 dj_2 \over (2\pi i)^2} {\pi \over  j_1^2\,\sin(\pi j_2) } \lr{z\over 1-z}^{-j_2}+ O(a^2) \,,
\end{align}
where the integration contours are the same as in \re{Kplus}. The main difference, however, is that the integrand has only a 
double pole at $j_1=0$ and, therefore, the $j_1-$integral vanishes,
\begin{align}\label{Fm1}
\mathcal F^{-}_{\mathcal {O O}}(z)  = O(a^2)\,.
\end{align}
Substituting \re{Fp1} and \re{Fm1} into the master formula \re{master} we notice that, firstly, the one-loop correction to the event shape function 
does not receive a contribution proportional to $\vev{S,S'}$ and, secondly, the coefficient  of $\vev{S}\vev{S'}$ coincides
with the one-loop result \re{funs} based on the amplitude calculation.

\section{Generalization of the master formulas}
\label{Sect:GenMasterFormulae}

In this section we extend our previous analysis to the general case of double-detector correlations \re{F(z)} and \re{diag} involving scalar, charge 
and energy flow operators.  As was explained in Sect.~\ref{se4.3}, these correlations are uniquely specified by the event shape functions 
depending on the scaling variable  $0<z<1$. 

To compute the event shape functions \re{F(z)} and \re{diag}, we can follow the same procedure as before. Namely, we start with the four-point 
correlation functions involving two half-BPS operators (for the source and sink) and various components of the stress-tensor supermultiplet \re{T}
(for the two detectors), analytically continue them to get the corresponding Wightman functions, and finally go to the detector limit to
compute the double-detector correlations. We recall that the various four-point functions of this type appear as particular components in the expansion
of the super correlation function $\mathcal G_E(1,2,3,4)$, Eq.~\re{superG}, in powers of the Grassmann variables.

According to Sect.~\ref{s5.1}, the super correlation function can be obtained from its lowest component $G_E(1,2,3,4)$ by applying a (complicated) 
differential operator. This operator does not depend on the coupling constant, so the perturbative corrections to $\mathcal G_E(1,2,3,4)$
are described by the unique scalar function $\Phi(u,v)$ from \re{1.1} and by its derivatives. Therefore, in order to get the Wightman function 
$\mathcal G_W(1,2,3,4)$ it suffices to replace $\Phi(u,v)$ by its Wightman counterpart $\Phi_W(u,v)$ defined in \re{Phi-W}. 
Then, using the Mellin representation \re{Mellin-limit} and going through the same steps as in the previous subsection, we obtain the 
following general representation for the event shape function \re{diag}%
\footnote{Here we give a short summary of the results, the details will be presented elsewhere. In obtaining them we have used the Mathematica package {\tt xAct} \cite{xAct}, especially the application package {\tt Spinors} \cite{GomezLobo:2011xv}.}
\begin{align}\label{Fgen}\notag
& \mathcal F_{\mathcal{AB}}(z) = \sum_R  \omega_{R}\mathcal F_{\mathcal{AB};R}(z)\,, 
\\
& \mathcal F_{\mathcal{AB}; R}(z) =   \int {dj_1 dj_2\over (2\pi i)^2} M(j_1,j_2;a)  K_{\mathcal{AB};R} (j_1,j_2;z)
\,.
\end{align}
Here the subscripts  $\mathcal{A,B}= \mathcal{O,Q,E}$ denote the type of detectors (scalar, charge or energy) and
$M(j_1,j_2;a)$ is the universal (detector independent) Mellin amplitude defined in \re{Mellin} and \p{5.15}. 

For the scalar and charge detectors, the event shape function \re{Fgen} also depends on the detector matrices $S_{IJ}$ and $Q_A^B$, respectively.
This dependence enters into \re{Fgen} through the $R-$symmetry invariant factors $\omega_R$ built from the matrices  $S, Q$ and from
the auxiliary variables $Y, \widebar Y$ defining the source and the sink (see Eqs.~\re{Q-factor} and \re{S-factor}). The set 
of such factors is in one-to-one correspondence with the set of $R-$symmetry structures that appear in the four-point
correlation function $\vev{0|O(1, \widebar Y) \mathcal A(2) \mathcal B(3) O(4, Y)|0}$, which we use to compute the event shape function \re{Fgen}. Denoting  
the $R-$symmetry representations of the two detectors by ${\bf A}$ and ${\bf B}$, respectively, (${\bf 20'}$ for scalar, ${\bf 15}$ for charge and ${\bf 1}$ for energy flow), 
we can identify the above mentioned structures with the overlap of  irreducible representations in the tensor products ${\bf A}\times {\bf B}$ and ${\bf 20'}\times {\bf 20'}$. 
They are labelled by the
index $R$ on the right-hand side of \re{Fgen}. In the detector limit, each irreducible component $R$ of the four-point correlation function gives rise to  the 
detector kernel $K_{\mathcal{AB};R} (j_1,j_2;z)$. As was already explained, this kernel depends on the Mellin variables $j_1$ and $j_2$ and the 
scaling variable $z$ but does not depend on the coupling constant. 

Let us first consider the kernels $K_{\mathcal{EE}}$, $K_{\mathcal{EQ}}$ and $K_{\mathcal{EO}}$ involving the energy flow. Since the energy flow 
is an $R-$symmetry singlet, the overlap of the two tensor products ${\bf 1}\times {\bf B}$ and ${\bf 20'}\times {\bf 20'}$ consists of a single irreducible structure
and, therefore,  the sum in \re{Fgen} reduces to a single term involving $\omega^{\mathcal{EE}}_{\mathbf {1}}$, $\omega^{\mathcal{EQ}}_{\mathbf{15}}$ and $\omega^{\mathcal{EO}}_{\mathbf{20'}}$   (see \p{E.7}). 
Quite remarkably, in these three cases the detector kernels turn out to be equal to the scalar-scalar kernel  $K$ defined in \p{KO}:  
 \begin{align}\label{K-pro}
K_{\mathcal{EE};\mathbf {1}}=    K_{\mathcal{EQ};\mathbf {15}}=   K_{\mathcal{EO};\mathbf {20'}}   = K(j_1,j_2;z) \,.
\end{align} 
As a consequence of these relations, the corresponding
event shape functions $\mathcal F_{\mathcal{EE}}(z)$, $\mathcal F_{\mathcal{EQ}}(z)$ and $\mathcal F_{\mathcal{EO}}(z)$  are proportional to each other for
any coupling constant. 
We interpret this as a corollary of $\cN=4$ superconformal symmetry which uniquely fixes the correlation functions $\vev{OTTO}$, $\vev{OTJO}$ and $\vev{OTOO}$, starting from $\vev{OOOO}$. 

The remaining kernels  $K_{\mathcal{OO}}$, $K_{\mathcal{QO}}$ and $K_{\mathcal{QQ}}$ have a more complicated structure. We found however that, 
similarly to \re{K-pro}, they can be expressed in terms of the function $\kernel$, Eq.~\re{KO}. 
For  ${\cal A}={\cal B}={\cal O}$ the range of irreducible representations $R$ in \re{Fgen} corresponds to the tensor product
\begin{align}
\label{E.1}
&\mathbf{20'} \times \mathbf{20'} = \mathbf{1} + \mathbf{15} + \mathbf{20'} + \mathbf{84} + \mathbf{105} + \mathbf{175}\,,
\end{align}
and the six structures $\omega^{\mathcal{OO}}_R$ are listed in  \p{6.17}. 
The kernel for each channel can be written as a differential operator acting on $\kernel$, 
\begin{align}\label{Shifts}
K_{\mathcal{O}\mathcal{O}; R}(j_1,j_2;z)= P_R^{\mathcal{O}\mathcal{O}}(\shifta,\shiftb) \  \kernel(j_1,j_2;z)\,,  
\end{align}
where $P_R^{\mathcal{O}\mathcal{O}}$ are polynomials in $t_{1,2}$ listed  in Table~\ref{Tab:ScalarScalar}
and $ t_i =\e^{\partial_{j_i}}$ (for  $i=1,2$) act as shifts of the arguments of the kernel, $j_i\to j_i+1$.
\begin{table}[H!tp]\centerline{\scalebox{1.0}{
\rotatebox{0}{$\begin{array}{|c|c|c|c|}\hline
&&&\\[-8pt]
\langle\mathcal{O}\mathcal{O}\rangle & P_R^{\mathcal{O}\mathcal{O}}(\shifta,\shiftb) & \mathcal{F}_{\mathcal{O}\mathcal{O};R}^{\,\text{(1)}} & \mathcal{F}_{\mathcal{O}\mathcal{O};R}^{\,(\infty)} \\[4pt]\hline\hline
&&&\\[-8pt]
\mathbf{1} & \shifta^2+\shiftb^2+4\shifta\shiftb-\frac{4}{5}(\shifta+\shiftb)+\frac{1}{10} &  \infty & \infty \\[4pt]\hline
&&&\\[-8pt]
\mathbf{15} &  \shifta^2-\shiftb^2-\frac{1}{2}(\shifta-\shiftb) &  0  & 0 \\[4pt]\hline
&&&\\[-8pt]
\mathbf{20'} &  (\shifta-\shiftb)^2-\frac{1}{2}(\shifta+\shiftb)+\frac{1}{10} &  -\frac{1}{10}  \frac{z\ln (1-z)+10(1-z)\ln z}{1-z}    & \infty \\[4pt]\hline
&&&\\[-8pt]
\mathbf{84} & 3(\shifta+\shiftb)-1  &    \frac{z\ln (1-z)}{ 1-z } &  \infty \\[4pt]\hline
&&&\\[-8pt]
\mathbf{175} & \shifta-\shiftb & 0 & 0 \\[4pt]\hline
&&&\\[-8pt]
\mathbf{105} & 1 &  \parbox{1.15cm}{$-\frac{z\ln (1-z)}{1-z}$\\[-4pt]} &  2z^3 \\[4pt]\hline
\end{array}$}}}
\caption{\small Scalar-scalar correlations. The first column lists the various channels $R$ in the tensor product \re{E.1}. The second column gives the polynomial of shift operators which acts on the kernel $\kernel(j_1,j_2;z)$, Eq.~\re{KO}, according to \re{Shifts}. The third and fourth columns list the event shape functions $ \mathcal{F}_{\mathcal{O}\mathcal{O};R} $  at one loop and at strong coupling, respectively. 
 They were obtained from \re{Fgen} using \re{Mellin-1loop} and \re{Mellin-strong}. 
 }
    \label{Tab:ScalarScalar}
\end{table}

Several comments are in order regarding this table. Each polynomial $P_R^{\mathcal{OO}}(\shifta,\shiftb)$ corresponds to a particular irreducible representation $R$ of  $SU(4)$ appearing in the decomposition \re{E.1}. Comparing these polynomials
with those listed in  \p{E.6}, we notice that $P_R^{\mathcal{OO}}(\shifta,\shiftb)$ coincide
with the eigenfunctions of the quadratic Casimir of $SU(4)$ in the `mirror'  representation 
\begin{align}
(P_{\mathbf{1}}^{\mathcal{OO}},P_{\mathbf{15}}^{\mathcal{OO}},P_{\mathbf{20'}}^{\mathcal{OO}},P_{\mathbf{84}}^{\mathcal{OO}},P_{\mathbf{175}}^{\mathcal{OO}},P_{\mathbf{105}}^{\mathcal{OO}})=(\cY_{\mathbf{105}},\cY_{\mathbf{175}},\cY_{\mathbf{84}},\cY_{\mathbf{20'}},\cY_{\mathbf{15}},\cY_{\mathbf{1}})\,.\label{NOlist}
\end{align}
In particular,  we have $P_{\mathbf{105}}^{\mathcal{OO}}=\cY_{\mathbf{1}}=1$, with the corollary
\begin{align}\label{6.9}
K_{\mathcal{O}\mathcal{O}; \mathbf{105}} = K(j_1,j_2;z) \,.
\end{align}

Secondly, from Table~1 we see that the event shape functions $\mathcal{F}_{\mathcal{OO}}(z)$ in the channels $\mathbf{15}$ and  $\mathbf{175}$ 
vanish identically. This is due to the antisymmetry of the kernels $K_{\mathcal{O}\mathcal{O}; \mathbf{15}}$ and $K_{\mathcal{O}\mathcal{O}; \mathbf{175}}$ 
under the exchange of $j_1\leftrightarrow j_2$, which is incompatible with the symmetries of $M(j_1,j_2)$, as can be seen in (\ref{Mellin-cross}).

Thirdly, Table~1 lists the contributions to $\mathcal{F}_{\mathcal{OO}}(z)$ in two regimes, at weak coupling (one loop) and at strong coupling\footnote{The 
correlation function and its Mellin amplitude at strong coupling were computed in \cite{Arutyunov:1999fb,Arutyunov:2000py,Arutyunov:2000ku}. More details 
about the event shapes at strong coupling will be presented elsewhere.}. We observe the presence of divergences, at one loop in the singlet channel, and at 
strong coupling in all non-vanishing channels but the $\mathbf{105}$. 
The appearance of divergences at weak coupling is not surprising and is related to the cross-talk between scalar detectors, as explained in Sect.~\ref{sect-OO}.
We recall that the cross-talk can be eliminated by imposing the additional condition on the detector matrices, $[S,S'] =0$, Eq.~\re{cross-talk}. 
As explained in Sect.~\ref{s5.2} (see also Eq.~\p{E5}), this condition has the effect of suppressing half of the $y-$structures in \p{8.15} and \p{1.1}, see \p{hat-Born} 
and \p{5.3}. Equivalently, half of the irreps on the right-hand side of \p{E.1} also drop out, namely $\mathbf{1}$, $\mathbf{15}$ and $\mathbf{20'}$. So,  we can 
reinterpret $[S,S']=0$ as a condition for removing some of the divergences in the Mellin integrals. In the absence of this condition the channel $\mathbf{20'}$ 
remains finite at one loop, but diverges at strong coupling\footnote{A mechanism to tame them is discussed at length in Refs.\ \cite{Hofman:2008ar,PaperI}.}.  
The two remaining channels  $\mathbf{84}$ and $\mathbf{105}$ give finite contributions at one loop, but the former diverges at strong coupling. We can avoid 
this divergence if we  impose a stronger condition, $Y_2=Y_3$ (see \p{E6}).  Its effect is to suppress all channels but the $\mathbf{105}$, which turns out to be 
finite both at one loop and at strong coupling. 

Our next choice of detectors is $\mathcal{A}=\mathcal{Q}$ and $\mathcal{B}=\mathcal{O}$. In this case, the sum in \re{Fgen} runs over the overlap of irreducible
representations in the tensor product
\begin{align}
\mathbf{15} \times \mathbf{20'} =  \mathbf{15} + \mathbf{20'} +\mathbf{45} + \mathbf{\overline{45}}  + \mathbf{175}\,,\label{E.3}
\end{align}
with those in \re{E.1}, i.e. $R=\mathbf{15}, \mathbf{20'}, \mathbf{175}$. We can represent the results in a  manner similar to  the scalar case discussed above, introducing 
\begin{align}
K_{\mathcal{Q}\mathcal{O},R}(j_1,j_2;z)=P_R^{\mathcal{Q}\mathcal{O}}(\shifta,\shiftb) \kernel(j_1,j_2;z)\,,
\end{align}
where the kernel $\kernel$ and  $\shifta, \shiftb$, are defined in Eqs.~(\ref{KO}) and below (\ref{Shifts}), respectively.  
The results are summarized in Table~\ref{Tab:CurrentScalar}.
\begin{table}[H!tp]\centerline{\scalebox{1.0}{
\rotatebox{0}{$\begin{array}{|c|c|c|c|}\hline
&&&\\[-8pt]
\langle\mathcal{Q}\mathcal{O}\rangle & P_R^{\mathcal{Q}\mathcal{O}}(\shifta,\shiftb) & \mathcal{F}_{\mathcal{Q}\mathcal{O};R}^{\,\text{(1)}} 
& \mathcal{F}_{\mathcal{Q}\mathcal{O};R}^{\,(\infty)} \\[4pt]\hline\hline
&&&\\[-8pt]
\mathbf{15} &\frac{1}{2} (\shifta+\shiftb-\frac{1}{4}) &  \frac{z\ln(1-z)}{8(1-z)} & \infty \\[4pt]\hline
&&&\\[-8pt]
\mathbf{20'} &  \frac{1}{2}(\shifta-\shiftb) &  0  & 0 \\[4pt]\hline
&&&\\[-8pt]
\mathbf{175} & 1 &    - \frac{z \ln (1-z)}{1-z}    & 2z^3 \\[4pt]\hline
\end{array}$}}}
\caption{Charge-scalar correlations.  The channel $\mathbf{20'}$ vanishes identically for symmetry reasons. The channel $\mathbf{15}$ is removed by the  condition \p{E6}.  }
    \label{Tab:CurrentScalar}
\end{table}
The channel $\mathbf{20'}$ is antisymmetric under the exchange of $(j_1,j_2)$ which results in an identically  vanishing contribution. The channel $\mathbf{15}$, while 
finite at one loop, diverges at strong coupling and only the $\mathbf{175}$ yields a finite contribution in this regime. The strong-coupling correlations can be rendered 
finite again by imposing the stronger condition \p{E6}, which eliminates all channels apart from the $\mathbf{175}$.

With the choice ${\cal A}={\cal B}={\cal Q}$ we are dealing with the overlap of  \p{E.1} and
\begin{align}
\mathbf{15} \times \mathbf{15} =  \mathbf{1} + \mathbf{15}_s +\mathbf{15}_a + \mathbf{20'} + \mathbf{45} + \mathbf{\overline{45}}  + \mathbf{84}\,,\label{E.2}
\end{align}
which consists of  the irreps $\mathbf{1} + \mathbf{15}_a +\mathbf{15}_s + \mathbf{20'} + \mathbf{84}$ (notice the degeneracy of the $\mathbf{15}$, appearing in 
a symmetric and an antisymmetric versions). Once again, the kernels of all channels can be written in the form 
\begin{align}
K_{\mathcal{Q}\mathcal{Q},R}(j_1,j_2;z)=P_R^{\mathcal{Q}\mathcal{Q}}(\shifta,\shiftb,z)  \kernel(j_1,j_2;z)\,, 
\end{align}
and we have tabulated the polynomials $P_R^{\mathcal{Q}\mathcal{Q}}$  in Table~\ref{Tab:CurrentCurrent}. An interesting new feature is that $P_{\mathbf{1}}^{\mathcal{QQ}}$ and $P_{\mathbf{20'}}^{\mathcal{QQ}}$ explicitly depend on $z$. 
\begin{table}[H!tbp]\centerline{\scalebox{1.0}{
\rotatebox{0}{$\begin{array}{|c|c|c|c|}\hline
&&&\\[-8pt]
\langle\mathcal{Q}\mathcal{Q}\rangle& P_R^{\mathcal{QQ}}(\shifta,\shiftb,z) & \mathcal{F}_{\mathcal{Q}\mathcal{Q};R}^{\,\text{(1)}} & \mathcal{F}_{\mathcal{Q}\mathcal{Q};R}^{\,(\infty)} \\[4pt]\hline\hline
&&&\\[-8pt]
\mathbf{1} & \frac{2}{3}\left(1+15(\shifta+\shiftb)+25\frac{1-z}{z}\right)  &  -\frac{2(25-24 z) \ln \left(1-z\right)}{3 (1-z)} & \infty \\[4pt]\hline
&&&\\[-8pt]
\mathbf{15_a} & \shifta-\shiftb &  0  & 0 \\[4pt]\hline
&&&\\[-8pt]
\mathbf{15_s} & 0 &  0   & 0  \\[4pt]\hline
&&&\\[-8pt]
\mathbf{20'} & \frac{5}{3}\left(3-\frac{4}{z}\right) &    -\frac{5(3 z-4) \ln \left(1-z\right)}{3(1-z)} &  \frac{10(3 z-4) z^2}{3} \\[4pt]\hline
&&&\\[-8pt]
\mathbf{84} & 1 & - \frac{z \ln \left(1-z\right)}{ 1-z } & 2{z^3} \\[4pt]\hline
\end{array}$}}}\caption{\small Charge-charge correlations.  Notice that the polynomial in this case explicitly depends on the variable $z$.  The channel $\mathbf{15_s}$ 
vanishes identically for 
symmetry reasons. The channels $\mathbf{1}$ and $\mathbf{20'}$  are removed by the  condition $Y_2=Y_3$. }
\label{Tab:CurrentCurrent}
\end{table}
At one loop we need no condition since all five channels are finite (the two  channels $\mathbf{15_s}$ and $\mathbf{15_a}$ give identically  vanishing contributions for 
symmetry reasons). However, at strong coupling we again observe a 
divergence in the singlet channel. The difference with the previous case is that now the singlet can only be removed by imposing a stronger condition $Y_2=Y_3$. Its effect is to 
suppress all channels but the $\mathbf{84}$, which is finite both at one loop and at strong coupling. 

Summarizing the various tables above, we remark the following interesting property, generalizing  \p{K-pro} and \p{6.9}. For all choices of the two detectors, the 
kernels corresponding to the  $R-$symmetry channel with the highest value of the quadratic Casimir of $SU(4)$,  are identical: 
\begin{align}\label{6.7}
K_{\mathcal{EE};\mathbf {1}}=    K_{\mathcal{EQ};\mathbf {15}}=  K_{\mathcal{EO};\mathbf {20'}}   
= K_{\mathcal{Q}\mathcal{Q}; \mathbf{84}}  = K_{\mathcal{Q}\mathcal{O}; \mathbf{175}}  = K_{\mathcal{O}\mathcal{O}; \mathbf{105}} \,.
\end{align} 
This is a non-obvious corollary of $\cN=4$ superconformal symmetry, which will be investigated in our future work \cite{paper3}.

Finally,  putting together the one-loop results  for the event shape functions \re{Fgen} from the different tables, we exactly reproduce the one-loop results of 
the amplitude calculations (without 
rearrangement into irreducible $R-$symmetry representations) listed in  \p{funs} and \p{4.40}.  For instance,
\begin{align}
 \mathcal F_{\mathcal{QQ}}(z) = \sum_{R=\mathbf{1}, \mathbf{15}_s ,\mathbf{15}_a, \mathbf{20'},  \mathbf{84}} \omega^\mathcal{QQ}_{R}\ \mathcal F_{\mathcal{QQ};R}(z)\,,
\end{align}
with $\omega^\mathcal{QQ}_{R}$ given by \re{D9}.
    
\section{Conclusions}

In this paper we studied the weighted cross sections describing the angular distribution of various global charges (energy, $R-$charge) in the final states of  $\mathcal{N}=4$ SYM created 
from the vacuum by a source. We applied the approach developed in the companion paper \cite{PaperI} and computed them starting from Euclidean correlation functions both at weak and strong coupling. 

The starting point of our analysis was a relation between the weighted cross sections and Wightman correlation functions involving various flow operators. We defined three different types of 
the flow operators based on certain components of the $\mathcal{N}=4$ stress-tensor multiplet: the half-BPS scalar operators, the $R-$symmetry currents and the energy-momentum tensor. In 
addition, we used the half-BPS scalar operator as a source that creates the physical state out of the vacuum. Then, the weighted cross section is determined by the Wightman correlation function 
involving different components of the $\mathcal{N}=4$ stress-energy multiplet in a particular detector limit which includes sending some of the operators to the null infinity and subsequently 
integrating over their light-cone coordinates. 

The Euclidean version of such correlation functions have a number of remarkable properties in $\mathcal{N}=4$ SYM. In particular, three- and four-point correlators of operators belonging to the 
$\mathcal{N}=4$ stress-energy multiplet are uniquely determined by analogous correlation functions of half-BPS scalar operators. The corresponding relations do not depend on the coupling 
constant and will be explained in more detail in an upcoming publication. The next step consisted in the analytical continuation of the Euclidean correlation functions to their Wightman counterparts  
following the L\"uscher-Mack procedure \cite{Luscher:1974ez}. We demonstrated that the most efficient and convenient way to do this is via the Mellin representation for the correlation functions. 

In this way, we derived a master formula which yields an all-loop expression for the weighted cross sections as a convolution of the Mellin amplitude, defined by the 
Euclidean correlation function, with a coupling-independent `detector kernel' determined by the choice of the flow operators. We performed thorough checks for single- and double-detector 
correlations to leading order at weak coupling and found perfect agreement with the conventional amplitude calculations of the corresponding observables in $\mathcal{N}=4$ SYM. 
We would like to emphasize that our approach is applicable to all orders in the weak coupling expansion as well as at strong coupling.  In the latter case, we made use of the prediction for the four-point 
correlation functions of half-BPS operators, obtained via the AdS/CFT correspondence, to compute the double correlations at strong coupling. 

There are several directions in which the construction presented in this paper can be extended even further. Notice that the Mellin amplitude for the four-point correlation function of half-BPS operators 
is known to two loops and it should be possible to extend it beyond using the recent progress in computing the correlation function up to six loops \cite{Eden:2011we}. This opens up a possibility to compute 
the flow correlations at higher orders of perturbative expansion and confront them with techniques based on Keldysh-Schwinger diagram technique \cite{Keldysh:1964ud}. We will address this 
question in a forthcoming publication. 
 
Another important issue that has to be developed further  concerns  remarkable relations between the detector kernels corresponding to different flow operators and in different $R-$symmetry
channels, Eqs.~\re{K-pro}, \re{6.9} and \re{6.7}. They lead to analogous relations between the weighted cross sections. We expect that there should exist a supersymmetric Ward identity that 
directly relates various weighted cross section studied in this paper.  Finally, infrared finiteness of the charge flow correlations at weak coupling, the origin of the divergences at strong coupling in 
some $R-$symmetry channels require a more detailed study. These and other issues will be addressed elsewhere.

\section*{Acknowledgments}

We would like to thank Vladimir Braun, Simon Caron-Huot, Lance Dixon, James Drummond, Gerhard Mack, Juan Maldacena, Gavin Salam, Raymond Stora and Ivan Todorov for useful discussions. 
The work of A.B.\ was supported by the U.S.\ National Science Foundation under the grant No.\ PHY-1068286 and CNRS. G.K.\ and E.S.\ acknowledge partial support 
by the French National Agency for Research (ANR) under contract StrongInt (BLANC-SIMI-4-2011). A.Z.\ was supported in part by the U.S.\ National Science 
Foundation under Grant No.\ PHY-0756966. A.B., S.H. and A.Z.\ are grateful to the Institut de Physique Th\'eorique (Saclay) for the hospitality extended to them at various stages of 
the project. S.H., G.K. and E.S. would like  to thank the Simons Center for Geometry and Physics (Stony Brook) for hospitality during completion of this paper.

\appendix

\section{Cross-talk between scalar detectors}\label{app-cross}

By construction, the flow operators should commute with each other. As was mentioned in Sect.~\ref{sect-wcs},
this property is not obvious from their operator definition, Eqs.~\re{E-flow}, \re{Q-flow} and \re{O-flow}, due to the possibility 
for the same particle to go subsequently through various detectors. For two detectors oriented along two different 
spatial directions $\vec n$ and $\vec n'$, the momentum of such a particle should vanish. This means that the commutator of the flow operators, say $[\mathcal  O (\vec n;S),\mathcal  O (\vec n';S')]$, can
receive a contribution from particles with zero momenta only. 
To investigate this, we use the representation
of the scalar flow operator \re{Os} in terms of annihilation/creation operators of scalars combined with 
the identity\footnote{To verify this identity it suffices to integrate both sides with a test function.}
\begin{align}\label{useful}
\int_0^\infty d \tau\, \tau^\ell \, \delta^{(4)}(k-n \tau) = 2 k_0^{\ell-1}\delta_+(k^2) \delta^{(2)}(\Omega_{\vec k} - \Omega_{\vec n}) 
\end{align}
with $n^\mu=(1,\vec n)$ being a light-like vector, $n^2=0$. Thus, we obtain an  equivalent representation for the flow
operator (for $\ell=0$),
\begin{align}\label{new-rep}
\mathcal  O(\vec n; S) =  {1\over 2(2\pi)^3} \sum_{i=1}^6 S_i \int_0^\infty d \tau\,  (\phi^{i} a^\dagger(n \tau))(\phi^{i} a(n \tau))\,,
\end{align}
where $(\phi^{i} a(n \tau))\equiv \sum_{I=1}^6\phi^i_I a^I(n \tau)$.
Notice that each additional factor of $k_0$ on the right-hand side of \re{useful} is translated into a factor of $\tau$ in
the integral on the left-hand side. This implies that the charge and energy flow operators, $\mathcal  Q (\vec n;Q)$ and $\mathcal  E (\vec n)$, respectively, admit a representation similar
to \re{new-rep} with the important difference that the $\tau-$integral involves additional factors of $\tau$ and $\tau^2$, respectively (see Eq.~\re{E-tau}).

Making use of the commutation relation 
\begin{align}\label{A3}
[a^I(k),a^{\dagger J } (p)] =  (2\pi)^3 2k_0\, \delta^{(3)}(\vec k - \vec p)\delta^{IJ} \,,
\end{align}
we find from \re{new-rep}
\begin{align}\notag\label{inte}
 [\mathcal  O (\vec n;S),\mathcal  O (\vec n';S')] = 
 {1\over 2(2\pi)^3} & \sum_{i,j=1}^6 S_{i} S'_{j}  (\phi^{i}\phi'{}^{j})\int_0^\infty d \tau_1 d\tau_2\, \tau_2\,
 \delta^{(3)}(\tau_1\vec n-\tau_2\vec n' )
 \\ & \times \left[(\phi^{i} a^\dagger(n \tau_1)) (\phi'{}^{j} a(n'\tau_2))-(\phi'{}^{j} a^\dagger(n' \tau_2))  (\phi^{i} a(n \tau_1))\right],
\end{align}
where $\phi^i, \phi'{}^{j}$ and $S,S'$ define the detector matrices \re{C} and
$(\phi^{i}\phi'{}^{j})=\sum_1^6 \phi_{I}^{i}\phi_{I}'{}^{j}$. The delta function localizes the $\tau-$integral
at $\tau_1=\tau_2=0$ but leads to an ambiguous expression of the form $0\times \delta(0)$. To carefully evaluate it
we approximate the delta function and examine the following test integral involving an arbitrary test function $f(\tau_1,\tau_2)$
\begin{align}\notag
\int_0^\infty d \tau_1 d\tau_2\, \tau_2\, &
 \delta^{(3)}(\tau_1\vec n-\tau_2\vec n' ) f(\tau_1,\tau_2)  
 \\ \notag
 &
 = \lim_{\epsilon\to 0} (\pi\epsilon)^{-3/2}
 \int_0^\infty d \tau_1 d\tau_2\, \tau_2\, \e^{-(\tau_1\vec n-\tau_2\vec n')^2/\epsilon} f(\tau_1,\tau_2)
\\
& =\pi^{-3/2}
 \int_0^\infty d \tau_1 d\tau_2\, \tau_2\, \e^{-(\tau_1\vec n-\tau_2\vec n')^2}  \lim_{\epsilon\to 0} f(\epsilon^{1/2} \tau_1,\epsilon^{1/2}\tau_2)
  ={1\over 4\pi} {f(0,0)\over 1-(\vec n\vec n')} ,
\end{align}
where in the second relation we rescaled the integration variables as $\tau_i\to \epsilon^{1/2} \tau_i$. Applying this identity to an  appropriately
chosen function $f(\tau_1,\tau_2)$ we find from \re{inte}
\begin{align}\label{noncom}\notag
 [\mathcal  O (\vec n;S),\mathcal  O (\vec n';S')] &= 
 {1\over 4(2\pi)^4}  \sum_{i,j=1}^6 {S_{i} S'_{j}  (\phi^{i}\phi'{}^{j}) \over 1-(\vec n\vec n')}
 \left[(\phi^{i} a^\dagger(0)) (\phi'{}^{j} a(0))-(\phi'{}^{j} a^\dagger(0))  (\phi^{i} a(0))\right]
 \\
& =  {1\over 4(2\pi)^4} {a^{\dagger I}(0) (SS'-S'S)_{IJ}a^J(0)\over 1-(\vec n\vec n')}\,,
\end{align}
where in the second relation we applied \re{C}.

We conclude from \re{noncom} that, for general detector matrices $S$ and $S'$, the scalar flow operators do not commute.
Moreover, in agreement with our expectations, the commutator receives a contribution from particles with zero momentum only.
We recall that for the charge and energy flow operator, the integral representation analogous to \re{new-rep} contains  additional
factors of $\tau$. Calculating $ [\mathcal  O (\vec n;S),\mathcal  Q (\vec n';Q)] $ and 
$[\mathcal  O (\vec n;S),\mathcal  E (\vec n')] $ as in \re{noncom}, we find that the additional factor
of $\tau$ makes both commutators vanish. The same applies to all remaining commutators involving charge and energy flow operators.

\section{$SU(4)$ versus $SO(6)$} \label{appA}

To simplify the $R-$symmetry structure of the correlation functions, throughout the paper we use two different but equivalent
sets of isotopic (or harmonic) variables, $Y^I$ and $(y_{AB},\tilde y^{AB})$. The former defines a complex $SO(6)$ null vector $\sum_{I=1}^6 Y^I Y^I=0$, 
whereas the latter carries a pair of  $SU(4)$
indices $A,B=1,\dots,4$ and satisfies the relations
\begin{align}\label{def1}
y_{BA}=-y_{AB}\,,\qquad \tilde y^{AB} = \frac12 \epsilon^{ABCD} y_{CD}\,.
\end{align}
The  variables $Y$ and $y$ are related to each other as follows:
\begin{align}\label{def2}
Y_I = {1\over\sqrt 2} (\Sigma_I)^{AB} y_{AB}\,,\qquad  y_{AB}= {1\over\sqrt 2}\epsilon_{ABCD}(\Sigma_I)^{CD} Y_I\,,
\end{align} 
where $(\Sigma_I)^{AB}$ are the (chiral)  Dirac matrices for $SO(6)$.
They satisfy the relations 
 \begin{align}\label{def3}
\sum_{I=1}^6 (\Sigma_I)^{AB} (\Sigma_I)^{CD} = \frac12 \epsilon^{ABCD}\,,\quad
\frac12\epsilon_{ABCD}(\Sigma_I)^{AB} (\Sigma_J)^{CD} = \delta_{IJ} 
\end{align}
and can be expressed in terms of the `t Hooft symbols $\Sigma^{AB}_I = (\eta_I^{AB}, i \bar\eta_I^{AB})$ \cite{BelDerKorMan03}. Combining $(\Sigma_I)^{AB}$ with their complex conjugates $(\bar\Sigma_I)_{AB} =\overline{(\Sigma_I)^{AB}}$, we obtain the generators of the fundamental representation of $SU(4)$:
\begin{align}\label{SM}
(\Gamma_{IJ})^A_C = -(\Gamma_{JI})^A_C = \frac1{2}(\Sigma_I)^{AB}(\bar\Sigma_J)_{BC}- (I\leftrightarrow J)\,, \qquad (\Gamma_{IJ})^A_A =0\,.
\end{align}
We then use \re{def2} and \re{def3} to get the following identity
\begin{align}\label{bas}
(Y_1  Y_2)\equiv \sum_I Y_1^I Y_2^I = {1\over 4} \epsilon^{ABCD}(y_1)_{AB}(y_2)_{CD}= {1\over 2}  (y_1)_{AB} (\tilde y_2)^{AB}  = {1\over 2}  (y_2)_{AB} (\tilde y_1)^{AB}\,.
\end{align}     
In the special case $Y_1^I=Y_2^I$, or equivalently $(y_1)_{AB}=(y_2)_{AB}$, it follows from  $(Y_1Y_1)=0$ that
\begin{align}
{1\over 2} \epsilon^{ABCD}(y_1)_{AB}(y_1)_{CD}=(y_1)_{AB} (\tilde y_1)^{AB} =0\,.
\end{align}
Making use of \re{def1} and \re{def2} we can obtain two equivalent representations for various operators in terms of $SO(6)$
and/or $SU(4)$ harmonic variables. 

In particular, the projected scalar field admits two representations,
\begin{align}\label{A.3}
(Y  \Phi) = \sum_I Y^I \Phi^I =  y_{AB}\phi^{AB} =  \tilde y^{AB}\tilde\phi_{AB}\,,
\end{align}   
where $\Phi^I$ and $\phi^{AB}$ are related to each other as
\begin{align}
\phi^{AB} = {1\over\sqrt 2} (\Sigma_I)^{AB}\Phi^I\,,\qquad \tilde\phi_{AB} ={1\over 2} \epsilon_{ABCD}\phi^{CD} 
\end{align}
and the scalar field satisfies the additional reality condition $ \tilde\phi_{AB} = (\phi^{AB})^*$. Notice that we do not impose a similar
condition on $y_{AB}$ and treat $(y_{AB})^*$ as independent variables
\begin{align}
\bar y^{AB} = (y_{AB})^* \neq \tilde y^{AB}\,.
\end{align}
Then, the half-BPS operator $O(x,Y) =  \tr[ (Y  \Phi)^2]$ and its Hermitian conjugate $[O(x,Y)]^\dagger  =  \tr[ (\widebar Y  \Phi)^2] = O(x,\widebar Y) $ take the form
\begin{align}\label{O-SU(4)}
O(x,Y) =  \tr \big[ (y_{AB}\phi^{AB})^2 \big] \,,\qqqquad
O(x,\widebar Y)= \tr \big[ (\bar y^{AB}\tilde \phi_{AB})^2\big]\,,
\end{align}
where $\phi^{AB} =  \phi^{AB,a} T^a$, $\tilde \phi_{AB}=\tilde \phi_{AB}^{\ a} T^a$ and the $SU(N_c)$ generators are normalized as
$\tr[T^a T^b]=\frac12 \delta^{ab}$ (with $a,b=1,\dots,N^2_c-1$).
The total transition probability involves the two-point correlation function of such operators,
\begin{align} 
\vev{O(x_1,\widebar Y) O(x_2,Y) } &= {N^2_c-1\over 2}   \left[ \frac12  y_{A_1B_1}\bar y^{A_1 B_1}D(x_1-x_2)  \right]^2
={N^2_c-1\over 2}   \left[ (Y\widebar Y)D(x_1-x_2)  \right]^2\,,
\end{align}
where $(Y\widebar Y) = \frac12  y_{A_1B_1}\bar y^{A_1 B_1}$ and $D(x)$ denotes the propagator of a free scalar field,
\begin{align}\notag
& \vev{\Phi^{I,a}(x_1) \Phi^{J,b}(x_2)}  =  \delta^{ab} \delta^{IJ} D(x_1-x_2)\,,
\\[2mm]
& \vev{ \phi^{A_1B_1,a_1}(x_1) \tilde \phi_{A_2B_2}^{\,a_2}(x_2)} =  \frac14\delta^{a_1 a_2} \lr{\delta^{A_1}_{A_2}\delta^{B_1}_{B_2} -\delta^{A_1}_{B_2}\delta^{B_1}_{A_2} } D(x_1-x_2)\,.
\end{align}
The explicit expression for $D(x)$ contains a prescription, 
\begin{align}
D_F(x) = - {1\over 4\pi^2} {1\over x^2 - i\epsilon} \,,\qquad D_W(x) =- {1\over 4\pi^2} {1\over x^2 - i\epsilon x_0}\,,
\end{align}
for Feynman (time-ordered) and Wightman correlation functions, respectively.

\section{Scalar, charge and energy correlations at one loop}\label{app:1loop}

In this Appendix, we compute scalar-scalar, charge-charge and energy-energy  correlations to lowest order in the coupling, using amplitude techniques. 

\begin{figure}[h!t]
\psfrag{a}[cc][cc]{(a)}  
\psfrag{b}[cc][cc]{(b)}  
\psfrag{c}[cc][cc]{(c)}  
\psfrag{d}[cc][cc]{(d)}  
\centerline{\includegraphics[width=.95\textwidth]{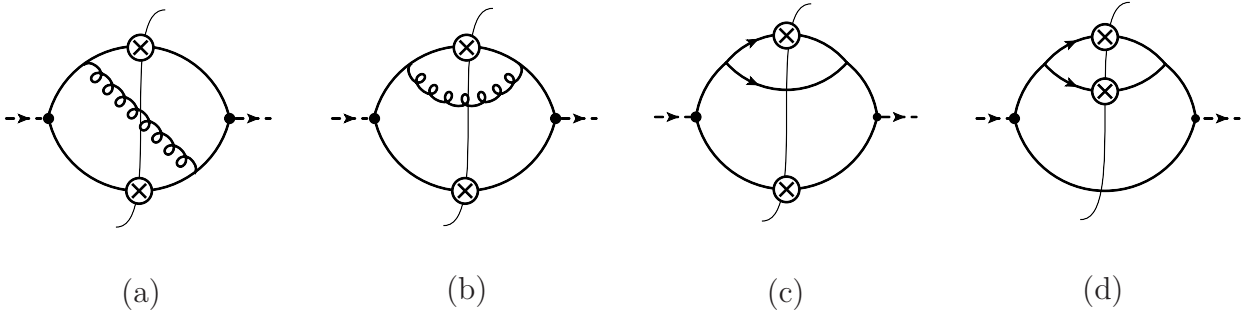}} 
\caption{\small One-loop correction to scalar-scalar and charge-charge correlations. Crosses denote detectors and the thin   line stands for the unitarity cut.}
\label{fig:QQC}
\end{figure}

To one-loop order, the scalar-scalar correlation $\vev{\mathcal O(\vec n) \mathcal O(\vec n')}$ receives contributions only 
from the final state $\ket{{\rm ssg}}$ involving a  pair of scalars and a gluon. The relevant Feynman diagrams are shown in Figs.~\ref{fig:QQC}(a) and (b).
For the charge-charge correlation $\vev{\mathcal Q(\vec n) \mathcal Q(\vec n')}$ there is an additional contribution from the state 
$\ket{{\rm s\lambda\lambda}}$ and it comes from the diagrams shown in Figs.~\ref{fig:QQC}(c) and (d). To compute their contribution, we have to 
combine the transition amplitudes given by \re{mat-el} with the corresponding $SU(4)$ weight factors and perform the integration over the 
phase space of the particles in the final state.

Let us start with the scalar-scalar correlation. It is simpler to do the calculation using the $SO(6)$ notation. To compute the $R-$symmetry
factor corresponding to the diagrams in Figs.~\ref{fig:QQC}(a) and (b) it suffices to perform a Wick contraction between the scalar fields
in the source $O(0,Y)=\tr[(Y\Phi)^2]$ and the sink $O(x,\widebar Y)=\tr[(\widebar Y\Phi)^2]$ and the creation/annihilation operators in the expression for the scalar flow operator \re{231},
$\mathcal{O}(\vec n,S)\sim S_{IJ} a^{\dagger I} a^J$:
\begin{align}
2\vev{(\widebar Y\Phi)\mathcal{O}(\vec n,S)(Y\Phi)}
\vev{(\widebar Y\Phi)  \mathcal{O}(\vec n',S')(Y\Phi)} \sim 2(Y^I S_{IJ} \widebar Y^J)(Y^{I'} S'_{I'J'} \widebar Y^{J'}) =2(Y\widebar Y)^2 \vev{S}\vev{S'}\,,
\end{align}
where the notation was introduced for $\vev{S} = (Y^I S_{IJ} \widebar Y^J)/(Y\widebar Y)$ and $(Y\widebar Y)=Y^I \widebar Y^I$. 
Here the overall factor of 2 takes into account the symmetry of the diagrams under the exchange of the detectors.
The two diagrams in Figs.~\ref{fig:QQC}(a) and (b)
have the same $R-$factor and, therefore, the one-loop correction to the scalar-scalar correlation has the factorized form  
\begin{align}\notag\label{app:SSC-1}
\vev{\mathcal O(\vec n)\mathcal O(\vec n')}=\sigma_{\rm tot}^{-1} \int   \text{dPS}_{3}  (k_1^0\, k_2^0)^{-1} \,
(Y S \widebar Y)(Y S' \widebar Y)  |\mathcal M_{O_{\bf 20'}\to {\rm s}(k_1){\rm s}(k_2){\rm g}(k_3)}|^2 
\\
 \times    
 \bigg[\delta^{(2)}(\Omega_{\vec k_1}-\Omega_{\vec n})\delta^{(2)}(\Omega_{\vec k_2}-\Omega_{\vec n'})
+\delta^{(2)}(\Omega_{\vec k_1}-\Omega_{\vec n'})\delta^{(2)}(\Omega_{\vec k_2}-\Omega_{\vec n})\bigg]
 \,,
\end{align}
where the matrix element is given by \re{mat-el} and the phase space integral is defined in \re{LIPS}. 

To perform the integration over the phase space of the three particles in \re{app:SSC-1} it is convenient to 
introduce the two-particle invariant masses $s_{ij}=(k_i+k_j)^2$ with  $k_1+k_2+k_3=q$.  In the rest frame 
of the source $q^\mu=(q_0,\vec 0)$ we have
\begin{align}\label{tau}
s_{12} = q_0^2(1-\tau_3)\,,\qquad s_{23} = q_0^2(1-\tau_1)\,,\qquad s_{13} = q_0^2(1-\tau_2)\,,
\end{align}
where the variables $\tau_i = 2 k_{i,0}/q_0$ are related to the energy of the particles and satisfy the conditions
 $0\le \tau_i\le 1$ and $\tau_1+\tau_2+\tau_3=2$.  Then, 
\begin{align}\label{PS3}
 \int   \text{dPS}_{3}\, \delta^{(2)}(\Omega_{\vec k_1}-\Omega_{\vec n})\delta^{(2)}(\Omega_{\vec k_2}-\Omega_{\vec n'}) = {q_0^2 \over 64 (2\pi)^5}  \int_0^1 d\tau_1 d\tau_2 \, \tau_1 \tau_2\  
 \delta(1-\tau_1-\tau_2 + \tau_1 \tau_2 z )\,,
\end{align}
where $z=(1-\cos\theta)/2$. In this way, we obtain from \re{app:SSC-1} 
\begin{align}
 \vev{\mathcal O(\vec n)\mathcal O(\vec n')} & =  {g^2 \over 32 \pi^4  }{(Y S \widebar Y)(Y S' \widebar Y) \over  q_0^{2} (Y\widebar Y)^2 } \int_0^1 d\tau_1 d\tau_2 \, { \tau_1+\tau_2-1
 \over (1-\tau_1)(1-\tau_2)} 
 \delta(1-\tau_1-\tau_2 + \tau_1 \tau_2 z )
 \end{align}
and arrive at \re{OO}.

The analysis of the charge-charge correlations goes along the same lines. To one-loop order, $\vev{\mathcal Q(\vec n) \mathcal Q(\vec n')}$ receives contributions from the diagrams shown in Fig.~\ref{fig:QQC}(a) -- (d). In this case, for computing the $R-$symmetry factor, it is convenient to use the $SU(4)$ representation for the source and sink operators, Eq.~\re{O-SU(4)}. For the diagrams in Fig.~\ref{fig:QQC}(a) and (b)  we 
use the scalar part of the charge flow operator 
$\mathcal Q(\vec n) \sim Q^A_B \,a_{AC}^\dagger a^{CB}$, Eqs.~\re{Q-osc} and \re{Qq}, and perform the Wick contractions  
\begin{align}\notag\label{R-factor}
Q^{A_3}_{B_3} & \vev{y_{A_1B_1} \phi^{A_1B_1}   a^\dagger_{A_3C_3}}  \vev{a^{C_3B_3} \bar y^{A_2B_2} \bar\phi_{A_2B_2}}
(Q')^{A_4}_{B_4}
\vev{y_{A_1'B_1'} \phi^{A_1'B_1'}  a^\dagger_{A_4C_4}}\vev{a^{C_4B_4} \bar y^{A_2'B_2'} \bar\phi_{A_2'B_2'}}
\\ \notag
& \qquad \sim Q^{A_3}_{B_3}(Q')^{A_4}_{B_4}y_{A_1B_1}  \bar y^{A_2B_2} y_{1,A_1'B_1'}\bar y^{A_2'B_2'}
\delta_{A_3}^{A_1}\delta_{C_3}^{B_1}\delta_{A_2}^{C_3}\delta_{B_2}^{B_3}\delta_{A_4}^{A_1'}\delta_{C_4}^{B_1'}
\delta_{A_2'}^{C_4}\delta_{B_2'}^{B_4}
\\ 
&\qquad  \sim Q^{A_3}_{B_3} y_{A_3B_1}  \bar y^{B_1B_3} (Q')^{A_4}_{B_4}y_{A_4B_1'}  \bar y^{B_1'B_4}
\equiv \tr \lr{\bar y Q y}\tr\lr{\bar y Q' y}\,.
\end{align}
For the diagrams in  Figs.~\ref{fig:QQC}(c) and (d) we have to use the gluino part of the charge flow operator, $\mathcal Q(\vec n) \sim Q^A_B\,a_{A,1/2}^\dagger a_{-1/2}^{B}$. Going through the calculation
of the $SU(4)$ factors,  we find that for the diagram in Fig.~\ref{fig:QQC}(c) it is given by \re{R-factor}
whereas for the diagram in Fig.~\ref{fig:QQC}(d) we have
\begin{align}\label{idyQQ}
Q^{A_3}_{B_3} (Q')^{A_4}_{B_4} y_{A_3A_4} \bar y^{B_3B_4}
\equiv
 \tr\left(\bar y  Q  y  Q' \right) \tr\left( \bar y  y \right) =  \tr\left( \bar y  Q' y Q \right) \tr\left( \bar y  y \right)=  \tr\left(y Q \bar y  Q'  \right) \tr\left( \bar y  y \right) \,,
\end{align}
where the second relation follows from antisymmetry of $y$ and $\bar y$ under exchange of the $SU(4)$ indices. There is one more diagram of the type~\ref{fig:QQC}(d) in which the fermion lines have opposite orientation. Its $SU(4)$
factor can be obtained by the substitution  $y_{AB} \to  \tilde {\bar y}_{AB}\equiv =\frac12 \epsilon_{ABCD} \bar y^{CD}$ and $\bar y^{AB} \to \tilde y^{AB} \equiv \frac12 \epsilon^{ABCD} y_{CD}$,
 and reads $ \tr\left(\tilde {\bar y} Q \tilde y  Q'  \right) \tr\left( \bar y  y \right)$. Then, the one-loop correction to $\vev{\mathcal Q(\vec n)\mathcal Q(\vec n')}$ takes the following form
\begin{align} \label{QQ-1loop}
\vev{\mathcal Q(\vec n)\mathcal Q(\vec n')}=   4  {\tr( y Q \bar y)\over \tr(y\bar y)}   {\tr( y Q' \bar y) \over \tr(y\bar y)}
 \left(I_{a+b} + I_c\right)
 + {\tr\left[y Q' \bar y Q {+}{\tilde {\bar y} Q' \tilde y  Q } \right] \over {2}\tr(y\bar y)} I_{d} \,,
\end{align}   
 where $I_{a+b}$ denotes the contribution of the diagrams shown in Figs.~\ref{fig:QQC}(a) and (b) 
 \begin{align}
I_{a+b}  ={2 [\tr(\bar y y)]^{2} \over \sigma_{\rm tot}} \int   \text{dPS}_{3}\,\delta^{(2)}(\Omega_{\vec k_1}-\Omega_{\vec n})\delta^{(2)}(\Omega_{\vec k_2}-\Omega_{\vec n'})
|\mathcal M_{O_{\bf 20'}\to {\rm s}(k_1){\rm s}(k_2){\rm g}(k_3)}|^2\,,
\end{align}
 and similarly for $I_{c}$ and $I_d$,
\begin{align}\notag
I_{c}  = {2 [\tr(\bar y y)]^{2} \over \sigma_{\rm tot}} \int   \text{dPS}_{3}\,\delta^{(2)}(\Omega_{\vec k_1}-\Omega_{\vec n})\delta^{(2)}(\Omega_{\vec k_2}-\Omega_{\vec n'})
|\mathcal M_{O_{\bf 20'}\to {\rm s}(k_1)\lambda(k_2)\lambda(k_3)}|^2\,,
\\
I_d = {2 [\tr(\bar y y)]^{2} \over \sigma_{\rm tot}} \int   \text{dPS}_{3}\,\delta^{(2)}(\Omega_{\vec k_2}-\Omega_{\vec n})\delta^{(2)}(\Omega_{\vec k_3}-\Omega_{\vec n'})
|\mathcal M_{O_{\bf 20'}\to {\rm s}(k_1)\lambda(k_2)\lambda(k_3)}|^2\,.
\end{align}
The evaluation of these integrals goes along the same lines as before: we go to the new variables \re{tau}, replace the matrix elements
by their explicit expressions  \re{mat-el} and integrate over the phase space using the relation \re{PS3}. In this way, we obtain
\begin{align}\notag
I_{a+b} &= {a \over 8 \pi^2}  { (z-2)\ln(1-z)-2 z\over   z^2(1-z)}\,,
\\\notag
I_c &=  {a \over 4 \pi^2}{ (1-z)\ln(1-z) + z \over  z^2(1-z)}\,,
\\
I_d &= - {a \over 4 \pi^2} {\ln(1-z) \over   z^2 }\,.
\end{align}
The substitution of these relations into \re{QQ-1loop} yields \re{C.6}.

Finally, the one-loop correction to the energy-energy correlation $\vev{\mathcal E(n) \mathcal E(n')}$ is given by \re{EE-aux}.
We replace the transition amplitudes  in \re{EE-aux} by their explicit expressions \re{mat-el} and take into account the symmetry of the integration measure under the exchange of any pair of particles to obtain  
\begin{align} 
\vev{\mathcal E(n) \mathcal E(n')} =  16\pi g^2 \int {\rm d PS_3} \, k_1^0 k_2^0 \,\delta^{(2)}(\Omega_{\vec k_1}-\Omega_{\vec n})\delta^{(2)}(\Omega_{\vec k_2}-\Omega_{\vec n'})
 {4(q^2)^2 \over s_{12}  s_{13}s_{23}}  \,,
\end{align}
where the last factor has the same origin as the one in \re{tot-zero}. Using \re{tau} and \re{PS3}, we find
\begin{align}\notag
\vev{\mathcal E(n) \mathcal E(n')} &= { g^2 q_0^2 \over  8 (2\pi)^4}  \int_0^1 { d\tau_1 d\tau_2 \, (\tau_1 \tau_2)^2  \over (1-\tau_1)(1-\tau_2)(\tau_1+\tau_2-1)}
 \delta(1-\tau_1-\tau_2 + \tau_1 \tau_2 z )
 \\
 &= { g^2 q_0^2 \over  8 (2\pi)^4} {1\over z(1-z)} \int_0^1 {d\tau_1\over 1- z\tau_1}\,.
\end{align}
Replacing $z=(1-\cos\theta)/2$ we arrive at \re{EE-1}.

\section{$R-$symmetry invariant structures}\label{appE}

In this appendix, we discuss explicit expressions for the $R-$symmetry factors $\omega_R$ in (\ref{Fgen}). We will use the same notation as in the amplitude computations, but in addition we also introduce
\begin{align}
&\vev{S}=\frac{ ({Y} S \overline Y)}{(Y\widebar Y)}\,,&&[S]=\frac{ (Y S Y)}{(Y\widebar Y)} \,,&&\overline{[S]}=\frac{ (\widebar Y S \widebar Y)}{(Y\widebar Y)}\,, \nonumber\\ 
&\vev{S S'} =\frac{({Y} SS' \overline Y)}{(Y\overline Y)}\,,&&\overline{\vev{S S'}} =\frac{({Y} S'S \overline Y)}{(Y\overline Y)} \,,&& (SS') = \tr(SS')\,.
\end{align}With this notation, we find the following explicit expressions for $\omega^{\mathcal{OO}}_R$:  
\begin{align} \label{6.17}
&  \omega^{\mathcal{OO}}_{\mathbf{1}}=\tfrac{1}{6}(SS')\,,\nonumber\\[1.5mm] 
& \omega^{\mathcal{OO}}_{\mathbf{15}}=\tfrac{1}{6}\big\{ { \vev{S S'}- \overline{\vev{S S'}}   } \big\}\,, \nonumber\\[1.5mm] 
& \omega^{\mathcal{OO}}_{\mathbf{20'}}=\tfrac{1}{6}\big\{{ \vev{S S'}+ \overline{\vev{S S'}}   } -\tfrac1{3}(SS')\big\}\,, \nonumber\\[1.5mm] 
& \omega^{\mathcal{OO}}_{\mathbf{84}}=\tfrac{1}{6}\big\{{  [S]  \overline{[S']} +    \overline{[S]} [S'] -2\vev{S} \vev{S' }}   -\tfrac1{2}\big( { \vev{S S'}+ \overline{\vev{S S'}}   }\big)  +\tfrac1{10}(SS') \big\}\,,\nonumber\\[1.5mm] 
& \omega^{\mathcal{OO}}_{\mathbf{175}}=\tfrac{1}{6}\big\{ {  [S]   \overline{[S']} -   \overline{[S]} [S']}   
 -\tfrac1{2} \big( {  \vev{S S'}- \overline{\vev{S S'}}}\big)\big\}\,, \nonumber\\[1.5mm] 
& \omega^{\mathcal{OO}}_{\mathbf{105}}=\tfrac{1}{6}\big\{
{  [S]  \overline{[S']} +    \overline{[S]} [S'] +4\vev{S} \langle S' \rangle}   -\tfrac4{5}\big({  \vev{S S'}+ \overline{\vev{S S'}}   }\big) +\tfrac1{10}(SS') \big\}\,.
\end{align}

We recall that in order to rewrite the basis structures $Y_R$ in the correlation function notation of Sect.~\ref{s5.2}, one has to make the replacements $S^{IJ} \to Y^I_2 Y^J_2$, $(S')^{IJ} \to Y^I_3 Y^J_3$, $Y\to Y_4$, $\overline Y \to Y_1$. Converting the structures \p{6.17} into correlation function ones, we get 
\begin{align}\label{}
 \omega^{\mathcal{OO}}_R \to (Y_2  Y_3)^2 \cY_R(t_1, t_2)\,, 
\end{align} 
with  $t_1=  (Y_1  Y_2) (Y_3  Y_4) /\lr{(Y_1  Y_4) (Y_2  Y_3)}$,   $t_2= {(Y_1  Y_3) (Y_2  Y_4)}/\lr{(Y_1  Y_4) (Y_2  Y_3)}$ and
\begin{align}\label{E.6}
&  \cY_{\mathbf{1}}=1 \notag\\[1.5mm]
& \cY_{\mathbf{15}}= t_1-t_2 \notag\\[1.5mm]
& \cY_{\mathbf{20'}}=t_1+t_2-\tfrac1{3} \notag\\[1.5mm]
& \cY_{\mathbf{84}}=(t_1-t_2)^2 -\tfrac1{2}(t_1+t_2) +\tfrac1{10} \notag\\[1.5mm]
& \cY_{\mathbf{175}}= t_1^2-t_2^2 -\tfrac1{2}(t_1-t_2) \notag\\[1.5mm]
& \cY_{\mathbf{105}}=t_1^2+t_2^2 +4t_1t_2 -\tfrac4{5}(t_1+t_2)+\tfrac1{10}\,.
\end{align}
The polynomials $\cY_R$ coincide with the eigenfunctions $Y_{nm}(t_1, t_2)$ of the quadratic Casimir operator of $SU(4)$ for the irreps with Dynkin labels $[n-m,2m,n-m]$, as listed in Appendix B in \cite{Nirschl:2004pa}.

Further, the condition \p{cross-talk} for absence of cross-talk between the two scalar detectors, $[S,S']=0$, is translated into $(Y_2  Y_3) (Y^I_2 Y^J_3 - Y^J_2 Y^I_3)=0$. This condition has two solutions, the weaker constraint $(Y_2  Y_3)=0$ or the stronger $Y_2 = Y_3$ \footnote{The variables $Y_i$ are projective, so without loss of generality we can set $Y_2 = Y_3$ instead of $Y_2 \propto Y_3$.}.  Inserting these constraints back into \p{6.17}, we find that the weaker one implies
\begin{align}\label{E5}
 & (Y_2  Y_3)=0 \ \rightarrow \ (SS')=\vev{SS'}=\vev{S'S}=0  \ \rightarrow \  \omega^{\mathcal{OO}}_{\mathbf{1}}= \omega^{\mathcal{OO}}_{\mathbf{15}}= \omega^{\mathcal{OO}}_{\mathbf{20'}}=0\,,
\end{align}
while the stronger constraint yields, in addition to \p{E5},
\begin{align}\label{E6}
&Y_2 =  Y_3\ \rightarrow \ {  [S]   \overline{[S']} =   \overline{[S]} [S']} = \vev{S} \vev{S' } \ \rightarrow \  \omega^{\mathcal{OO}}_{\mathbf{84}}= \omega^{\mathcal{OO}}_{\mathbf{175}}=0\,. 
\end{align}
Thus, the strong version \p{E6} of the condition for absence of cross-talk results in only one surviving $R-$symmetry structure, $\omega_{\mathbf{105}}^{\mathcal{OO}} \neq 0$. 

Inversely, we can ask the question which conditions on the projection variables $Y$ and on the detector matrices $S$ and $S'$ eliminate all $R-$symmetry structures but $\omega_{\mathbf{105}}^{\mathcal{OO}}$. Assuming that the detector matrices satisfy the stronger form of the no-cross-talk condition, $SS'=0$ (instead of $[S,S']=0$), from \p{6.17} we derive the additional conditions
\begin{align}\label{}
[S]   \overline{[S']} =  \overline{[S]} [S']=\vev{S} \vev{S' }\,.
\end{align}
These can be solved by, e.g., the following choice of the auxiliary variables (used in Ref.~\cite{PaperI}):
\begin{align}\label{}
Y=(1,0,1,0,i,i)\,,  \qquad S = {\rm diag} (1, - 1, 0, 0, 0 ,0) \, , \qquad S' = {\rm diag} (0, 0, 1, -1, 0 ,0)\,.
\end{align}

In a similar fashion, we can deal with basis structures in the case of charge-scalar correlations. Indeed, we can write (up to normalization)
\begin{align}\notag
\omega_{\mathbf{15}}^{\mathcal{QO}}&= -4\big[\vev{QS} + \overline{\vev{QS}}\big] \,,\\[1.5mm]
\notag
\omega_{\mathbf{20'}}^{\mathcal{QO}}&= 4\big[ \vev{QS} - \overline{\vev{QS}}\big]  \,,\\[1.5mm] 
\omega_{\mathbf{175}}^{\mathcal{QO}}&=  4\big[ \vev{S} \vev{Q}-\tfrac{1}{8}\big\{\vev{QS} + \overline{\vev{QS}}\big\}\big] \,.
\end{align}
Here $\vev{QS} = Y^I Q^{IJ} S^{KL} \widebar Y^L$, and we have rewritten the detector matrix $Q^{IJ}=-Q^{JI} = (\Gamma^{IJ})_A^B Q^A_B$ in $SO(6)$ notation, with the help of the $SO(6)$ gamma matrices defined in \p{SM}. 

In the case of the charge-charge correlations, the relevant $R-$invariant building blocks can be written in the form
\begin{align}\label{D8}
&\mathcal{Z}_1= \tr(Q Q') \,,&&\mathcal{Z}_2=\frac{ \tr(y Q \bar y) \tr(y Q' \bar y)}{[\tr(y\bar y)]^2} \,,\nt
&\mathcal{Z}_3=\frac{ \tr(y Q Q'  \bar y)}{\tr(y\bar y)}\,,&&\mathcal{Z}_4=\frac{ \tr(y Q' Q  \bar y)}{\tr(y\bar y)}\,, &&\mathcal{Z}_5=\frac{ \tr(y Q \bar y Q')}{\tr(y\bar y)}\,.
\end{align} 
The additional structure appearing in the second term in  \p{Q-factor} is not independent: 
\begin{align}\label{}
\frac{\tr\left(\tilde {\bar y} Q \tilde y  Q'  \right)}{\tr(y\bar y)} = \mathcal{Z}_5-\tfrac1{2} \mathcal{Z}_1 -(\mathcal{Z}_3 + \mathcal{Z}_4)\quad \rightarrow \quad \vev{Q,Q'} = 2\mathcal{Z}_5-\tfrac1{2} \mathcal{Z}_1 -(\mathcal{Z}_3 + \mathcal{Z}_4)\,.
\end{align}
The $R-$symmetry factors $\omega_R$ are constructed from them as follows (up to normalization): 
\begin{align}\label{D9}
&\omega_{\mathbf{1}}^{\mathcal{QQ}}=-\ft15 \mathcal{Z}_1  \,,\notag \\[1mm]
&\omega_{\mathbf{15_a}}^{\mathcal{QQ}}=-\ft15\big[ \mathcal{Z}_3-\mathcal{Z}_4 \big]\,, \notag  \\[1mm]
&\omega_{\mathbf{15_s}}^{\mathcal{QQ}}= -\ft15\big[{\cal Z}_1-4 ({\cal Z}_3 + {\cal Z}_4) \big]\,,\notag  \\[1mm]
&\omega_{\mathbf{20'}}^{\mathcal{QQ}}= -\ft15\big[{\cal Z}_1 - 3 ({\cal Z}_3 + {\cal Z}_4) +6 {\cal Z}_5 \big]\,, \notag  \\[1mm]
&\omega_{\mathbf{84}}^{\mathcal{QQ}}=-\ft15\big[ {\cal Z}_1 - 40  {\cal Z}_2 -5 ({\cal Z}_3 + {\cal Z}_4) +10 {\cal Z}_5 \big] \,.
\end{align}
Finally, the two correlations involving energy detectors have particularly simple $R-$symmetry factors:
\begin{align}\label{E.7}
\omega_{\mathbf{1}}^{\mathcal{EE}}=   1\,, \qquad \omega_{\mathbf{15}}^{\mathcal{EQ}}=  8\vev{Q}\,, \qquad \omega^{\mathcal{EO}}_{\mathbf{20'}} =  2\vev{S}\,.
\end{align}

\section{Scalar detector in Mellin space}\label{app:K}

In this Appendix we compute the function $K (j_1,j_2;z)$ defined in \re{K-fun}.  Replacing the function $f(j_1,j_2;x_1,x_{2-},x_{3-})$ in \re{K-fun}
with its explicit expression \re{f} we find  
\begin{align}\notag\label{K-start}
 K (j_1,j_2;z) & = {{1 \over 16 \pi ^5}} q^2  ( (nn')/2 )^{-j_1-j_2+1}  \int { d^4x_1 \e^{iqx_1} \over (-x_{1}^2+i\epsilon x_{1}^0)^{j_1+j_2}} \\[1.5mm] \notag
& \times  \int_{-\infty}^\infty  dx_{2-}  ((x_1n)-x_{2-}-i\epsilon\big)^{j_1-1}  (-x_{2-}+i\epsilon\big)^{j_2-1}
\\[1.5mm] &\times \int_{-\infty}^\infty  dx_{3-}    
  ((x_1n')-x_{3-}-i\epsilon )^{j_2-1}    (-x_{3-}+i\epsilon )^{j_1-1}\,.
\end{align}
The evaluation of the $x_--$integrals is straightforward, making use of the Schwinger representation for the factors involved
\begin{align}\nonumber
&\int_{- \infty}^{\infty} dx_{2-} ( (x_1 n) - x_{2-}  - i0)^{j_1 - 1}(-x_{2-} + i0)^{j_2 - 1} 
\\
&\qquad\qquad\qquad
=
  \frac{2 \pi \,i^{j_2 - j_1}}{\Gamma (1 - j_1)\Gamma (1 - j_2)} 
\int_0^\infty d \omega_1 \omega_1^{- j_1 - j_2} {\rm e}^{-i \omega_1  (x_1 n)}
\, . 
\end{align}
Here we left the final integration intact, which helps in evaluating the $x_{1}-$integral in \re{K-start},
\begin{align}\label{inter}
K (j_1,j_2;z) & =  {q^2( (nn')/2 )^{-j_1-j_2+1} \over 4 \pi^3 [\Gamma (1 - j_1)\Gamma (1 - j_2)]^2}\int_0^\infty d \omega_1 d \omega_2(\omega_1\omega_2)^{- j_1 - j_2} 
D_{j_1+j_2}(q-\omega_1 n - \omega_2 n')
\end{align}
where the notation was introduced for
\begin{align}
D_j(q) = \int { d^4x_1 \e^{iqx_1} \over (-x_{1}^2+i\epsilon x_{1}^0)^{j}}  = 
2\pi^3 {(q^2/4)^{j-2} \theta(q^0)\theta(q^2) \over \Gamma(j)\Gamma(j-1)}\,.
\end{align}
Computing the $\omega-$integrals in \re{inter} we find
\begin{align}
K (j_1,j_2;z) & = \lr{z\over 1-z}^{1-j_1-j_2} {{2\pi} \over \sin(\pi(j_1+j_2))[\Gamma(j_1+j_2)\Gamma (1 - j_1)\Gamma (1 - j_2)]^2}\,.
\end{align}



\begin{thebibliography}{100}
%
\bibitem{Sterman:1977wj} 
G.F.~Sterman, S.~Weinberg,
``Jets from Quantum Chromodynamics",
Phys.\ Rev.\ Lett.\  {\bf 39} (1977) 1436.
%
\bibitem{Kunszt:1989km}
Z.~Kunszt, P.~Nason, G.~Marchesini, B.R.~Webber,
``QCD at LEP,''
In Geneva 1989, Proceedings, Z physics at LEP~1, vol.~1, pp.~373-453.
%
\bibitem{Biebel:2001dm}
O.~Biebel,
``Experimental tests of the strong interaction and its energy dependence in electron positron annihilation,'' 
Phys.\ Rept.\  {\bf 340} (2001) 165.
%
\bibitem{Basham:1978bw}
 C.L.~Basham, L.S.~Brown, S.D.~Ellis, S.T.~Love,
``Energy correlations in electron-positron annihilation: testing QCD,''
Phys.\ Rev.\ Lett.\  {\bf 41} (1978) 1585;
%
``Electron-positron annihilation energy pattern in Quantum Chromodynamics: asymptotically free perturbation theory,''
Phys.\ Rev.\ D {\bf 17} (1978) 2298; 
%
``Energy correlations in electron-positron annihilation in Quantum Chromodynamics: asymptotically free perturbation theory,''
Phys.\ Rev.\ D {\bf 19} (1979) 2018.
%
\bibitem{Dasgupta:2003iq}
  M.~Dasgupta and G.~P.~Salam,
  ``Event shapes in e+ e- annihilation and deep inelastic scattering,''
  J.\ Phys.\ G {\bf 30} (2004) R143
  [hep-ph/0312283].
%
\bibitem{Kinoshita:1962ur}  
 T.~Kinoshita,
``Mass singularities of Feynman amplitudes,''
J.\ Math.\ Phys.\  {\bf 3} (1962) 650.
%
\bibitem{Lee:1964is}
T.D.~Lee, M.~Nauenberg,
``Degenerate Systems and Mass Singularities,''
Phys.\ Rev.\  {\bf 133} (1964) B1549.
 %
\bibitem{vanNeerven:1985ja}
  W.L.~van Neerven,
  ``Infrared behavior of on-shell form-factors in  N=4 supersymmetric Yang-Mills field theory,''
  Z.\ Phys.\ C {\bf 30} (1986) 595.
%
\bibitem{Sveshnikov:1995vi}  
N.A. Sveshnikov, F.V.~Tkachov,
``Jets and quantum field theory,'' 
Phys.\ Lett.\ B {\bf 382}  (1996) 403 [hep-ph/9512370].
%
\bibitem{Korchemsky:1997sy} 
G.P.~Korchemsky, G.~Oderda, G.F.~Sterman,
``Power corrections and nonlocal operators,''
hep-ph/9708346.
%
\bibitem{Korchemsky:1999kt}
G.P.~Korchemsky, G.F.~Sterman,
  ``Power corrections to event shapes and factorization,''
  Nucl.\ Phys.\ B {\bf 555} (1999) 335
  [hep-ph/9902341].
%
\bibitem{BelKorSte01}
A.V. Belitsky, G.P. Korchemsky, G. Sterman,
``Energy flow in QCD and event shape functions,''
Phys.\ Lett.\ B {\bf 515} (2001) 297 [hep-ph/0106308].
%
\bibitem{Hofman:2008ar}
  D.M.~Hofman, J.~Maldacena,
  ``Conformal collider physics: Energy and charge correlations,''
  JHEP {\bf 0805} (2008) 012
  [arXiv:0803.1467 [hep-th]].
%
\bibitem{PaperI}
A.V.~Belitsky, S.~Hohenegger, G.P.~Korchemsky, E.~Sokatchev, A.~Zhiboedov,
  ``From correlation functions to event shapes,''
  arXiv:1309.0769 [hep-th].
%
\bibitem{Howe:1996rb}
P.S.~Howe, P.C.~West,
``Nonperturbative Green's functions in theories with extended superconformal symmetry,''
Int.\ J.\ Mod.\ Phys.\ A {\bf 14} (1999) 2659 [hep-th/9509140];
%
``Operator product expansions in four-dimensional superconformal field theories,''
Phys.\ Lett.\ B {\bf 389} (1996) 273 [hep-th/9607060].
%
\bibitem{D'Hoker:1998tz}
E.~D'Hoker, D.Z.~Freedman, W.~Skiba,
``Field theory tests for correlators in the AdS / CFT correspondence,''
Phys.\ Rev.\ D {\bf 59} (1999) 045008 [hep-th/9807098].
%
  \bibitem{Howe:1998zi}
  P.S.~Howe, E.~Sokatchev, P.C.~West,
  ``Three point functions in N=4 Yang-Mills,''
  Phys.\ Lett.\ B {\bf 444} (1998) 341
  [hep-th/9808162].
%
\bibitem{Lee:1998bxa}
  S.~Lee, S.~Minwalla, M.~Rangamani, N.~Seiberg,
  ``Three point functions of chiral operators in D = 4, N=4 SYM at large N,''
  Adv.\ Theor.\ Math.\ Phys.\  {\bf 2} (1998) 697
  [hep-th/9806074].
%
\bibitem{Penati:1999ba}
S.~Penati, A.~Santambrogio, D.~Zanon,
``Two point functions of chiral operators in N=4 SYM at order $g^4$,''
JHEP {\bf 9912} (1999) 006 [hep-th/9910197]; 
%
 ``More on correlators and contact terms in N=4 SYM at order g**4,''
  Nucl.\ Phys.\ B {\bf 593} (2001) 651
  [hep-th/0005223].
  %
\bibitem{Collins:1989bt}
J.C.~Collins,
``Sudakov form-factors,''
Adv.\ Ser.\ Direct.\ High Energy Phys.\  {\bf 5} (1989) 573 [hep-ph/0312336].
%
\bibitem{Kunszt:1992tn}
  Z.~Kunszt, D.E.~Soper,
  ``Calculation of jet cross sections in hadron collisions at order $\alpha_s^3$,''
  Phys.\ Rev.\ D {\bf 46} (1992) 192.
%
\bibitem{Galperin:1984av}
  A.~Galperin, E.~Ivanov, S.~Kalitsyn, V.~Ogievetsky, E.~Sokatchev,
  ``Unconstrained N=2 Matter, Yang-Mills and Supergravity Theories in Harmonic Superspace,''
  Class.\ Quant.\ Grav.\  {\bf 1} (1984) 469.
%
  \bibitem{Howe:1995md}
  P.S.~Howe, G.G.~Hartwell,
  ``A Superspace survey,''
  Class.\ Quant.\ Grav.\  {\bf 12} (1995) 1823.  
 %
\bibitem{Salam}
A.~Banfi, G.P.~Salam, G.~Zanderighi,
``Infrared safe definition of jet flavor,''
Eur.\ Phys.\ J.\ C {\bf 47} (2006) 113
[hep-ph/0601139].
%
\bibitem{Collins:1981uk}
J.C.~Collins, D.E.~Soper,
``Back-To-Back Jets in QCD,''
Nucl.\ Phys.\ B {\bf 193} (1981) 381 [Erratum-ibid.\ B {\bf 213} (1983) 545, Nucl.\ Phys.\ B {\bf 213} (1983) 545];
  %
``Back-To-Back Jets: Fourier Transform from B to K-Transverse,''
Nucl.\ Phys.\ B {\bf 197} (1982) 446.
 %
\bibitem{Engelund:2012re}
  O.~T.~Engelund, R.~Roiban,
  ``Correlation functions of local composite operators from generalized unitarity,''
  JHEP {\bf 1303} (2013) 172
  [arXiv:1209.0227 [hep-th]].
%
\bibitem{Eden:1999gh}
  B.~Eden, P.S.~Howe, P.C.~West,
  ``Nilpotent invariants in N=4 SYM,''
  Phys.\ Lett.\ B {\bf 463} (1999) 19
  [hep-th/9905085].
 %
\bibitem{Eden:1999kw}
  B.~Eden, P.S.~Howe, C.~Schubert, E.~Sokatchev, P.C.~West,
  ``Extremal correlators in four-dimensional SCFT,''
  Phys.\ Lett.\ B {\bf 472} (2000) 323
  [hep-th/9910150].
 %
\bibitem{GonzalezRey:1998tk}
  F.~Gonzalez-Rey, I.Y.~Park, K.~Schalm,
  ``A Note on four point functions of conformal operators in N=4 superYang-Mills,''
  Phys.\ Lett.\ B {\bf 448} (1999) 37
  [hep-th/9811155].
%
\bibitem{Eden:1998hh}
  B.~Eden, P.S.~Howe, C.~Schubert, E.~Sokatchev, P.C.~West,
  ``Four point functions in N=4 supersymmetric Yang-Mills theory at two loops,''
  Nucl.\ Phys.\ B {\bf 557} (1999) 355
  [hep-th/9811172].
  %
 \bibitem{Eden:1999kh}
  B.~Eden, P.S.~Howe, C.~Schubert, E.~Sokatchev, P.C.~West,
  ``Simplifications of four point functions in N=4 supersymmetric Yang-Mills theory at two loops,''
  Phys.\ Lett.\ B {\bf 466} (1999) 20
  [hep-th/9906051].
%
\bibitem{Eden:2000bk}
B.~Eden, A.C.~Petkou, C.~Schubert, E.~Sokatchev,
``Partial nonrenormalization of the stress tensor four point function in N=4 SYM and AdS / CFT,''
Nucl.\ Phys.\ B {\bf 607} (2001) 191 [hep-th/0009106].
%
\bibitem{Eden:2000mv}
  B.~Eden, C.~Schubert, E.~Sokatchev,
  ``Three loop four point correlator in N=4 SYM,''
  Phys.\ Lett.\ B {\bf 482} (2000) 309
  [hep-th/0003096].
%
\bibitem{Bianchi:2000hn}
  M.~Bianchi, S.~Kovacs, G.~Rossi, Y.S.~Stanev,
  ``Anomalous dimensions in N=4 SYM theory at order g**4,''
  Nucl.\ Phys.\ B {\bf 584} (2000) 216
  [hep-th/0003203].
 %
 \bibitem{Usyukina:1992wz}
  N.I.~Usyukina, A.I.~Davydychev,
  ``Exact results for three and four point ladder diagrams with an arbitrary number of rungs,''
  Phys.\ Lett.\ B {\bf 305} (1993) 136.
  N.I.~Usyukina, A.I.~Davydychev,
  ``Some exact results for two loop diagrams with three and four external lines,''
  Phys.\ Atom.\ Nucl.\  {\bf 56} (1993) 1553
   [Yad.\ Fiz.\  {\bf 56N11} (1993) 172]
  [hep-ph/9307327].
 %
\bibitem{Luscher:1974ez}
  M.~Luscher and G.~Mack, ``Global conformal invariance in Quantum Field Theory,''
  Commun.\ Math.\ Phys.\  {\bf 41} (1975) 203.
 %
\bibitem{Mack:2009mi}
G.~Mack, 
``D-independent representation of Conformal Field Theories in D dimensions via transformation to auxiliary Dual Resonance Models. Scalar amplitudes,'' 
arXiv:0907.2407 [hep-th].
%
\bibitem{Arutyunov:1999fb}
  G.~Arutyunov, S.~Frolov,
  ``Scalar quartic couplings in type IIB supergravity on AdS(5) x S**5,''
  Nucl.\ Phys.\ B {\bf 579} (2000) 117
  [hep-th/9912210].
%
 \bibitem{Arutyunov:2000py} 
 G.~Arutyunov, S.~Frolov,
 "Four point functions of lowest weight CPOs in N=4 SYM(4) in supergravity approximation",
  Phys.\ Rev.\ D {\bf 62} (2000) 064016
  [hep-th/0002170].
%
\bibitem{Arutyunov:2000ku}
G.~Arutyunov, S.~Frolov, A.C.~Petkou, ``Operator product expansion of the lowest weight CPOs in N=4 SYM(4) at strong coupling,''
  Nucl.\ Phys.\ B {\bf 586} (2000) 547
   [Erratum-ibid.\ B {\bf 609} (2001) 539]
  [hep-th/0005182].
%
\bibitem{xAct} J. M. Martin-Garcia, xAct: efficient tensor computer algebra, http://www.xact.es.    
%
\bibitem{GomezLobo:2011xv}
  A.G.-P.~Gomez-Lobo, J.M.~Martin-Garcia,
  ``Spinors: a Mathematica package for doing spinor calculus in General Relativity,''
  Comput.\ Phys.\ Commun.\  {\bf 183} (2012) 2214
  [arXiv:1110.2662 [gr-qc]].
 %
 \bibitem{paper3}
A.V.~Belitsky,  S.~Hohenegger, G.P.~Korchemsky, E.~Sokatchev, A.~Zhiboedov (to appear). 
%
 \bibitem{Eden:2011we}
  B.~Eden, P.~Heslop, G.P.~Korchemsky, E.~Sokatchev,
  ``Hidden symmetry of four-point correlation functions and amplitudes in N=4 SYM,''
  Nucl.\ Phys.\ B {\bf 862} (2012) 193
  [arXiv:1108.3557 [hep-th]].
 %
\bibitem{Keldysh:1964ud}
  L.V.~Keldysh,
  ``Diagram technique for nonequilibrium processes,''
  Zh.\ Eksp.\ Teor.\ Fiz.\  {\bf 47} (1964) 1515
   [Sov.\ Phys.\ JETP {\bf 20} (1965) 1018].
%
\bibitem{BelDerKorMan03}
A.V.~Belitsky, S.E.~Derkachov, G.P.~Korchemsky, A.N.~Manashov,
``Superconformal operators in N=4 superYang-Mills theory,''
Phys.\ Rev.\ D {\bf 70} (2004) 045021 [hep-th/0311104].
%
\bibitem{Nirschl:2004pa}
M.~Nirschl, H.~Osborn,
``Superconformal Ward identities and their solution,''
Nucl.\ Phys.\ B {\bf 711} (2005) 409 [hep-th/0407060].

\end{thebibliography}
\end{document}